\documentstyle[12pt,epsf]{article}
\newcommand{\beq}{\begin{equation}}
\newcommand{\eeq}{\end{equation}}
\newcommand{\beqa}{\begin{eqnarray}}
\newcommand{\eeqa}{\end{eqnarray}}

\newcommand{\ve}{\varepsilon}
\newcommand{\krig}[1]{\stackrel{\circ}{#1}}

\begin{document}


{\small \hfill FZJ-IKP(TH)-1998-08

\hfill TK 98 04}


\bigskip\bigskip

\begin{center}

{{\Large\bf Renormalization of the chiral pion--nucleon\\[0.3cm]
    Lagrangian beyond next--to--leading order}}\footnote{Work supported
    in part by DFG (grants Me 864-11/1,
    Schu 439-10/1), by funds provided by the Graduiertenkolleg "Die Erforschung 
    subnuklearer Strukturen der Materie" at Bonn University and by DAAD.   
}

\end{center}

\vspace{.3in}

\begin{center}
{\large
Ulf-G. Mei{\ss}ner$^\ddagger$\footnote{email: Ulf-G.Meissner@fz-juelich.de},
Guido M\"uller$^\dagger$\footnote{email: 
mueller@itkp.uni-bonn.de}\footnote{Address 
after Nov. 1$^{st}$, 1998: Institut f\"ur Theoretische
Physik, Universit\"at Wien, A--1090 Wien, Austria.},
Sven Steininger$^{\ddagger,\dagger}$\footnote{email:
S.Steininger@fz-juelich.de}}

\bigskip

\bigskip

$^\ddagger${\it Forschungszentrum J\"ulich, Institut f\"ur Kernphysik 
(Theorie)\\ D-52425 J\"ulich, Germany}

\bigskip

$^\dagger${\it Universit\"at Bonn, Institut f{\"u}r Theoretische Kernphysik\\
Nussallee 14-16, D-53115 Bonn, Germany}\\

\end{center}

\vspace{.7in}

\thispagestyle{empty} 

\begin{abstract}
\noindent
The complete renormalization of the generating functional for Green
functions of  quark currents between one--nucleon  states
in two flavor heavy baryon chiral perturbation theory is performed 
to order $q^4$. We show how the heat kernel method has to be extended
for operators orthogonal to the heavy fermion four--velocity. A method
is developed to treat the multi--coincidence limit arising from
insertions of dimension two (and higher) operators on internal baryon 
propagators in self--energy graphs.  As examples, we study the
divergences in the isoscalar magnetic moment and the scalar form
factor of the nucleon.
\end{abstract}

\vspace{1in}


\vfill

\pagebreak



\section{Introduction}

Chiral perturbation theory (CHPT) allows
to systematically investigate the consequences of the spontaneous and
explicit chiral symmetry breaking QCD is believed to undergo.
In the presence of nucleons, such studies can be extended and related to 
a large variety of precisely measured  processes. The basic 
degrees of freedom are the three (almost) massless pseudoscalar Goldstone 
bosons, i.e. the pions, and the spin--1/2 fields, the nucleons, treated
as very heavy, static sources. The corresponding effective field theory
is subject to an expansion in small momenta and quark (meson) masses or
equivalently an expansion in the number of pion loops.
To one loop order, divergences appear. Some of these were treated e.g. in
\cite{bkkm}. A systematic treatment of the leading divergences of the
generating functional for Green functions of quark currents between
one--nucleon states was given in ref.\cite{ecker}. This allows for a
chiral invariant renormalization of {\it all} two--nucleon Green
functions of the pion--nucleon system to order $q^3$ in the
low--energy expansion, were $q$ denotes the various expansion parameters,
in our case the pion energy ($E_\pi$) and mass ($M_\pi$) 
with respect to the scale of chiral
symmetry breaking and with respect to the nucleon mass $m$. Since these
two scales are of comparable size, one effectively has a double expansion
in powers of $E_\pi /m$ and $M_\pi /m$. The study of ref.\cite{ecker}
was extended to the three flavor case to the same order in the chiral
expansion in \cite{guido}.\footnote{An alternative method to work out
  the divergence structure at third order has recently been developed
  in ref.\cite{neu}.} 
However, a series of precise calculations of
single nucleon processes like Compton scattering, neutral and charged
pion photoproduction off nucleons and deuterium
or the chiral corrections to Weinberg's predictions
for the S--wave pion--nucleon scattering lengths have shown that it is 
mandatory to go order $q^4$, for a review see~\cite{bkmrev}. At that order,
one has to construct one loop graphs with exactly one insertion from the
dimension two pion--nucleon Lagrangian and local counterterms with a priori
unknown coupling constants. These allow to absorb the divergences appearing
in the loop diagrams. In this paper, we extend the calculation of \cite{ecker}
to order $q^4$, i.e. we investigate the divergences of the 
generating functional for Green functions of quark currents 
between one--nucleon states using heat--kernel techniques. We do not treat
nucleon--anti-nucleon S--matrix elements, which formally start to appear
at this order. We perform  a chiral invariant renormalization of all 
two--nucleon Green functions of the pion--nucleon system to ${\cal O}(q^4)$ 
in the low--energy (chiral) expansion. Another argument why one has to
perform the renormalization at fourth order is the observation that
the effective pion--nucleon Lagrangian consists of terms with odd and even 
chiral dimension starting at orders one and two, respectively. Therefore, the first
corrections to the dimension two tree graphs (which are sometimes
large) appear at one loop and order four~\cite{juerg}.
This investigation is only a first
step in a systematic evaluation of isospin--violating effects at low energies
and to eventually gain a deeper insight into the mechanism of this isospin
violation.  In a next step, virtual photons have to be included in the 
generating functional. This will then allow us to separate the hadronic
(QCD) isospin violating effects $\sim m_d - m_u$, i.e.  the ones due to
the light quark mass difference, from the purely electromagnetic ones,
which are of the same size and are e.g. the main contribution to the charged
to neutral pion mass difference. The novel data on pion photoproduction
from MAMI and SAL as well as the level shift measurements for pionic hydrogen
and deuterium at PSI have now reached such a precision that a clean separation
of electromagnetic from the purely hadronic contributions based on a
{\it consistent} machinery is called for.

The manuscript is organized as follows. In section~\ref{sec:HBCHPT}
we review the path integral formalism of heavy nucleon CHPT, following
closely ref.\cite{bkkm}. This section
is mostly relevant to define our notations. In particular, we write down
the dimension two  pion--nucleon Lagrangian in the form that is particularly
useful for our purpose. In section~\ref{sec:genone}, we work out the
generating functional to order $q^4$. Here and in what follows, our work 
closely parallels the one of ref.\cite{ecker}. Since we work to one order
higher, our emphasis is on discussing the novel contributions appearing
beyond ${\cal O}(q^3)$. To be specific, instead of the two irreducible
graphs at ${\cal O}(q^3)$, we have to deal with four (tadpole and three
types of self--energy graphs). 
In section~\ref{sec:tadpole} we work out the renormalization of the 
irreducible tapdole graph based on standard heat kernel techniques.
The much more involved renormalization of the irreducible self--energy graphs
is spelled out in section~\ref{sec:selfenergy}. We split this into three
subsections. First, we consider the vertex--corrected self--energy diagrams
which can be evaluated straightforwardly by the method developed in 
ref.\cite{ecker} for such type of operators. We then consider the self-energy diagram
with a dimension two insertion on the intermediate nucleon line
(the so--called ``eye graph''), which
formally involves a triple coincidence limit in the proper time. 
Its contributions are worked out making use of an n--fold
coincidence technique which we develop in one part of the
section. In that context, we have to deal with operators which are
no longer projected in the direction of the nucleons' four--velocity.
We show how to extend the heat--kernel methods used in
\cite{ecker,guido} to  handle such type of operators.
We proceed and work out the pertinent singularity structure. 
In section~\ref{sec:LCT}, we write down the full
counterterm Lagrangian at order $q^4$ and tabulate the pertinent
operators and their $\beta$--functions. This table constitutes the
main result of this investigation. 
Section~\ref{sec:sample} contains a few sample calculations. We consider
the isoscalar magnetic moment of the nucleon
and the scalar form factor of the nucleon to order $q^4$. We
extract the pertinent divergences by straightforward Feynman diagram
evaluation  and we show how to use table~1. 
A summary and a discussion of the various checks on our calculation is given 
in section~\ref{sec:summ}. The appendices contain sufficiently detailed 
technicalities to check the calculation at various intermediate steps.
In particular, in app.~\ref{app:seeley} we list all Seeley--deWitt coeffcients for
a certain class of elliptic differential operators up to dimension four. Also
given are all products of singular operators with one meson and one baryon propagator
as well as for one meson and two baryon propagators, see app.~\ref{app:singops} 
and app.~\ref{app:triple}, respectively. Furthermore, the contributions
from the most complicated diagram, the eye graph, are separately listed in
app.~\ref{app:eye}.  In appendix~\ref{app:eyecheck}, 
we discuss an alternative way of treating parts of the eye graph which 
allows for a good check on parts of the rather involved calculations.
We also consider it useful to give the
divergent operators and their $\beta$--functions from the tadpole, self--energy
and eye graphs in separate tables. This is spelled 
out in app.~\ref{app:tables}. 
  

\section{Brief expos{\'e} of the heavy nucleon effective field 
theory and its path integral formulation}
\label{sec:HBCHPT}
\def\theequation{\arabic{section}.\arabic{equation}}
\setcounter{equation}{0}

To keep the manuscript self--contained, we briefly develop the
path--integral formulation of the chiral effective pion--nucleon 
system. This follows largely the original work of \cite{bkkm},
which was reviewed in \cite{bkmrev}. The reader familiar with these
methods can skip this section. Most important is the definition of the
dimension two $\pi N$ Lagrangian given at the end of this section because
it will be used extensively later on.

\medskip 

The interactions of the pions with the nucleons
are severely constrained by chiral symmetry. The
generating functional for Green functions of quark currents 
between single nucleon states, $Z[j,\eta, \bar \eta$], is defined via
\beq
\exp \, \bigl\{ i \, Z[j,\eta, \bar \eta] \bigr\} 
= {\cal N} \int [du] [dN] [d{\bar N}] \, \exp 
\, i \biggl[ S_{\pi\pi} + S_{\pi N} + 
\int d^4x (\, \bar \eta N  + \bar N \eta \, ) \biggr] \, \, \, ,
\label{defgenfun}
\eeq
with $S_{\pi\pi}$ and $S_{\pi N}$ denoting the pionic and the pion--nucleon
effective action, respectively, to be discussed below. $\eta$ and
$\bar \eta$ are fermionic sources coupled to the baryons and $j$
collectively denotes the external fields of vector ($v_\mu$), 
axial--vector ($a_\mu$), scalar ($s$) and pseudoscalar ($p$) type. 
These are coupled in the standard chiral
invariant manner. In particular, the scalar source contains the quark
mass matrix ${\cal M}$, $s(x) = {\cal M} + \ldots$. 
The underlying effective Lagrangian
can be decomposed into a purely mesonic ($\pi\pi$) and a pion--nucleon
($\pi N$) part as follows (we only consider processes with exactly one
nucleon in the initial and one in the final state)
\beq
{\cal L}_{\rm eff} = {\cal L}_{\pi\pi} + {\cal L}_{\pi N}
\eeq
subject to the following low--energy expansions
\beq
{\cal L}_{\pi\pi} =  {\cal L}_{\pi\pi}^{(2)} 
 + {\cal L}_{\pi\pi}^{(4)} + \ldots  \, \, ,
\quad {\cal L}_{\pi N} =  {\cal L}_{\pi N}^{(1)} + {\cal L}_{\pi N}^{(2)}
+ {\cal L}_{\pi N}^{(3)} + {\cal L}_{\pi N}^{(4)}  \ldots \,
\eeq
where the superscript denotes the chiral dimension. 
The pseudoscalar Goldstone fields, i.e. the pions, are collected in
the  $2 \times 2$ unimodular, unitary matrix $U(x)$, 
$ U(\phi) = u^2 (\phi) = \exp \lbrace i \phi / F \rbrace$
with $F$ the pion decay constant (in the chiral limit).
The external fields appear  in the following chiral invariant
combinations:
$r_\mu = v_\mu +a_\mu \, ,$ $l_\mu = v_\mu -a_\mu \, ,$ and
$\chi = 2 B_0 \,(s+ip)$. Here, $B_0$ is related to the quark condensate 
in the chiral limit, $B_0 = |\langle 0|\bar q q|0 \rangle|/F^2$. 
We adhere to the standard chiral counting, i.e. $s$ and $p$ are
counted as ${\cal O}(q^2)$, with $q$ denoting a small momentum or meson mass.
The effective meson--baryon Lagrangian starts with terms of dimension one,
\beqa
{\cal L}_{\pi N}^{(1)} = \bar \Psi \, \biggl(\, i \nabla\!\!\!\!/  - m 
+ {g_A \over 2} \, u\!\!\!/ \gamma_5  \, \biggr) \Psi \,\,\, ,
\label{LMB1}
\eeqa
with $m$ the nucleon mass in the chiral limit and $u_\mu = i [ u^\dagger 
(\partial_\mu-ir_\mu) u - u (\partial_\mu-il_\mu) u^\dagger ]$. 
The nucleons, i.e. the proton and the neutron, are collected in the 
iso--doublet $\Psi$, 
\begin{eqnarray}
\Psi  =  \left(
\matrix  { p \nonumber \\ n \nonumber \\}
\!\!\!\!\!\!\!\!\!\!\!\!\!\!\!\!\! \right)  \, \, \, .
\end{eqnarray}
Under $SU(2)_L \times SU(2)_R$, $\Psi$  transforms as any matter field.
$\nabla_\mu$ denotes the covariant derivative,
$\nabla_\mu  \Psi = \partial_\mu \, \Psi +  \Gamma_\mu \Psi$
and $\Gamma_\mu$ is the chiral connection,
$\Gamma_\mu = \frac{1}{2}\, [ u^\dagger (\partial_\mu-ir_\mu) u +
u (\partial_\mu-il_\mu) u^\dagger ]$. Note that the first term in
Eq.(\ref{LMB1}) is of dimension one since $( i \nabla \!\!\!\!/ - m )\, 
\Psi = {\cal O}(q)$ \cite{krause}. The lowest order
pion--nucleon Lagrangian contains two parameters, the nucleon mass $m$ and
the axial--vector coupling constant $g_A$, both taken at their values
in the chiral limit.\footnote{We omit the often used superscript 
'$\krig{}$' to keep our notation compact.} Treating the nucleons as
relativistic spin--1/2 fields, the chiral power counting is no more
systematic due to the large mass scale $m$, $\partial_0 \, \Psi 
\sim m \, \Psi \sim \Lambda_\chi \, \Psi$, with $\Lambda_\chi \sim 1\,$GeV
the scale of chiral symmetry breaking. This problem can be overcome in
the heavy mass formalism proposed in \cite{jm}. 
We follow here the path integral approach  developed in \cite{bkkm}.
Defining velocity--dependent spin--1/2 fields by a particular choice 
of Lorentz frame and decomposing the fields into their velocity
eigenstates (sometimes called 'light' and 'heavy' components),
\beq
H_v (x) = \exp \{ i m v \cdot x \} \, P_v^+ \, N(x) \, , \quad
h_v (x) = \exp \{ i m v \cdot x \} \, P_v^- \, N(x) \,\, , 
\label{Bheavy}
\eeq
the mass dependence is shuffled from the fermion propagator into a
string of $1/m$ suppressed interaction vertices. The projection operators
appearing in Eq.(\ref{Bheavy}) are given by 
$P_v^\pm = (1 \pm v \!\!\!/)/2$, with $v_\mu$ the four--velocity subject to
the constraint $ v^2 = 1$. In this basis, the 
effective pion--nucleon action takes the form
\beq
S_{\pi N} = \int d^4x \, \biggl\{ \bar{H}_v \, A \, H_v 
- \bar{h}_v \, C\, h_v + \bar{h}_v \, B \, H_v
+ \bar{H}_v \, \gamma_0 \,  {B}^\dagger \, \gamma_0 \, h_v 
\biggr\}\,\,\, .
\eeq
The matrices $A$, $B$ and $C$ admit low energy expansions, 
e.g.\footnote{Notice that for these matrices the chiral dimensions
are given as subscripts.}
\beq
A = A_{(1)}+ A_{(2)} + A_{(3)} + A_{(4)} + \ldots \,\, .
\eeq
Explicit expressions for the various contributions can be found in 
\cite{bkmrev}.  Similarly, we split the
baryon source fields $\eta (x)$ into velocity eigenstates,
\beq
R_v (x)= \exp \{ i m v \cdot x \} \, P_v^+ \, \eta(x) \, , \quad
\rho_v (x) = \exp \{ i m v \cdot x \} \, P_v^- \, \eta(x) \,\, , 
\label{sourceheavy}
\eeq
and shift variables, $h_v = h_v - C^{-1} \, ( B \, H_v + \rho_v )$,
so that the generating functional takes the form
\beq
\exp[iZ] = {\cal N} \, \Delta_h \, \int [dU][dH_v][d\bar{H}_v] \, \exp 
\bigl\{iS_{\pi\pi} + i S_{\pi N}' \, \bigr\}
\label{Zinter}
\eeq
in terms of the new pion--nucleon action $S_{\pi N}'$,
\beq
S_{\pi N}' = \int d^4x \, \bar{H}_v \bigl( A^{} + \gamma_0
B^\dagger \gamma_0 \, C^{-1} B \, \bigr) H_v
+ \bar{H}_v \bigl( R_v + \gamma_0 B^\dagger \gamma_0 
C^{-1} \rho_v \bigr) + \bigr( \bar{R}_v + \bar{\rho}_v
C^{-1} \, B \bigr) H_v \, \, .
\eeq       
The determinant $\Delta_h$ related to the 'heavy' components is
identical to one, i.e. the positive and negative velocity sectors are 
completely separated. The generating
functional is thus entirely expressed in terms of the Goldstone bosons
and the 'light' components of the spin--1/2 fields. The action is,
however, highly non--local due to the appearance of the inverse of 
the matrix $C$. To render it local, one now expands $C^{-1}$ in powers
of $1/m$, i.e. in terms of increasing chiral dimension. 
To any finite power in $1/m$, one can now perform the integration of
the 'light' baryon field components $N_v$ by again completing the
square,
\beq
H_v' = H_v + T^{-1} \, \bigl( R_v + \gamma_0 \, B^\dagger
\, \gamma_0 \, C^{-1} \, \rho_v \, \bigr) \, , \quad
T = A + \gamma_0 \, B^\dagger \, \gamma_0 \, C^{-1} \, B \,\,\, .
\eeq
Notice that the second term in the expression for $T$ only starts
to contribute at chiral dimension two. To be more
precise, we give the chiral expansion of $T$ up to and including
all terms of order $q^4$,
\beqa
T &=& A_{(1)} + A_{(2)} + A_{(3)} + A_{(4)} + \frac{1}{2m}
\, \gamma_0 \, B^\dagger_{(1)} \, \gamma_0 \, B_{(1)}
\nonumber \\
&+& \frac{1}{2m} \biggl(  \gamma_0 \, B^\dagger_{(1)} \, 
\gamma_0 \, B_{(2)} +  \gamma_0 \, B^\dagger_{(2)} \, 
\gamma_0 \, B_{(1)} \biggr) - \frac{1}{(2m)^2} 
\, \gamma_0 \, B^\dagger_{(1)} \, \gamma_0 \, (C_{(1)} - 2m) \,
B_{(1)} \nonumber \\ 
&+& \frac{1}{2m} \biggl( \gamma_0 \, B^\dagger_{(1)} \, 
\gamma_0 \, B_{(3)} +  \gamma_0 \, B^\dagger_{(3)} \, 
\gamma_0 \, B_{(1)} +  \gamma_0 \,  B^\dagger_{(2)} \, \gamma_0 \,
 B_{(2)}   \biggr) \nonumber \\ 
&-& \frac{1}{(2m)^2}\biggl( 
\, \gamma_0 \, B^\dagger_{(1)} \, \gamma_0 \, (C_{(1)} -2m) \, B_{(2)}
+ \gamma_0 \, B^\dagger_{(2)} \, \gamma_0 \, (C_{(1)} -2m) \, B_{(1)}
+ \gamma_0 \, B^\dagger_{(1)} \, \gamma_0 \, C_{(2)} \, B_{(1)}
 \biggr) \nonumber \\
&+&  \frac{1}{(2m)^3} \, \gamma_0 \, B^\dagger_{(1)} \, \gamma_0 \, 
 (C_{(1)} -2m)   \, (C_{(1)} -2m) \,
B_{(1)}+ {\cal O}(q^5) \,\,\, .
\eeqa
We thus arrive at
\beq
\exp[iZ] = {\cal N}' \, \int [dU] \, \exp \bigl\{ iS_{\pi\pi} 
+ i Z_{\pi N} \, \bigr\} \,\,\, ,
\label{Zfinal}
\eeq
with ${\cal N}'$ an irrelevant normalization constant. The generating 
functional has thus been reduced to the purely mesonic functional. 
$Z_{\pi N}$ is given by
\beqa
Z_{\pi N} = - \int d^4x &\biggl\{& 
 \bar{\rho}_v \, \bigl( C^{-1} \, B
\, T^{-1} \, \gamma_0 \, B^\dagger \, \gamma_0 C^{-1} -
C^{-1} \, \bigr) \, \rho_v
\nonumber \\
&+& 
 \bar{\rho}_v \, \bigl( \, C^{-1} \, B \, T^{-1} 
\, \bigr) \, R_v + \bar{R}_v \, \bigl( T^{-1} \, \gamma_0 \,
B^\dagger \, \gamma_0 \, C^{-1} \, \bigr) \, \rho_v
\nonumber \\
&+&  \quad \bar{R}_v \, T^{-1} \, R_v  \quad \biggr\} \,\,\,\,\, . 
\label{ZMB}
\eeqa
At this point, some remarks are in order. First, physical matrix
elements are always obtained by differentiating the generating
functional with respect to the sources $\eta$ and $\bar\eta$. The
separation into the velocity eigenstates is given by the projection
operators as defined above. As shown in ref.\cite{eckmojwf}, the
chiral dimension of the `heavy' source $\rho_v \sim P_v^- \eta$ is larger
by one order than the chiral dimension of the `light' source, $R_v \sim
P_v^+ \eta$. Based on that observation, we can determine the chiral
dimension with which the various terms in Eq.(\ref{ZMB}) start to 
contribute. Consider first the term in the last line. It is
proportional to the propagator of the `light' fields and thus starts
at order $q^{-1}$. Interactions and loops related to this term
start at ${\cal O}(q)$ and  ${\cal O}(q^3)$, respectively. 
Consequently, to order $q^3$, only this last 
term in Eq.(\ref{ZMB}) generates the Green functions related to the
'light' fields. The second line in Eq.(\ref{ZMB}) starts at two orders
higher compared to the term just discussed. It thus affects tree
graphs at order $q^3$ and $q^4$ and loops only at  ${\cal O}(q^5)$,
which is beyond the accuracy of our calculation. 
The terms in the
first line in Eq.(\ref{ZMB}) contribute only at three chiral orders
higher than the leading term and thus lead to wave--function
renormalization at  ${\cal O}(q^4)$. We stress again that
the operator $C^{-1}$ is
related to the opposite--velocity--nucleon propagator. 
While the propagator of the nucleon moving in the direction of $v$ 
does not contain the mass any more, the anti--velocity nucleon
propagator picks up exactly the factor $2m$ which is nothing but the 
gap between the two  sectors.

It is straightforward to construct the dimension two effective chiral
effective Lagrangian from this action. We use the
definitions of \cite{bkmrev} but introduce some more compact notation for
the later use. The dimension one and two Lagrangians thus take the form:
\beqa
{\cal L}_{\pi N}^{(1)} &=& \bar{H}_v \biggl\lbrace i v \cdot \nabla +
g_A \, S \cdot u \, \biggr\rbrace H_v \,\,\, , \nonumber \\
{\cal L}_{\pi N}^{(2)} &=& \bar{H}_v \biggl\lbrace
\Theta_0^{\mu\nu} \, \nabla_\mu \nabla_\nu
+ \tilde{\Theta}_0 v^{\mu} \, S^\nu  \, \{\nabla_\nu , u_\mu \}
+ \Theta_{1,5} \, \langle \chi_+ \rangle + \Theta_{5} \, \chi_+
   \nonumber \\ 
&& + \Big( \Theta_{2,3}^{\mu\nu}
+  \Theta_{4}\, [S^\mu, S^\nu ]  \Big) \,u_\mu u_\nu 
 + \Theta_{6}\, [S^\mu, S^\nu ] \, F_{\mu\nu}^+ 
+ \Theta_{7} \, [S^\mu, S^\nu ]\, \langle  F_{\mu\nu}^+ \rangle 
\biggr\rbrace H_v \,\,\, ,
\eeqa
with 
\beqa
\Theta_0^{\mu\nu} &=& {1\over 2m} \, \Big( v^\mu  v^\nu - v^2 \,g^{\mu\nu}\Big)
\,\, , \quad  \tilde{\Theta}_0 = - {i g_A \over 2m} 
\,\, , \quad \Theta_{1,5} = \Big(c_1 - {c_5 \over 2}  \Big)\,\, ,
\nonumber \\
\Theta_{2,3}^{\mu\nu} &=& \Big( c_2 - {g_A^2 \over 8m} \Big)\,  v^\mu v^\nu
+ c_3 \,  g^{\mu\nu} \,\, , \quad \Theta_{4}
 = \Big( c_4 + {1 \over 4m} \Big) \,\, , \quad 
\Theta_{5} = c_5 \,\, , \nonumber \\
\Theta_{6} &=& -i {1\over 4m} \, (1 + c_6)\, \,\, , \quad
\Theta_{7}  = -i {1\over 4m} \, c_7\, \,\,\,\, .
\eeqa
and $S^\mu$ is the covariant spin--operator {\`a} la Pauli--Lubanski,
$S^\mu = \frac{i}{2} \, \gamma_5 \, \sigma^{\mu \nu} \, v_\nu$
subject to the constraint $S \cdot v = 0$.
Traces in flavor space are denoted by $\langle...\rangle$. 
Notice that the spin--matrices appearing in the operators have all to be 
taken  in the appropriate order. 
We also have $F_{\mu\nu}^\pm = u F_{\mu\nu}^L u^\dagger \pm u^\dagger
F_{\mu\nu}^R u$, with $ F_{\mu\nu}^{L,R}$ the field strength tensors
related to $l_\mu$ and $r_\mu$, respectively. The explicit symmetry 
breaking is encoded in the matrices $\chi_\pm = u^\dagger \chi
u^\dagger  \pm  u  \chi^\dagger  u $.

\noindent


\section{Generating functional to one loop}
\label{sec:genone}
\def\theequation{\arabic{section}.\arabic{equation}}
\setcounter{equation}{0}

In this section, we turn to the calculation of $S_{\pi\pi}[j]+
Z_{\pi N}[j,R_v]$ to one loop beyond leading order, i.e. to order 
$q^4$ in the small momentum expansion. The method has been exposed
in some detail by Ecker \cite{ecker} for SU(2) to ${\cal{O}}(q^3)$ and
for SU(3) in more detail to the same order in \cite{guido}. Here,
we just outline the pertinent steps following essentially the method 
used in ref.\cite{gss} and discuss the additional terms appearing to
the order we are working. To be specific, one has to expand
\beq
{\cal L}_{\pi\pi}^{(2)} +  {\cal L}_{\pi\pi}^{(4)} - \bar{R}_v  \,
[T_{(1)}+ T_{(2)}]^{-1} \, R_v
\label{Zexp}
\eeq
in the functional integral Eq.(\ref{Zfinal}) around the classical
solution, $u_{\rm cl} = u_{\rm cl}[j]$. This leads to a set of
reducible and irreducible one--loop diagrams to be discussed below.
We chose the fluctuation variables $\xi$ in a symmetric form,
$\xi_R = u_{\rm cl} \, \exp\{i \, \xi /2\}$,  
$\xi_L = u_{\rm cl}^\dagger \, \exp\{-i \, \xi /2\}$,
with $\xi^\dagger = \xi$ traceless 2$\times$2 matrices. Consequently,
we have also
$U =  u_{\rm cl} \, \exp\{i \, \xi \} \, u_{\rm cl}$.
To second order in $\xi$, the covariant derivative $\nabla_\mu$, 
the chiral connection $\Gamma_\mu$ and the axial--vector $u_\mu$
take the form
\beqa
\Gamma_\mu &=& \Gamma_\mu^{\rm cl} + \frac{1}{4} \,[\, u_\mu^{\rm cl}, \xi
\, ] + \frac{1}{8} \, \xi 
\stackrel{\leftrightarrow}{\nabla}_\mu^{\rm cl}\, \xi + {\cal O}(\xi^3)
\nonumber \\ 
{} \nabla_\mu^{\rm cl}  \xi  &=& \partial_\mu \, \xi +  \Gamma_\mu^{\rm
  cl} \xi  \,\, ,\,\, \xi \stackrel{\leftrightarrow}{\nabla}_\mu^{\rm
  cl} \xi = \xi 
[ \nabla_\mu^{\rm cl} , \xi ] - [ \nabla_\mu^{\rm cl} , \xi ] \xi
\nonumber \\
u_\mu &=& u_\mu^{\rm cl} - [ \nabla_\mu^{\rm cl} , \xi ] + \frac{1}{8} \,
[\, \xi, [\, u_\mu^{\rm cl}, \xi\,]\,] + {\cal O}(\xi^3) \,\,\, ,
\nonumber \\
\chi_\pm &=& \chi_\pm^{\rm cl} - {i \over 2} \{\chi_\mp^{\rm cl} , \xi\}
- {1\over 8} \{ \xi, \{ \chi_\pm^{\rm cl} , \xi\} \}+ {\cal O}(\xi^3) \,\,\, ,
\nonumber \\
F_{\mu\nu}^\pm &=& F_{\mu\nu}^{\pm , {\rm cl}}  - {i \over 2} 
[ F_{\mu\nu}^{\mp , {\rm cl}} , \xi ] + {1\over 8} [ \xi , [
F_{\mu\nu}^{\pm , {\rm cl}} , \xi ] ]+ {\cal O}(\xi^3) \,\,\, .
\label{nablacl}
\eeqa
Notice that while $\nabla_\mu^{\rm cl}$ defined here acts on the
fluctuation variables (fields) $\xi (x)$, the covariant derivative
$\nabla_\mu$ defined in Eq.(\ref{LMB1}) acts on the baryon fields.
We are now in the position to expand the fermion propagator $S$ to
quadratic order in the fluctuations making use of the relation
\beq
[A_{(1)} + A_{(2)}] \cdot S = {\bf 1} \quad .
\eeq
To the order we are working, the expansion of the propagator around 
the classical solution $S^{\rm cl}$ takes the form (there is at most one
insertion from the dimension two $\pi N$ Lagrangian)
\beq
S^{\rm cl} = S^{\rm cl}_{(1)} - S^{\rm cl}_{(1)} \,  A_{(2)}^{\rm cl}  \,
 S^{\rm cl}_{(1)} \,\, ,
\eeq
with $S_{(1)}^{\rm cl}$  the full lowest order classical fermion
propagator, i.e. with all possible tree structures of the external
sources attached, and we expand the matrix valued operators  $A_{(1)}$
and $A_{(2)}$ 
around their respective classical solutions,
\beq
A_{(i)} = A_{(i)}^{\rm cl} + A_{(i)}^1 + A_{(i)}^2 \,\, , 
\quad i=1,2 \,\, ,
\eeq
where the terms are of order $\xi^{0,1,2}$, in order.
We have tentatively assumed the existence of the inverse of the free fermion
propagator. Consequently, the fermion propagator to order $\xi^2$ reads
\beqa \label{Sexp}
A^{-1} &=& (A^{-1}_{(1)})^{\rm cl} -  (A^{-1}_{(1)})^{\rm cl} \,
A_{(2)}^{\rm cl} \,  (A^{-1}_{(1)})^{\rm cl} \nonumber \\
&-& (A^{-1}_{(1)})^{\rm cl} \, A^{1}_{(1)}\, (A^{-1}_{(1)})^{\rm cl}
-  (A^{-1}_{(1)})^{\rm cl} \, A^{1}_{(2)}\, (A^{-1}_{(1)})^{\rm cl}
\nonumber \\
&+& (A^{-1}_{(1)})^{\rm cl} \, A^{1}_{(1)}\, (A^{-1}_{(1)})^{\rm cl}
\, A_{(2)}^{\rm cl} \, (A^{-1}_{(1)})^{\rm cl}
+ (A^{-1}_{(1)})^{\rm cl} \, A_{(2)}^{\rm cl} \, 
(A^{-1}_{(1)})^{\rm cl} \, A^{1}_{(1)}\, (A^{-1}_{(1)})^{\rm cl}
\nonumber \\
&-& (A^{-1}_{(1)})^{\rm cl} \, A^{2}_{(1)}\, (A^{-1}_{(1)})^{\rm cl}
+ (A^{-1}_{(1)})^{\rm cl} \, A^{1}_{(1)}\, (A^{-1}_{(1)})^{\rm cl}
 \, A^{1}_{(1)}\, (A^{-1}_{(1)})^{\rm cl}
\nonumber \\
&+& (A^{-1}_{(1)})^{\rm cl} \, A^{2}_{(1)}\, (A^{-1}_{(1)})^{\rm cl}
\, A_{(2)}^{\rm cl} \, (A^{-1}_{(1)})^{\rm cl}
+ (A^{-1}_{(1)})^{\rm cl} \,  A_{(2)}^{\rm cl} \, (A^{-1}_{(1)})^{\rm
  cl} \,  A^{2}_{(1)}\, (A^{-1}_{(1)})^{\rm cl}
\nonumber \\
&-&  (A^{-1}_{(1)})^{\rm cl} \, A^{1}_{(1)}\, (A^{-1}_{(1)})^{\rm cl}
\,  A^{1}_{(1)}\, (A^{-1}_{(1)})^{\rm cl}\,  A_{(2)}^{\rm cl} \,
 (A^{-1}_{(1)})^{\rm cl}
\nonumber \\
&-&  (A^{-1}_{(1)})^{\rm cl}\, A_{(2)}^{\rm cl} \,(A^{-1}_{(1)})^{\rm cl}
\,  A^{1}_{(1)}\, (A^{-1}_{(1)})^{\rm cl}\,  A^{1}_{(1)}\, 
(A^{-1}_{(1)})^{\rm cl}
\nonumber \\
&-&  (A^{-1}_{(1)})^{\rm cl} \, A^{1}_{(1)}\, (A^{-1}_{(1)})^{\rm cl}
\,  A_{(2)}^{\rm cl} \, (A^{-1}_{(1)})^{\rm cl}\,  A^{1}_{(1)}\, 
(A^{-1}_{(1)})^{\rm cl}
\nonumber \\
&+& (A^{-1}_{(1)})^{\rm cl} \, A^{1}_{(2)}\, (A^{-1}_{(1)})^{\rm cl}
\, A^{1}_{(1)}\, (A^{-1}_{(1)})^{\rm cl}
+ (A^{-1}_{(1)})^{\rm cl} \, A^{1}_{(1)}\, (A^{-1}_{(1)})^{\rm cl}
\, A^{1}_{(2)}\, (A^{-1}_{(1)})^{\rm cl}
\nonumber \\
&-&  (A^{-1}_{(1)})^{\rm cl} \, A^{2}_{(2)}\, (A^{-1}_{(1)})^{\rm cl}
+ {\cal O}(\xi^3,q^4) \,\, .
\eeqa
The expanded fermion propagator in Eq.(\ref{Sexp}) consists of
irreducible and reducible parts as depicted in Fig.~1, which we now
discuss briefly. The first line of Eq.(\ref{Sexp}) corresponds to the
terms of order $\xi^0$. The second and third line, corresponding to
the second row in Fig.~1, comprise the terms of order $\xi$, which vanish
by use of the equations of motion. For this to happen, one must chose
a consistent parametrization of the fields in the pion and the
pion--nucleon sector \cite{eckerm} and account for the pertinent
tadpole graphs with exactly one insertion from ${\cal L}_{\pi\pi}^{(4)}$.
The fourth line (third row in Fig.~1)
corresponds to the irreducible self--energy and tadpole diagrams first
worked out by Ecker. In the next three lines, corresponding to the fourth
row in Fig.~1, the reducible graphs at ${\cal O}(q^4)$ are shown. As
stated before, these graphs together with similar ones from the
meson sector can be made to vanish for a consistent set of field definitions.
In the lowest row of the figure, the novel irreducible self--energy and tadpole
diagrams are shown. The eight line in  Eq.(\ref{Sexp}) 
is the novel self--energy
diagram with three propagators, i.e. the insertion from 
${\cal L}_{\pi N}^{(2)}$ on the nucleon line while the pion is in the
air.  The next two graphs are vertex corrections to the order $q^3$
self--energy graph and the last diagram, corresponding to the last line
in Eq.(\ref{Sexp}). All the diagrams starting with the third row in Fig.~1 
are of ${\cal O}(\xi^2)$. The corresponding generating functional reads 
\beqa
Z_{\rm irr}[j,R_v] &=& \int d^4x \, d^4x' \, d^4y \, d^4y' \, 
\bar{R}_v (x) \, S_{(1)}^{\rm cl} (x,y) \, \bigl[ \,
\Sigma_2^{(1)+(2)}(y,y') \,  \delta(y-y') \nonumber \\
&& \qquad + \, \Sigma_1^{(1)+(2)} (y,y') + \Sigma_3^{(2)} (y,y') \, \bigr] \, 
S_{(1)}^{\rm cl}(y',x') \, R_v (x')
\label{Zirr}
\eeqa
in terms of the self--energy functionals $\Sigma_{1,2,3}$. $\Sigma_1$
refers to the self-energy graphs at order $q^3$ and the same diagram
with one dimension two insertion on the nucleon line. $\Sigma_2$
collects the tadpoles at orders $q^3$ and $q^4$ and  $\Sigma_3$ refers
to the dimension two vertex corrected self--energy diagrams.  Consider
first $\Sigma_1^{(1)}$ and  $\Sigma_3^{(2)}$. These take the form (we
only display the one for $\Sigma_1^{(1)}$)
\beq
\label{Sig1}
\Sigma_1^{(1)} = -\frac{2}{F^2} \, V_i \, G_{ij} \, 
[A_{(1)}^{\rm cl}]^{-1} \, V_j 
=  -\frac{2}{F^2} \, V_i \, G_{ij} \, 
S_{(1)}^{\rm cl} \, V_j \, \, , 
\eeq
with vertex functions of dimension one\footnote{Note that for the $V_i$ 
the chiral dimension is again given as subscript '(i)'.}
\beqa \label{Sig1V}
V_{i(1)} & = &  V_{i(1)}^{(1)} + V_{i(1)}^{(2)} \,\, , 
\nonumber \\
V_{i(1)}^{(1)} & = & \frac{i}{4\sqrt{2}} \, [ v\cdot u^{cl} , \tau_i
\,] \, \, , \quad
V_{i(1)}^{(2)}  =  - \frac{g_A}{\sqrt{2}} \, \tau_j \, S \cdot d_{ji}^{\rm cl}
\,\, ,
\eeqa
with $i,j,k= 1, 2 ,3$ and the $\tau^i$ denote the conventional 
Pauli (isospin) matrices.  
Similarly, the vertex--corrected self--energy contribution $\Sigma_3^{(2)}$
is given by the same form as in Eq.(\ref{Sig1}) with exactly one of the
following dimension two vertices
\beqa \label{Vi2}
V_{i(2)} &=& \sum_{k=1}^{11} \, V_{i(2)}^{(k)} \,\, , \nonumber \\
V_{i(2)}^{(1)} & = & \frac{\Theta_0^{\mu \nu}}{4\sqrt{2}} \,[[
\nabla_\mu^{\rm cl} , u_\nu^{\rm cl}] , \tau_i ] \, \, , \quad
V_{i(2)}^{(2)} = \frac{\Theta_0^{\mu \nu}}{4\sqrt{2}} \, [
u_\nu^{\rm cl} , \tau_j ] \, d_{ji}^{\mu, \, {\rm cl}} \,\, , \nonumber \\
V_{i(2)}^{(3)} & = &  \frac{\Theta_0^{\mu \nu}}{4\sqrt{2}} \, \biggl\{
[  u_\nu^{\rm cl} ,  \tau_i ] \, \nabla_\mu^{\rm cl} +
[  u_\mu^{\rm cl} ,  \tau_i ] \, \nabla_\nu^{\rm cl} \biggr\}
 \,\, ,\nonumber \\
V_{i(2)}^{(4)} & = &  \frac{\tilde{\Theta}_0}{4\sqrt{2}} \,
\{  u_\mu^{\rm cl} , [ u_\nu^{\rm cl} ,  \tau_i ] \} v^\mu \, S^\nu
\,\, , \quad 
V_{i(2)}^{(5)} = -\frac{2\tilde{\Theta}_0}{\sqrt{2}} \,
\tau^j \, v \cdot d_{ji}^{{\rm cl}} \, S \cdot \nabla^{\rm cl}
\,\, ,  \nonumber \\
V_{i(2)}^{(6)} &=& -\frac{\tilde{\Theta}_0}{\sqrt{2}} \,
\tau^k \, S \cdot d_{kj}^{{\rm cl}} \, v \cdot d_{ji}^{{\rm cl}} \,
 \,\, \quad V_{i(2)}^{(7)} =  -\frac{\Theta_{2,3}^{\mu \nu}}{\sqrt{2}} \,
\{ u_\mu^{\rm cl} , \tau^j \} \, d_{ji}^{\nu, \, {\rm cl}} \,\, ,
\nonumber \\
V_{i(2)}^{(8)} &=&  -\frac{\Theta_{4}}{\sqrt{2}} \,
[ u_\mu^{\rm cl} , \tau^j \, d_{ji, \, \nu}^{{\rm cl}}]\, [S^\mu ,
S^\nu ] \,\, , \quad V_{i(2)}^{(9)} =-\frac{i\Theta_{1,5}}{2\sqrt{2}}
\, \langle \{ \chi_- , \tau^j \} \rangle \,\, , \nonumber \\
V_{i(2)}^{(10)} &=& -\frac{i\Theta_{5}}{2\sqrt{2}}\, \{ \chi_- , \tau^j \}
\,\, , \quad V_{i(2)}^{(11)} = -\frac{i\Theta_{6}}{2\sqrt{2}}\, [ F_{\mu
  \nu}^- , \tau^i ] [S^\mu , S^\nu ] \,\, .
\eeqa
Notice that there is no contribution 
$\sim \Theta_7$ because $\langle  [ F_{\mu  \nu}^- , \tau^i ] \rangle = 0$. 
The new contribution at order $q^4$ with exactly
one dimension two insertion on the intermediate nucleon line,
$\Sigma_{1}^{(2)}$, has the  form
\beq
\label{Sig12}
\Sigma_1^{(2)} = \frac{2}{F^2} \, V_i \, G_{ij} \, 
\bigl\{ [A_{(1)}^{\rm cl}]^{-1} \,  [A_{(2)}^{\rm cl}] \,
 [A_{(1)}^{\rm cl}]^{-1} \bigr\} \, V_j 
\eeq
with the fermion propagator 
$A^{-1}$ properly expanded around its classical solution. 

\medskip

\noindent The tadpole contributions $\Sigma_2^{(1),(2)}$ read 
(from here on, we drop the superscript 'cl')  
\beqa
\Sigma_2^{(1)} &=& \frac{1}{8F^2} \biggl\{ g_A
 \, [ \, \tau_i \, , \, [ \, S \cdot u , \tau_j \,] \, ]\,  G_{ij} 
+ i \tau_i [G_{ij} v \cdot \stackrel{\leftarrow}{d}_{jk} 
- v \cdot d_{ij} G_{jk} ] \tau_k  \biggr\} \,\, , \\
\Sigma_2^{(2)} &=& {1\over F^2} \sum_{k=1}^{15} \, \Sigma_{2}^{(2,k)} 
\,\, , \nonumber \\ 
\Sigma_2^{(2,1)} &=& -\frac{\Theta_0^{\mu \nu}}{8} \, \biggl\{ \tau_i \bigl[
d_{ij}^\mu G_{jk} - G_{ij}  \stackrel{\leftarrow}{d}^\mu_{jk} \bigr]
\tau_k \biggr\} \nabla_\nu + (\mu \leftrightarrow \nu )  \,\, , \quad
\Sigma_2^{(2,2)} =  \frac{\Theta_0^{\mu \nu}}{16} \, [ u_\mu , \tau_i ]
\, [ u_\nu , \tau_j ] \, G_{ij} \,\, , \nonumber \\
\Sigma_2^{(2,3)} &=& -\frac{\Theta_0^{\mu \nu}}{8} \, \biggl\{ \tau_i \biggl[
d_{ij}^\mu d_{jl}^\nu  G_{lk} -  G_{ij} \stackrel{\leftarrow}{d}^\nu_{jl}
\stackrel{\leftarrow}{d}^\mu_{lk} + d_{ij}^\nu  G_{jl} 
\stackrel{\leftarrow}{d}^\mu_{lk} - d_{ij}^\mu  G_{jl}
\stackrel{\leftarrow}{d}^\nu_{lk} \biggr] \tau_k  \biggr\}\,\, , \nonumber \\
\Sigma_2^{(2,4)} &=& -\frac{\tilde{\Theta}_0}{8} \, \{ v \cdot u ,  \tau_i
[ d_{ij}^\nu  G_{jk}-G_{ij} \stackrel{\leftarrow}{d}^\nu_{jk} ]  \tau_k \}
S^\nu  \,\, , \nonumber \\
\Sigma_2^{(2,5)} &=& -\frac{\tilde{\Theta}_0}{4} \, \biggl\{ \tau_i
d_{ij}^\mu G_{jk} [  u_\nu , \tau_k ] +  [  u_\nu , \tau_i ] G_{ij}  
\stackrel{\leftarrow}{d}^\mu_{jk}\tau_k \biggr\} S^\nu v^\mu \,\, ,
\nonumber \\ 
\Sigma_2^{(2,6)} &=& \frac{\tilde{\Theta}_0}{4} \,  S^\nu \,
[ \tau_i , [ v \cdot u ,  \tau_j ]] G_{ij}  \nabla_\nu  \,\, , \nonumber \\
\Sigma_2^{(2,7)} &=& \frac{\tilde{\Theta}_0}{8} \,  S^\nu \,  \biggl\{
G_{ij} \stackrel{\leftarrow}{d}^\nu_{jk} [\tau_k, [ v \cdot u , \tau_i ]] +
[\tau_i, [ v\cdot u , \tau_j ]] {d}^\nu_{jk}  G_{ki} \biggr\}  
\,\, , \nonumber \\ 
\Sigma_2^{(2,8)} &=& \frac{\tilde{\Theta}_0}{8} \,  S^\nu \,
[ \tau_i , [[\nabla_\nu, v \cdot u] ,  \tau_j ]] G_{ij}  \,\, , \nonumber \\
\Sigma_2^{(2,9)} &=&  \frac{\Theta^{\mu\nu}_{2,3}}{8}
 \,   \{ u_\mu , [ \tau_i ,
 [ u_\nu , \tau_j ]]\}  G_{ij}\,\, , \quad 
\Sigma_2^{(2,10)} =  \Theta^{\mu\nu}_{2,3} \, \biggl\{  \tau_i {d}^\nu_{ij}
G_{jl} \stackrel{\leftarrow}{d}^\mu_{lk} \tau_k \biggr\}\,\, , \nonumber \\
\Sigma_2^{(2,11)} &=&  \frac{\Theta_{4}}{8} \,[S^\mu , S^\nu] [u_\mu , 
[ \tau_i , [ u_\nu , \tau_j ]]]  G_{ij}\,\, , \quad 
\Sigma_2^{(2,12)} = -\frac{\Theta_{4}}{2} \, [ S_\mu, S_\nu ]
 \biggl\{  \tau_i {d}^\nu_{ij}
G_{jl} \stackrel{\leftarrow}{d}^\mu_{lk}  \tau_k - (\mu \leftrightarrow \nu )
\biggr\}\,\, , \nonumber \\
\Sigma_2^{(2,13)} &=&  -\frac{\Theta_{1,5}}{8} \,\langle \{  \tau_i ,
\{ \chi_+ , \tau_j \}\}\rangle  G_{ij}\,\, , \quad 
\Sigma_2^{(2,14)} = -\frac{\Theta_{5}}{8} \, \{  \tau_i ,
\{ \chi_+ , \tau_j \}\} G_{ij}\,\, , \nonumber \\
\Sigma_2^{(2,15)} &=&  \frac{\Theta_{6}}{8} \,[S^\mu , S^\nu] [ \tau_i ,
[ F_{\mu\nu}^+ , \tau_j ]] G_{ij}\,\, . 
\eeqa  
In these equations, $G_{ij}$ is the full meson propagator \cite{gl85}
\beq
G_{ij} =  ( d_\mu \, d^\mu \, \delta^{ij} + \sigma^{ij} \, )^{-1}
\label{Mesprop}
\eeq
with
\beqa
\nabla^\mu_{\rm cl}\, \xi &=& \frac{1}{\sqrt{2}} \tau_j \,
d^\mu_{jk} \, \xi_k \,\,  \, , 
\quad \xi = \frac{1}{\sqrt{2}} \, \tau_i \, \xi_i \,\, , \quad
d^\mu_{ij} = \delta_{ij} \, \partial^\mu + \gamma^\mu_{ij} \,\, ,
\quad \stackrel{\leftarrow}{d}^\mu_{ij} = \delta_{ij}\,
\stackrel{\leftarrow}{\partial}^\mu - \gamma_{ij}^\mu \,\, , 
\nonumber \\
\gamma^\mu_{ij} &=& -\frac{1}{2} \langle \Gamma^\mu_{\rm cl} \, 
[ \, \tau_i, \tau_j \, ] \rangle \,\,\, , \quad
\sigma^{ij} = \frac{1}{8} \langle[ \, u^{\rm cl}_\mu , \tau_i \, ][ \,
\tau_j , u_{\rm cl}^\mu \, ] + \chi_+ \, \{ \, \tau_i ,\tau_j \,\} \rangle
 \,\,\, .
\label{MesHK}
\eeqa
Note that the differential operator $d_{ij}$ is related to the
covariant derivative $\nabla^\mu_{\rm cl}$ and it acts on the meson
propagator $G_{ij}$. 
The connection $\gamma_\mu$ defines a field strength tensor,
\beq
\gamma_{\mu \nu} = \partial_\nu \, \gamma_\mu - \partial_\mu \,
\gamma_\nu + [ \gamma_\mu , \gamma_\nu ] \,\,\, , \quad
\, [d_\lambda \,, \, \gamma_{\mu \nu}]= \partial_\lambda \, \gamma_{\mu \nu}
+ [ \gamma_\lambda , \gamma_{\mu \nu} ] \,\,\, ,
\label{mesfs}
\eeq
where we have omitted the flavor indices.

\section{Renormalization of the tadpole graph}
\label{sec:tadpole}
\def\theequation{\arabic{section}.\arabic{equation}}
\setcounter{equation}{0}

In this section, we consider the renormalization of the tadpole
contribution $\Sigma_2^{(1)+(2)} (y,y')$. This is done in Euclidean space
letting $x^0 \to -i \, x^0$, $v^0 \to  - i \, v^0$, $v \cdot \partial \to 
v \cdot \partial$, $S^0 \to i \, S^0$ and $ S \cdot u \to  - S \cdot u$.
We remark that the sense of the Wick rotation is not uniquely defined,
one only has to perform it consistently.
In the coincidence limit $y \to y'$, the functionals
$\Sigma_2^{(1),(2)} (y,y')$ are divergent. The divergences can be
extracted by using standard heat kernel techniques since they appear as
simple poles in $d = 4 - \ve$. The corresponding residua are local
polynomials in the fields of ${\cal O}(q^3)$ and  ${\cal O}(q^4)$ for
$\Sigma_2^{(1)}$ and $\Sigma_2^{(2)}$, respectively. They can easily
be transformed back to Minkowski space. The tadpole graphs are
proportional to the meson propagator $G_{ij}$, which is an elliptic
second--order differential operator of the type $A(x) = -d_\mu d_\mu +
\sigma (x)$, and derivatives thereof. The methods to treat such type
of operators are spelled out in detail in \cite{guido}. Here, we only
add some additional remarks. The singular function related to the 
meson propagator in the coincidence limit takes the form~\cite{jacko}\cite{guido}
\beq
G(x,x) = \frac{2}{\ve} \frac{\mu^{-\ve}}{16\pi^2} \, a_1 (x,x) + 
{\rm finite} \,\,\, ,
\eeq
i.e. it singularities are given by the Seeley--deWitt coefficient $a_1$.
For operators with one derivative acting on the meson propagator, one
gets
\beq
d_\mu \, G(x,y)|_{x \to y} = \frac{2}{\ve}
  \frac{\mu^{-\ve}}{16\pi^2} \,  d_\mu \, a_1 (x,y)|_{x \to y} + 
{\rm finite} \,\,\, ,
\eeq
whereas in the case of two derivatives, an extra $\delta$--function
appears,
\beq
d_\mu \,d_\nu \, G(x,y)|_{x \to y} = \frac{2}{\ve}
 \frac{\mu^{-\ve}}{16\pi^2} \, \delta_{\mu \nu} \, (-2) \,
 a_2(x,y)|_{x \to y} + \frac{2}{\ve}
  \frac{\mu^{-\ve}}{16\pi^2} \,  d_\mu \, d_\nu \, a_1 (x,y)|_{x \to y} + 
{\rm finite} \,\,\, .
\eeq
These are the basic structures we have to deal with. Notice that we
need the pertinent Seeley--deWitt coefficients and their derivatives
(in the coincidence limit) to higher order than given in~\cite{guido}.
We remark that the coefficient $a_n (x,x)$ has chiral dimension $2n$.
In appendix~\ref{app:seeley}, the necessary formulas are collected.
After some algebra, the divergent part of the tadpoles can be cast in the form 
(rotated back to Minkowski space) 
\beq
\Sigma_2^{(1)+(2), {\rm div}} (y,y) = \frac{1}{(4 \pi F )^2}  
\frac{2}{\ve} \, \biggl[ \, \hat{\Sigma}_2^{(1)} (y,y) \, + \,
\sum_{k=1}^{15} \hat{\Sigma}_2^{(2,k)} (y,y) \biggr]
\,\, ,
\label{rentad}
\eeq
with  $\hat{\Sigma}_2^{(1),(2)} (y,y)$  finite monomials in the
fields of chiral dimension three and four, in order. The explicit form
of  $\hat{\Sigma}_2^{(1)} (y,y)$ is given in \cite{ecker}. For the 
new contributions at order $q^4$, we get
\beqa 
\hat{\Sigma}_2^{(2,1)} &=& \frac{\Theta_0^{\mu \nu}}{6} \, \biggl[
[ \nabla^\tau , \Gamma_{\mu\tau} ] \nabla_\nu + (\mu \leftrightarrow
\nu ) \biggr] \,\,\, , \,\,\,\,
\hat{\Sigma}_2^{(2,2)} =\frac{-\Theta_0^{\mu \nu}}{32} 
\langle u_\mu J(u_\nu) \rangle
\,\,\, 
\hat{\Sigma}_2^{(2,3)} =  \frac{\Theta^{\mu\nu}}{6} [  \nabla^\gamma ,
[ \nabla_\mu,  \Gamma_{\nu\gamma} ] ]  
\nonumber \\
\hat{\Sigma}_2^{(2,4)} &=& \frac{\tilde{\Theta}_0}{6} \,
 \, \{ v \cdot u , [   \nabla^\tau , S^\nu \Gamma_{\nu\tau} ] \} \,\, , \quad
\hat{\Sigma}_2^{(2,5)} = -\frac{\tilde{\Theta}_0 \, v^{\mu}}{4}
\,S^\nu \, \biggl[ {2 \over 3} \langle u_\nu [ \nabla_\tau ,
\Gamma_{\mu\tau} ] \rangle + {1\over 2} [ \eta_\mu (1) , u_\nu] \biggr]
\,\, , \nonumber \\
\hat{\Sigma}_2^{(2,6)} &=& -\frac{\tilde{\Theta}_0}{4} \,
{\cal J}(v \cdot u) S \cdot  \nabla \,\, , \quad
\hat{\Sigma}_2^{(2,8)} = -\frac{\tilde{\Theta}_0}{8} \,
{\cal J}([S \cdot \nabla , v \cdot u ] )   \,\, , 
\hat{\Sigma}_2^{(2,9)} = -\frac{{\Theta}^{\mu\nu}_{2,3}}{8} \,
\{ u_\mu , {\cal J}(  u_\nu  )  \}  \,\, , 
\nonumber \\
\hat{\Sigma}_2^{(2,7)} &=&  -\frac{\tilde{\Theta}_0}{8} \,
\, \biggl[ \langle v \cdot u [ S \cdot \nabla, (u^2 + \chi_+)]\rangle -
v \cdot u \langle [ S \cdot \nabla , (u^2 + \chi_+)]\rangle - \{
v \cdot u , [ S \cdot \nabla , (u^2 + \chi_+)]\} \nonumber \\
&& \qquad\qquad- 2 [S \cdot \nabla , u_\tau ] 
\langle v \cdot u u^\tau \rangle - 
2 u_\tau  \langle v \cdot u  [S \cdot \nabla , u^\tau  ]
 \rangle \biggr] \nonumber \\
\hat{\Sigma}_2^{(2,10)} &=& \Theta_{2,3}^{\mu\nu} \biggl( -\frac{1}{6}
\eta_{\mu\nu} (1) + \frac{2}{3} \langle \Gamma_{\mu\gamma}
\Gamma_\nu^{\,\, \gamma} \rangle \biggl)
\nonumber \\
\hat{\Sigma}_2^{(2,11)} &=&  -\frac{\Theta_4}{8} \, [S^\mu,
S^\nu ] \, [ u_\mu ,{\cal J}(u_\nu) ]  \,\,\, ,
\nonumber \\
\hat{\Sigma}_2^{(2,12)} &=& - \Theta_4 [S^\mu , S^\nu ]  \biggl(
-\frac{1}{2} \eta(\Gamma_{\mu\nu}) + \frac{1}{4} \{ \eta(1),
\Gamma_{\mu\nu} \} + \frac{1}{3} [\nabla^\gamma , [ \nabla_\gamma ,
\Gamma_{\mu\nu} ]] - \frac{2}{3} [ \Gamma_{\mu\gamma} ,
\Gamma_\nu^{\,\, \gamma} ] \biggl)   
\nonumber \\
\hat{\Sigma}_2^{(2,13)} &=& \frac{\Theta_{1,5}}{8} \, \langle {\cal C}
(\chi_+) \rangle \,\, , \quad \hat{\Sigma}_2^{(2,14)} =  \frac{\Theta_{5}}{8} \,
 {\cal C}(\chi_+ ) \,\, , \quad  \hat{\Sigma}_2^{(2,15)} =
- \frac{\Theta_{6}}{8} \, [S^\mu , S^\nu ] \, {\cal J}( F_{\mu\nu}^+ ) \,\, ,
\eeqa
where we used the abbreviations
\beqa
\eta (X) &=& {1\over 2} (\chi_+ + u^2) \langle X \rangle + {1\over 2}
 \langle X (\chi_+ + u^2) \rangle + {1\over 4} X  \langle \chi_+
 \rangle -  {1\over 2} \{ X , \chi_+ \} - u_\tau \langle
u^\tau \, X \rangle \,\,\, , \nonumber \\
\eta^\mu (X) &=&  {1\over 2} [ \nabla^\mu , (u^2+\chi_+)]  \langle X
\rangle  + {1\over 2}  \langle X [ \nabla^\mu , (u^2+\chi_+)]\rangle
 + {1\over 4} X \langle  [ \nabla^\mu ,  \chi_+] \rangle \nonumber \\  
&-& {1\over 2} \{ X,[ \nabla^\mu , \chi_+ ] \} - [ \nabla^\mu , u_\tau
] \langle u^\tau X \rangle - u_\tau \langle X [ \nabla^\mu , u^\tau ]
\rangle \,\,\, ,
\eeqa
so that
\beqa
\eta (1) &=& {1\over 2} \langle u^2  \rangle + u^2 + {3 \over
  4}\langle \chi_+ \rangle \,\,\, , \, \,
 \eta^\mu (1) = [\nabla^\mu , u^2] + {1\over 2}  \langle [\nabla^\mu ,
 u^2] \rangle
+{3\over 4} \langle [\nabla^\mu , \chi_+] \rangle \,\,\, , \nonumber \\
\eta^{\mu\nu} (1) &=& {1\over 2} \langle [\nabla^\mu , [ \nabla^\nu ,
u^2]] \rangle + [ \nabla^\mu , [ \nabla^\nu ,u^2] ] +{3\over 4} 
\langle [\nabla^\mu , [ \nabla^\nu , \chi_+ ] ] \rangle  \,\,\, ,
\eeqa
and
\beq
{\cal J}(X) = -\{(u^2+\chi_+) ,X\} + \langle  (u^2 + \chi_+) X
\rangle - X \langle u^2 + \chi_+ \rangle - 2 u_\mu \langle u^\mu X
\rangle  + (u^2 + \chi_+ ) \langle X \rangle \,\, ,
\eeq
for any  2$\times$2 matrix $X$, and
\beq
{\cal C}(\chi_+) = u^2 \langle \chi_+\rangle + \langle  (u^2 + \chi_+)
\chi_+ \rangle + \chi_+ \langle u^2 \rangle + 3 \chi_+ \langle  \chi_+
\rangle + \{ \chi_+, u^2 \} - 2 \chi_+ \chi_+  -2 u_\mu \langle 
u^\mu \chi_+ \rangle  \, .
\eeq
Furthermore, we set $u^2 = u_\mu u^\mu$.

\section{Renormalization of the self--energy graphs}
\label{sec:selfenergy}
\def\theequation{\arabic{section}.\arabic{equation}}
\setcounter{equation}{0}

This section is split into three paragraphs. In the first one, we
renormalize the self--energy graph at order $q^3$ together with
the two vertex--corrected self--energy diagrams at order $q^4$.
This can be done straightforwardly by the method proposed by 
Ecker~\cite{ecker}. In the next paragraph, we extend the heat kernel
methods to deal with operators which are orthgonal to the direction
given by the four--velocity of the heavy baryon. Such type of operators
start to appear at fourth order, specifically in the self--energy graph with
a dimension two insertion on the internal fermion line. We then 
consider this particular diagram, which we call the ``eye graph'' 
from here on. Formally, such a graph involves a triple
coincidence limit.  The eye graph  has thus to be worked
out making use of the novel technique to evaluate n--fold coincidences which
is presented and discussed in the third paragraph of this section. 
In addition, we show in appendix~\ref{app:eyecheck} that as long as one 
works to ${\cal O}(q^4)$, one can still use the  method developed by Ecker to
extract the singularities based on the heat kernel expansion for
part of the eye graph contribution.  Therefore, part of the divergent 
structure of the eye graph can be evaluated
by two different methods, which serves as an excellent check on the rather
involved algebra.

\subsection{${\cal O}(q^3)$ and vertex corrected self--energy graphs at
${\cal O}(q^4)$}
\label{sec:selfenergy0}

In this paragraph, we consider the renormalization of the self--energy
contributions $\Sigma_1^{(1)} (y,y)$ and $\Sigma_3^{(2)} (y,y)$. 
The pertinent divergences are due to the
singular behavior of the product of the meson and the baryon
propagators $ G_{ij} (x,y) \, [A_{(1)}]^{-1} (x,y)$
in the coincidence limit $x \to y$. This expression is directly 
proportional to the full
classical fermion propagator $S_{(1)}^{{\rm cl}}$.
However, this differential operator is not
elliptic and thus not directly amenable to the standard heat--kernel
expansion. Consider therefore the object \cite{ecker}
\beq \label{D1}
S_{(1)} = i \,  D_{(1)}^\dagger \, [ D_{(1)}  \, D_{(1)}^\dagger ]^{-1} 
\, \, , \quad D_{(1)} \equiv i \, A_{(1)} \,\,\, .
\label{D}
\eeq
In fact, the operator in the square brackets in Eq.(\ref{D}) is
positive definite and hermitian. Furthermore, it is a one--dimensional
operator in the direction of $v$ and we can use the heat kernel
methods for such type of operators invented by Ecker \cite{ecker} 
and discussed in detail in \cite{guido}. 
For completeness, we give the basic definitions.
Consider an differential operator of the type,\footnote{Note that in
  this section and in what follows, the symbol ``$d$'' is sometimes used for
  the derivative acting on the nucleon fields instead of ``$\nabla$''.}
\beq
\Delta  = - (v \cdot d)^2 + a(x) + \mu^2 \,\,   , \quad d_\mu = 
\partial_\mu + \gamma_\mu \,\,\, 
\eeq
and
\beq
a(x) = - g^2_A \, (S \cdot u)^2 \, +\,  g_A \, [i v \cdot \nabla, S
\cdot u ]  \,\,\, . 
\eeq 

\noindent
The heat kernel $J(t) = \exp\{-\Delta t\}= J_0 \, K$ 
can be split again into its free part,
\beq
\label{J0}
J_0 (t) =  g(x,y) \, \frac{1}{\sqrt{4\pi t}} \exp
  \biggl\{ -\mu^2\, t - \frac{[v \cdot (x-y)]^2}{4 t} \biggr\} \,\, , 
\eeq
and the interaction part $K$, which satisfies the equation,
\beq
\biggr[ \frac{\partial}{\partial \,t} - (v \cdot d)^2+ a(x) 
+ \frac{1}{t}\, v \cdot (x-y) \, v \cdot d \, \biggr] \, K(x,y,t)
= 0 \,\,\, ,
\label{H2eq}
\eeq
using $ v^4 =1$. 
The function $g(x,y)$ is introduced so that one can fulfill the
boundary condition  in the coordinate--space representation,
\beq
\label{gnuc}
g(x) = \int \frac{d^d p}{(2\pi)^{d-1}} \, \delta(k \cdot v)  \, 
{\rm e}^{-ipx} \,\, \, ,
\eeq
where the explicit form of the function $g(x)$ is not needed to
derive the recurrence relations for the heat kernel, but we demand
that $\partial g(x,y)$ is orthogonal to the direction of $v$,
\beq
v \cdot \partial \, g(x,y) = 0 \,\,\, .
\eeq
Later
when products of singular operators are constructed (see appendix~B)
we make use of this special form in Eq.(\ref{gnuc}).
It is important to stress the difference to the
standard case. Because $v \cdot d$ is a scalar, one has
essentially reduced the problem to a one--dimensional one, i.e.
$v \cdot d$ is a one--dimensional differential operator in the
direction of $v$. Using the heat kernel expansion for $K$,
\beq
K(x,y,t) =  \sum_{n=0}^\infty k_n(x,y) \, t^n \,\, ,
\eeq
where $t$ is the proper time, we derive the Seeley--deWitt
coefficients and their derivatives in the coincidence 
limit (denoted by the ``$|$''),
\beqa
k_0| &=& 1 \,\, , \, \,  (v \cdot d)^m \, k_0|=0 \,\, ,\,\,
(v \cdot d)^m \, k_0 \, (v \cdot
\stackrel{\leftarrow}{d})^n| = 0 \, \, ,\,\,
k_1|=-a\,\, , \nonumber \\
(v \cdot d) \, k_1| &=&-\frac{1}{2} \, [ \, v \cdot d , a \, ]
\,\, , \, \, 
(v \cdot d)^2 \, k_1|=-\frac{1}{3} \, [ \, v \cdot d , 
\, [ \, v \cdot d ,a \, ]] \,\, , \nonumber \\
k_2| &=&\frac{1}{2} \, a^2 - \frac{1}{6} \, [ \, v \cdot d, [ \, 
v \cdot d , a \, ] \, ] \,\, , \, \, k_1 \, v \cdot
\stackrel{\leftarrow}{d}| = v \cdot d \, k_1| \,\,\, .
\eeqa
The propagator is given as the integral in terms of the pertinent
Seeley--DeWitt coefficients,
\beqa
\label{Jn}
J(x,y) &=& \Delta^{-1}(x,y) = \int_0^\infty dt \, J(x,y,t) \nonumber
\\ J(x,y) &=& \sum_{n=0}^\infty J_n(x,y) \, k_n (x,y) \nonumber \\
J_n (x,y) &=&  g(x,y)  \, 
\int_0^\infty  \frac{dt}{\sqrt{4\pi t}} \exp
  \biggl\{ -\mu^2\, t - \frac{[v \cdot (x-y)]^2}{4 t} \biggr\} \, 
 t^n  \,\,\, , \nonumber \\
&=& \frac{\Gamma (n+1)}{v^2} \, \int \frac{dk^d}{(2\pi )^d} \,
\biggl[ {1 \over (v\cdot k)^2 + \mu^2} \biggr]^{n+1} \, e^{-ik\cdot
(x-y)} \,\,\,\, , 
\eeqa
where in the last line we have made use of the rest--frame
representation of $g(x,y)$, $g(x,y) = \delta^{d-1} (x-y)$ for 
$v = (1,0,\ldots,0)$.  The physical
interpretation of this result is rather simple. It is the Fourier
transform of $n+1$ massive propagators since we have $n$ insertions
and the $\Gamma$--function simply takes care about the combinatorics. 
Since the particle propagator $A_{(1)}$ is massless, one must
keep $\mu^2 \ne 0 $ in intermediate steps to get a well defined heat
kernel representation without infrared singularities. In case of the
vertex--corrected self--energy $\Sigma_3^{(2)}$, one has exactly
one of the dimension two vertices collected in Eq.(\ref{Vi2}).
For more details, in particular the recurrence relations between the 
Seeley--DeWitt coefficients and so on,  we refer the reader to 
refs.\cite{ecker}\cite{guido}. We add the important remark that the
coefficients $k_m$ have chiral dimension ${2m}$.
After the same type of algebra spelled  out in section~6 of \cite{guido},
one finds (notice that in intermediate steps covariance is destroyed by this 
method but a particular recombination of the terms allows to restore it)
\beqa \label{S11local}
\Sigma_1^{(1) , {\rm div}} (y,y) &=& \frac{1}{(4 \pi F )^2}  
\frac{2}{\epsilon} \, \, \delta^4 (x-y) \, \sum_{i=1}^{16}\,\,
 \hat{\Sigma}_{1,i}^{(1)} (y) \, \,\, ,\\
\label{S32local}
\Sigma_3^{(2) , {\rm div}} (y,y) &=& \frac{1}{(4 \pi F )^2}
\frac{2}{\epsilon} \, \, \delta^4 (x-y) \, \sum_{i=1}^{157}\,\,
 \hat{\Sigma}_{3,i}^{(2)} (y) \, \,\, 
\eeqa
where the 16 terms corresponding to the ${\cal O}(q^3)$ self--energy
can be extracted from appendix~C of ref.\cite{guido} if one 
transforms the SU(3) operators given there into their SU(2) 
counterparts\footnote{These terms have first been worked out in
\cite{ecker}. However,
since we are organizing the various contributions in a different
way than Ecker, the number of terms at this intermediate stage is
different.} and the 157 new 
${\cal O}(q^4)$ operators and their corresponding contributions to the
self--energy are collected in \cite{thesis}.

\subsection{Extension of the heat kernel methods}
\label{sec:hkex}

\noindent
The extraction of divergences in the meson--baryon sector is based
essentially on the method described in the previous section. The complete
information about the interaction is hidden in the Seeley--DeWitt
coefficients. Note that up to now in the meson--baryon sector we only
have considered the projection of a differential operator in direction
of $v$, e.g. we have calculated operators like $ v \cdot d $. This
procedure ensures the renormalization to third order as pointed out in
refs.\cite{ecker,guido}. We will see that the renormalization of
a part of the self--energy diagrams can be done by  applying the same methods. 
However, to fourth order we also have to renormalize insertions on the
intermediate baryon line of the form 
\beq
 (v \cdot d)^2  - d^2 
\eeq
which is orthogonal to $ v $. For the complete renormalization we thus
have to calculate the Seeley--DeWitt coefficients not only in the direction of
$v$. It is clear that we can not calculate 
\beq
 d^x_\mu k_m(x,y)
\eeq
from the recurrence relation in Eq.(\ref{H2eq}) in the coincidence
limit, because this relation always projects on the direction of $v$.
A direct calculation of  $ (d_\mu k_m(x,y)) |$  does not seem to be
possible but fortunately is not needed. The renormalization can be done 
by considering the sum of left and right covariant derivatives, 
\beqa
&&
(d^x_\mu k_m(x,y) +  k_m(x,y)  \stackrel{\leftarrow}{d}^y_\mu)
|_{x=y} \nonumber  \\ &&   
= ( \partial^x_\mu  k_m(x,y) + \Gamma_\mu(x) k_m(x,y) +
k_m(x,y) \stackrel{\leftarrow}{\partial}^y_\mu - k_m(x,y) \Gamma_\mu(y) )
|_{x=y}  \nonumber \\
&&=  ( \partial^x_\mu  (k_m(x,y) |_{x=y} ) + [  \Gamma_\mu(x) ,
(k_m(x,y) |_{x=y} ) ]  \nonumber \\
&&= [ d^x_\mu  , \,  (k_m(x,y) |_{x=y} ) ]  \\ && 
 \equiv \,\, 
 [ d^x_\mu  , \,  k_m | ]  \label{seeleynew}  \,\,\, .
\eeqa
While in the first line we differentiate before taking the
coincidence limit, in the last line we first perform the
coincidence limit. Since we know the coincidence limit of $ k_m |$,
this expression is well defined as pointed out above. 
Using this trick, we find for example 
\beqa
(d^x_\mu k_0(x,y) +  k_0(x,y)  \stackrel{\leftarrow}{d}^y_\mu)
|_{x=y}  
 &=&  [ d_\mu  , \,  k_0 | ]  = 0 \\
(d^x_\mu k_1(x,y) +  k_1(x,y)  \stackrel{\leftarrow}{d}^y_\mu)
|_{x=y}  
 &=&  [ d_\mu  , \,  k_1 | ]  = - [  d_\mu , a ] \\
(d^x_\mu  (v \cdot d k_1(x,y)) +  ( v \cdot d k_1(x,y))
\stackrel{\leftarrow}{d}^y_\mu) 
|_{x=y}  
 &=&  [ d_\mu  , \, v \cdot d  k_1 | ]  = - \frac{1}{2} [
 d_\mu ,  [ v \cdot d , a ]  ] 
\eeqa
This method allow us to calculate the complete eye graph as detailed
in the next paragraph.

\subsection{Eye graph at ${\cal O}(q^4)$ and treatment of n--fold
coincidences}
\label{sec:selfenergy2}

In this section, we consider the eye graph at ${\cal O}(q^4)$. It is
not obvious that the method described in paragraph~\ref{sec:selfenergy0} can be
applied to such type of diagram because if one dissects it into its 
lowest order (chiral dimension one) pieces, it formally involves a 
triple coincidence limit. We thus have develop a method to treat such terms.
So far, we only dealt with the product of one meson and one baryon
propagator, i.e. double coincidences to extract the pertinent
short--distance singularities. Formally, one Fourier--transforms
\beq
\int d^dx \, G_m(x) \, J_n (x) \, e^{i k \cdot x} \,\, \, ,
\eeq
and obtains
\beq
 \frac{\Gamma (n+1) \Gamma (m+1)}{v^2} \,  \, \int d\ell^d 
\biggl[{1 \over \ell^2} \biggr]^{m+1} \, \biggl[ {1\over 
[ (k - \ell ) \cdot v ]^2 + \mu^2 } \biggr]^{n+1} \,\, .
\eeq
The physical interpretation of this last formula is again very simple.
For $n$ insertions of external fields on the pion line and $m$
insertions on the nucleon one, we get the Fourier--transform of 
$n+1 \, (m+1)$ pion (nucleon)  propagators with the appropriate 
combinatorial factors given by the $\Gamma$--functions. For the eye
graph, we have to deal with triple coincidences of the type
\beq\label{triplebasic}
\int d^dx \, d^d y \, G_l(x+y) \, J_m (x) \, J_n (y) 
\, e^{i k \cdot x} \, e^{i q \cdot y}   \,\, \, .
\eeq
Again, in momentum--space this takes a simple form,
\beq \label{GJJ}
{\Gamma (l+1) \Gamma (m+1) \Gamma (n+1) \over (v^2)^{2}} \, 
\int {d\ell^d \over (2\pi)^d } 
\biggl[{1 \over \ell^2} \biggr]^{l+1} \, \biggl[ {1\over 
[(k -\ell ) \cdot v]^2 + \mu^2} \biggr]^{m+1} \, 
\biggl[{1 \over [(q- \ell) \cdot v]^2 + \mu^2 } \biggr]^{n+1}
\,\, .
\eeq
The interpretation of this formula is obvious. Furthermore, it can
easily be extended to n--fold coincidences which makes it useful for
many other applications. It is important to note that Eq.(\ref{GJJ})
can be reduced to already known integrals.
To be specific, introduce the abbreviations $\omega = v \cdot k$ and
 $\omega' = v \cdot q$. We can simplify the integral, which has
$(m+1)$ and $(n+1)$ insertions on the two intermediate nucleon
lines, with the standard trick,
\beq
\biggl[{1 \over \omega - v\cdot \ell}\biggr]^{2m+2} = 
{(-)^{2m+1} \over (2m+1) ! } \biggl({\partial \over \partial 
\omega}\biggr)^{2m+1} \,  {1 \over \omega - v\cdot \ell} \,\,\, ,
\eeq
which allows us to write  Eq.(\ref{GJJ}) in the form
\beqa \label{GJJ1}
&& {\Gamma (m+1) \Gamma (n+1) \over v^2} \, 
{(-)^{2m+1} \over (2m+1) ! }{(-)^{2n+1} \over (2n+1) ! } \times \nonumber \\
&& \biggl({\partial \over \partial \omega}\biggr)^{2l+1} \,
\biggl({\partial \over \partial \omega}\biggr)^{2m+1} \, 
\underbrace{
{\Gamma (l+1) \over v^2} \, \int {d\ell^d \over (2 \pi)^d} 
\biggl[{1 \over \ell^2} \biggr]^{l+1} \,  {1\over \omega - 
v \cdot \ell } \, 
{1 \over \omega' - v \cdot \ell   }
}_{\displaystyle = I(\omega,\omega')   } 
\,\, .
\eeqa
The remaining loop integral, which we denote as $I(\omega , \omega')$,
has only insertions on the pion line and none any more on the internal
nucleon one.  It can be further simplified by partial fractions,
\beqa 
I(\omega , \omega') &=&
\int d^dx \, d^d y \, G_l(x+y) \, v \cdot \partial J_0 (x) 
\,  v \cdot \partial J_0 (y) \, e^{i k \cdot x} \, e^{i q \cdot y}
 \nonumber \\ &=& 
{\Gamma (l+1) \over v^2} \,
\int {d\ell^d  \over (2 \pi)^d} \,   
\biggl[{1 \over \ell^2} \biggr]^{l+1} \, \biggl[
 {1\over \omega - v \cdot \ell } -  {1 \over \omega' - v \cdot \ell }
\biggr]\biggl( { -1 \over \omega - \omega'} \biggr) \nonumber \\
&=& - { I_0(\omega ) - I_0(\omega' ) \over \omega - \omega' } \\
I_0 (\omega)  &=&  {\Gamma (l+1) \over v^2} \,
 \int {d\ell^d  \over (2 \pi)^d} \,   
\biggl[{1 \over \ell^2} \biggr]^{l+1} \,  {1\over \omega - v \cdot
  \ell }  \,\,\, = \,\,
\int d^dx \, G_l(x) \, v \cdot \partial J_0 (x) \, e^{i k \cdot x} 
\,\,\,\, . \nonumber 
\eeqa
The integral $I_0 (\omega)$ is, however, already known from the
calculation of the leading order self--energy graph. Since  $I_0
(\omega)$ is a polynom in $\omega$, the numerator of last equation is
a polynom in $\omega - \omega'$. Consequently, the same factor
appearing in the denominator can always be canceled so that $I
(\omega ,\omega')$ is a rational polynom in $\omega$ and $\omega'$.
This is the reason why the method of differentiation allows to treat
such type of integrals, even in the  case of arbitrary many
coincidences. Let us briefly discuss this general case, i.e. a
one--loop diagram with $n$ insertions from ${\cal L}_{\pi N}^{(k)}$
$(k \ge 2)$ on the nucleon line. Denoting by $x$ and $x'$ the meson--nucleon
interaction points and by $x_i$ ($i=1,\ldots , n$) the points of the
insertions, we have to deal with a structure of the type
\beq
G_k (x-x') \, J_{m_{1}} (x-x_1) \,  J_{m_{2}} (x_1-x_2) \, \ldots
J_{m_{n}} (x_{n-1}-x_n) \, J_{m_{n+1}} (x_{n}- x') \, \,\,\, .
\eeq
Redefining $u_1 = x - x_1, u_2 = x_1-x_2, \ldots , u_n = x_{n-1} -
 x_n , u_{n+1} = x_n -x'$, this takes the form
\beq
G_k (u_1+u_2+\ldots +u_n+u_{n+1}) \, J_{m_{1}} (u_1) \,  J_{m_{2}} (u_2) \, \ldots
J_{m_{n}} (u_n) \, J_{m_{n+1}} (u_{n+1}) \, \,\,\, ,
\eeq
and  can be evaluated in momentum--space as described before, i.e. one
forms
\beq
\int \prod_{i=1}^{n+1} d^du_i \,\, G_k \biggl( \sum_{i=1}^{n+1} u_i
 \biggr) \,\prod_{i=1}^{n+1} \biggl[ J_{m_{i}} (u_i) \, {\rm e}^{i q_i \,
 u_i } \biggr] \,\,\, ,
\eeq
which is a rational polynom of the momenta $q_i$, 
$P(q_1, \ldots , q_{n+1} )$. This can now 
be treated exactly along the same lines as shown above for the special
case of a triple coincidence, i.e. with exactly one insertion from
${\cal L}_{\pi N}^{(2)}$.
The  results for the pertinent products of three singular
operators are collected in app.~\ref{app:triple}.
The whole eye graph can be evaluated using the
method described here. 
\medskip

\noindent Putting pieces together, the singularities
related to the eye graph can be extracted,
\beq \label{S12local}
\Sigma_1^{(2) , {\rm div}} (y,y) = \frac{1}{(4 \pi F )^2}  
\frac{2}{\epsilon} \, \, \delta^4 (x-y) \, \sum_{i=1}^{181}\,\,
 \hat{\Sigma}_{1,i}^{(2)} (y) \, \,\, ,
\eeq
with the  new 181 ${\cal O}(q^4)$ operators and their corresponding 
contributions to the self--energy are collected in 
appendix~\ref{app:eye}. More precisely, these terms are already partially summed
in that appendix and given are the contributions with zero, one and two
covariant derivatives acting on the nucleon fields.

\section{The counterterm Lagrangian}
\label{sec:LCT}
\def\theequation{\arabic{section}.\arabic{equation}}
\setcounter{equation}{0}

We are now in the position to enumerate the full counterterm
Lagrangian at order $q^4$. The $q^3$ terms can be found in~\cite{ecker}.
To bring it in a more compact form, we use the same relations as
discussed in ref.~\cite{guido}.
To separate the finite parts in dimensional regularization, we follow
the conventions of \cite{gl85} to decompose the irreducible one--loop
functional into a finite and a divergent part. Both depend on the
scale $\mu$:
\beqa
\label{split}
&& \delta^4(x-y) \, \Sigma^{(1)}_2 (x,y) + \Sigma^{(1)}_1 (x,y) 
 + \delta^4(x-y) \, \Sigma^{(2)}_2 (x,y) + \Sigma^{(2)}_1 (x,y) 
                                         + \Sigma^{(2)}_3 (x,y) 
\nonumber \\
&& = \delta^4(x-y) \, \Sigma^{(1)+(2), {\rm fin}}_2 (x,y, \mu) 
+ \Sigma^{(1)+(2), {\rm fin}}_1 (x,y, \mu) 
+ \Sigma^{(2), {\rm fin}}_3 (x,y, \mu) \nonumber \\
&& \quad 
-\frac{2 L(\mu )}{F^2} \, \delta^4 (x-y) \, [ \hat{\Sigma}_2^{(1)+(2)} (y) +
\hat{\Sigma}_1^{(1)+(2)} (y) + \hat{\Sigma}_3^{(2)} (y) \, ] \,\,\, ,
\eeqa
with
\beq\label{defL}
L(\mu ) = \frac{\mu^{d-4}}{(4 \pi)^2} \biggl\{ \frac{1}{d-4} -
\frac{1}{2} [ \log(4\pi) + 1 - \gamma  ] \biggr\} \,\, .
\eeq
The first two terms on the left--hand--side of Eq.(\ref{split}) are
of dimension three whereas the remaining three are of dimension four.
The generating functional can then be renormalized by introducing the
counterterm Lagrangian
\beqa \label{LCT}
{\cal L}^{\rm ct}_{\pi N} &=&
{\cal L}_{\pi N}^{(3)\, {\rm ct}} (x) + {\cal L}_{\pi N}^{(4)\,{\rm ct}} (x)
\nonumber \\ &=&
\frac{1}{(4 \pi F )^2} \, \sum_i
b_i \, \bar{N}_v (x) \, \tilde{O}^{(3)}_i (x) \, N_v (x)
+ \frac{1}{(4 \pi F )^2} \, \sum_i
d_i \, \bar{N}_v (x) \, \tilde{O}^{(4)}_i (x) \, N_v (x)
\eeqa
where the $b_i$ and $d_i$ are dimensionless coupling constants and the field
monomials $\tilde{O}^{(3)}_i (x)$ and $\tilde{O}^{(4)}_i (x)$ 
are of order $q^3$ and $q^4$, respectively. The low--energy 
constants $b_i$ and $d_i$ are decomposed in analogy to Eq.(\ref{split}),
\beqa
b_i &=& b_i^r (\mu ) + (4 \pi)^2 \, \beta_i \, L(\mu ) \,\,\, , 
\nonumber\\
d_i &=& d_i^r (\mu ) + (4 \pi)^2 \, \delta_i \, L(\mu ) \,\,\, .
\eeqa
The $\beta_i$ are dimensionless functions of $g_A$ constructed
such that they cancel the divergences of the one--loop functional to
order $q^3$. They have been listed by Ecker~\cite{ecker}.
The $\delta_i$ not only depend on $g_A$ but also on the finite
dimension two LECs $c_{1, \ldots,7}$. Their numerical values have
recently been determined in ref.~\cite{bkmlec} and independently
in ref.~\cite{moj}. There is by now yet another determination
available, see ref.~\cite{FMS}. The $\delta_i$ are listed in table~1 
together with the corresponding operators $\tilde{O}^{(4)}_i (x)$. 
For easier comparison, we also give the separate tables for the tadpole,
self--energy and eye graph counterterms and respective $\beta$--functions
in app.~\ref{app:tables}. Note that one can make the explicit
mass--dependence disappear from the $\beta$--functions if one makes
the $c_i$ dimensionless via~\cite{bkmlec}
\beq
c_i ' = 2m \, c_i \quad (i=1, \ldots ,7)~,
\eeq
so that the $\delta_i$ only depend on $g_A$ and the $c_i '$.
The operators listed in table~1 
constitute a complete set for the 
renormalization of the irreducible tadpole and self--energy functional
for {\it off-shell} baryons. These are the terms where the covariant 
derivative acts on the nucleon fields. There are 44 terms of this
type. As long as one is only
interested in Green functions with {\it on-shell} nucleons, the number
of terms can be reduced considerably by invoking the baryon equation
of motions. In particular, all equation of motion terms of the form
\beq
 i v \cdot \nabla \, N  = -g_A S \cdot u \, N 
\label{Eqmotion}
\eeq
can be eliminated by appropriate field redefinitions in complete
analogy to the order $q^3$ calculation \cite{eckerm}.
Also, many of the terms given in the
table refer to processes with at least three Goldstone bosons. These
are only relevant in multiple pion  production by photons or
pions off nucleons \cite{bkmpipin}.
As an illuminating example
we mention the elastic pion--nucleon scattering amplitude. At
second and third order, one has four and five counterterms,
respectively, if one also counts the deviation from the
Goldberger--Treiman relation (GTR), i.e. the difference between the
physical value of the pion--nucleon coupling constant and its
value as given by the GTR. Although the fourth order amplitude has
not yet been worked out in full detail, one can enumerate the possible
new counterterms from the structure of the amplitude. This leads to
only five new operators~(for details, see~\cite{ulfaus}). 
Certainly, there are much more low energy 
data points for $\pi N$ scattering. This pattern is in marked
contrast to the total number of terms, which are seven~\cite{bkkm},
31~\cite{FMS} and 199 for the second, third and fourth order, respectively.
Similar remarks apply to single 
pion production off nucleons. An analogous situation happens in the
mesonic sector. While there are about 130 counterterms at sixth (two
loop) order, only two new operators contribute to the elastic $\pi\pi$
scattering amplitude (and four others amount to quark mass 
renormalizations of dimension four LECs).

The renormalized LECs $b_i^r (\mu)$ and $d_i^r (\mu)$ are measurable 
quantities. They satisfy the renormalization group equations
\beqa \label{betai}
\mu \, \frac{d}{d \mu} b_i^r (\mu ) &=& -\beta_i \,\,\, ,
\nonumber \\
\mu \, \frac{d}{d \mu} d_i^r (\mu ) &=& -\delta_i \,\,\, .
\eeqa
Therefore, the choice of another scale $\mu_0$ leads to modified
values of the renormalized LECs,
\beqa
 b_i^r (\mu_0 ) &=&  b_i^r (\mu ) + \beta_i \, \log \frac{\mu}{\mu_0}
 \,\, , \nonumber \\ 
 d_i^r (\mu_0 ) &=&  d_i^r (\mu ) + \delta_i \, \log \frac{\mu}{\mu_0}
 \,\, .
\eeqa
We remark that the scale--dependence in the counterterm Lagrangian is,
of course, balanced by the scale--dependence of the renormalized
finite one--loop functional for observable quantities.
In addition to the terms listed in table~1, there are also a few more
finite counterterms, i.e. terms with $\delta_i = 0$. These are not
needed for the renormalization discussed here. We are presently
working
on setting up the minimal fourth order Lagrangian containing also
these finite terms.
%
%
\begin{table}
\caption{Counterterms to fourth order and their $\beta$-functions.
Here, $\tilde{A} = A - \langle A \rangle/2$.}
\vskip -0.7truecm
$$
\begin{tabular}{|r|c|c|} \hline  \label{tabq4}
i  & $\tilde{O}_i^{(4)}$ & $\delta_i$ \\ \hline
1  & $ \langle u_\mu \, u_\nu \rangle \, \langle u^\mu \, u^\nu \rangle
$ & $  -(c_2-g_A^2/8m)/6  + c_3 (g_A^2 +g_A^4/6)  $ \\
&& $ + (c_4 + 1/4m) ( -49 g_A^4/48)  $  \\
&& $ +(-1/8 + 5g_A^2/24 + 17g_A^4/24 + 9g_A^6/32)/2m $ \\
2  & $ \langle u^2 \rangle \, \langle u^2 \rangle
$ & $ -(c_2-g_A^2/8m)/12 + c_3 (-7g_A^2/4 -5g_A^4/16)  $ \\
&& $ +(c_4+1/4m) (g_A^2 + 19g_A^4/16) $ \\
 && $ + (-1/8 -55g_A^2/48 - 21g_A^4/16  -27g_A^6/64) / 2m $ \\
3  & $ \langle (v\cdot u)^2 \rangle \langle u \cdot u \rangle $  &
$  (c_2-g_A^2/8m) (1/3 -g_A^2/4 - 9g_A^4/16) $  \\
\hline
\end{tabular}
$$
\end{table}

$\,$

\addtocounter{table}{-1}

$\,$
\begin{table}
\caption{continued}
\vskip -0.7truecm
$$
\begin{tabular}{|r|c|c|} \hline
i  & $\tilde{O}_i^{(4)}$ & $\delta_i$ \\ \hline
23 &  $ i v_\rho \, \epsilon^{\rho\mu\nu\sigma} \, [u_\mu,u_\nu] 
 \langle u_\sigma  \, S \cdot u \rangle $ & 
$  +(c_4 +1/4m) (-g_A^2/2 + g_A^4/24)  $ \\
24 &  $ i v_\rho \, \epsilon^{\rho\mu\nu\sigma} \, 
 \langle  [u_\mu,u_\nu] \, S \cdot u \rangle \,u_\sigma   $ & 
$  (c_4 +1/4m) (-g_A^2 - 3g_A^4/8)  + (-3g_A^6/48)/2m   $ \\
25 &  $ i v_\rho \, \epsilon^{\rho\mu\nu\sigma} \, 
 \langle  [u_\mu,u_\nu] \,u_\sigma   \rangle \,  S \cdot u    $ & 
$  c_3 g_A^4/6 - 7 (c_4 +1/4m) g_A^4/24  +(-g_A^4/12 +  3g_A^6/48)/2m   $ \\
26 &  $  v_\rho \, \epsilon^{\rho\mu\nu\sigma} \,
 \tilde{F}_{\mu\nu}^+ \langle u_\sigma  \, S \cdot u \rangle $ & 
$  - (c_4 +1/4m) g_A^2 + (1+ c_6) (g_A^4/24)/2m  $ \\
27 &  $  v_\rho \, \epsilon^{\rho\mu\nu\sigma} \,
\langle {F}_{\mu\nu}^+ \rangle \, 
\langle u_\sigma  \, S \cdot u \rangle $ & 
$ + (1+ c_6+2 c_7) (g_A^4/16)/2m  $ \\
28 &  $  v_\rho \, \epsilon^{\rho\mu\nu\sigma} \,
 \langle \tilde{F}_{\mu\nu}^+  \,  S \cdot u \rangle  \,  u_\sigma  $ & 
$ - (c_4 +1/4m) g_A^2  -(1+ c_6) (g_A^4/8)/2m  $ \\
29 & $  v_\rho \, \epsilon^{\rho\mu\nu\sigma} \,
 \langle  \tilde{F}_{\mu\nu}^+ \, u_\sigma  \rangle \,   S \cdot u  $ & 
$ (1+ c_6) (g_A^4/24)/2m   $ \\
\hline
\end{tabular}
$$
\end{table}

\addtocounter{table}{-1}

$\,$
\begin{table}
\caption{continued}
\vskip -0.2truecm
$$
\begin{tabular}{|r|c|c|} \hline
i  & $\tilde{O}_i^{(4)}$ & $\delta_i$ \\ \hline
50 &  $ i v_\rho \, \epsilon^{\rho\mu\nu\sigma} \, S_\sigma \,
 \langle  [u_\mu, u_\nu] \,  \tilde\chi_+   \rangle $ & 
$  c_5 (-g_A^2 /2 - g_A^4/8)   $ \\
51 &  $ i v_\rho \, \epsilon^{\rho\mu\nu\sigma} \, S_\sigma \,
  [u_\mu, u_\nu] \langle   \chi_+   \rangle $ & 
$  c_1 (g_A^2 + g_A^4/4)  + (c_4+1/4m) (1/4 -g_A^2/16) $ \\
 && $ + c_3 g_A^2/2  + (5g_A^2/8 + 3 g_A^4/32 ) / 2m  $ \\
52 &  $ i  v_\rho \, \epsilon^{\rho\mu\nu\sigma} \, S_\sigma \,
 [  \tilde{F}_{\mu\kappa}^+ \, , \,   \tilde{F}_{\nu\kappa}^+ ] $ & 
$   - (c_4+1/4m)/3   + (5 g_A^2/6 ) / 2m      $ \\
53 &  $ i  v_\rho \, \epsilon^{\rho\mu\nu\sigma} \, S_\sigma \,
 [  \tilde{F}_{\mu\kappa}^+ v^\kappa \, , \,
 \tilde{F}_{\nu\kappa '}^+ v^{\kappa '} ] $ & 
$   (g_A^2 ) / 2m      $ \\
54 &  $  v_\rho \, \epsilon^{\rho\mu\nu\sigma} \, S_\sigma \,
  \tilde{F}_{\mu\nu}^+ \langle   \chi_+   \rangle $ & 
$ + 2 c_1 g_A^2  - (c_4+1/4m)/2    $ \\
&& $  + (1+c_6) ( 1/4 - g_A^2/16) /2m  + (5g_A^2 / 4 ) /2m   $ \\
\hline
\end{tabular}
$$
\end{table}

\addtocounter{table}{-1}

$ \, $
\begin{table}
\caption{continued}
\vskip -0.7truecm
$$
\begin{tabular}{|r|c|c|} \hline
i  & $\tilde{O}_i^{(4)}$ & $\delta_i$ \\ \hline
79 & $ \langle [  u_\mu , S \cdot u ] \, [ i  \nabla^\mu ,  v \cdot u
] \rangle $ & 
$ 5(c_4+1/4m)g_A/3 + (7g_A/12  - g_A^3/24 ) / 2m $ \\
80 & $ \langle [ u_\mu , v \cdot u ] \, [ i S \cdot \nabla ,  u^\mu  ]
 \rangle $ & 
  $ - (c_4+1/4m)g_A/3  + ( g_A/6 - g_A^3/6 + g_A^5/8) / 2m  $ \\
81 & $ \langle [ v \cdot u ,  u_\mu  ] \, [ i \nabla^\mu  , S
\cdot u ] \rangle $ & 
  $   (c_4+1/4m)g_A/3  + (5g_A/12 - 5g_A^3/3 - g_A^5/8 ) / 2m  $ \\ 
82 & $ \langle [ v \cdot u , S \cdot u ] \, [ i v \cdot \nabla , v
\cdot u ] \rangle $ & 
  $ -4 g_A (c_2 -g_A^2/8m) + c_3 (-4g_A -2g_A^3/3)  $ \\
&& $ + (c_4+1/4m) (-4g_A/3 + g_A^3/2 ) $ \\
&& $  +( -73g_A/6 -17g_A^3/12 -9g_A^5/8 ) / 2m  $ \\
83  &  $ \langle  u \cdot u   \rangle  \,  v \cdot u  \, i S \cdot
  \nabla $ + h.c.  &  
$ + ( g_A^5/8 ) / 2m $ \\
\hline
\end{tabular}
$$
\end{table}

\addtocounter{table}{-1}

$ \, $
\begin{table}
\caption{continued }
\vskip -0.7truecm
$$
\begin{tabular}{|r|c|c|} \hline
i  & $\tilde{O}_i^{(4)}$ & $\delta_i$ \\ \hline
108 & $  [ v^\mu \tilde{F}_{\mu\nu}^+ \, S^\nu ,   v \cdot u ] \,  v \cdot
\nabla  $   + h.c.  &
$  2 (c_2 -g_A^2/8m) g_A +  c_3 (4g_A)  - (c_4 +1/4m) g_A  $ \\
&& $ - (1+c_6) (g_A^3/4)/2m  - (6g_A) /2m  $ \\
109 & $  [  \tilde{F}_{\mu\nu}^+ \, S^\nu ,   v \cdot u ] \,  
\nabla^\mu  $   + h.c.  &
$    (g_A) /2m  $ \\
110 & $  [  \tilde{F}_{\mu\nu}^+ \, v^\nu ,   S \cdot u ] \,  
\nabla^\mu  $   + h.c.  &
$   +  (g_A^3/12) /2m  $ \\
111 & $  [ v^\mu \tilde{F}_{\mu\nu}^+ \, S^\nu ,    u_\kappa ]
\,  \nabla^\kappa  $   + h.c.  &
 $ (g_A)/2m  $ \\
112  & $  \langle  [\nabla^\kappa , v^\mu \tilde{F}_{\mu\nu}^+ \, S^\nu ]
 u_\kappa  \rangle  $ & 
$ -2 g_A(c_4+1/4m)/3  + (-5g_A/3 +2 g_A^3/3)/2m   $ \\
113 & $  \langle   \tilde{F}^+_{\mu\nu} v^\nu  \, [ \nabla^\mu , \,  S \cdot u
] \rangle      $ & $ (g_A^3)/2m $ \\
\hline
\end{tabular}
$$
\end{table}

\addtocounter{table}{-1}

$ \, $
\begin{table}
\caption{continued}
\vskip -0.7truecm
$$
\begin{tabular}{|r|c|c|} \hline
i  & $\tilde{O}_i^{(4)}$ & $\delta_i$ \\ \hline
139   & $  [ [ v \cdot \nabla , v  \cdot u] , \,  u_\mu  ]  \, \nabla^\mu
$  + h.c.
  & $ (1/2 + 3 g_A^2/16  - g_A^4/8 ) /2m   $\\   
140  & $  [ [ v \cdot \nabla ,  u_\mu ] , \,  v \cdot u  ]  \, \nabla^\mu
$  + h.c.
  & $ (1/2 - 3g_A^2/16  - g_A^4/8 ) /2m   $\\   
141  & $  [ [ \nabla^\mu ,  v \cdot u ] , \,  v \cdot u  ]  \, \nabla_\mu
$  + h.c.
  & $ (+1/2  -g_A^4/8 ) /2m   $\\   
142  & $  [ [ v \cdot \nabla  ,  u_\mu ] , \,   u^\mu  ]  \, v \cdot \nabla
$  + h.c.
  & $  - c_3 g_A^2  + (-7g_A^4/8 ) /2m   $\\   
143  & $  [ [ \nabla^\mu  ,  u^\nu ] , \,   u_\nu  ]  \,  \nabla_\mu
$  + h.c.
  & $ (g_A^4/8 ) /2m   $\\   
144  & $  [u_\mu , \,  [ \nabla^\mu  ,  v \cdot u ] ]  \,  v \cdot \nabla
$  + h.c.
  & $ (-7/12  - 89g_A^2/48) /2m   $\\   
145  & $  [ u_\mu ,  [ \nabla^\mu  ,  u_\nu ] ]  \, \nabla^\nu
$  + h.c.
  & $ (1/12 + 5g_A^2/12 ) /2m   $\\
\hline
\end{tabular}
$$
\end{table}

\addtocounter{table}{-1}

$ \, $
\begin{table}
\caption{continued }
$$
\begin{tabular}{|r|c|c|} \hline
i  & $\tilde{O}_i^{(4)}$ & $\delta_i$ \\ \hline
172 & $i  v_\rho \epsilon^{\rho\mu\nu\sigma} S_\sigma \,  \langle \, [ 
v \cdot \nabla  ,  u_\nu  ] \, v \cdot  u \rangle  \,  \nabla_\mu $  + h.c. & 
$ (- g_A^4/4 ) /2m  $  \\
173 & $i  v_\rho \epsilon^{\rho\mu\nu\sigma} S_\sigma \,  \langle \, [ 
\nabla_\mu  ,  u_\nu  ] \, v \cdot  u \rangle  \,  v \cdot \nabla $  + h.c. & 
$ (-2 g_A^2 - 3g_A^4/4 ) /2m  $  \\
174 & $i  v_\rho \epsilon^{\rho\mu\nu\sigma} S_\sigma \,  \langle \, [ 
v \cdot \nabla  ,  u_\mu  ] \,  u_\nu  \rangle  \,  v \cdot \nabla $  + h.c. & 
$  2 c_3 g_A^2  + ( g_A^2  + 7 g_A^4/4 ) /2m  $  \\
175 & $i  v_\rho \epsilon^{\rho\mu\nu\sigma} S_\sigma \,  \langle \, [ 
 \nabla_\kappa  ,  u_\mu  ] \,  u_\nu  \rangle  \,  \nabla^\kappa $  + h.c. & 
$ ( -g_A^4/4 ) /2m  $  \\
176 & $i  v_\rho \epsilon^{\rho\mu\nu\sigma} S_\sigma \, v \cdot
\stackrel{\leftarrow}{\nabla}  [v \cdot u , u_\mu ] \,  \nabla_\nu  $
+ h.c.  & 
 $ ( 6g_A^2 +g_A^4/2 ) / 2m $  \\
177 & $i  v_\rho \epsilon^{\rho\mu\nu\sigma} S_\sigma \, v \cdot
\stackrel{\leftarrow}{\nabla}  [u_\mu,u_\nu ] v \cdot \nabla  $ & 
$ (c_4+1/4m) g_A^2/2 -4 c_3 g_A^2 $ \\
&& $   + ( -6 g_A^2 - 7g_A^4/4 ) / 2m  $ \\
\hline
\end{tabular}
$$
\end{table}

\vfill\eject

\section{Sample calculations and checks}
\label{sec:sample}
\def\theequation{\arabic{section}.\arabic{equation}}
\setcounter{equation}{0}

In this section, we perform a few sample calculations. On one hand,
this shows how to use table~1 (or tables~2-4) for extracting the divergences for a
process under consideration, and on the other hand, since already
a large body of $q^4$ calculations exists~\cite{bkmrev}, it serves
as a good check on the rather involved manipulations leading to the
final results in table~1 (or tables~2-4). 

\subsection{Example I: The isoscalar nucleon magnetic moment}
\label{sec:isomm}
 
As a first example, we consider the isoscalar nucleon magnetic moment.
To order $q^3$, it is finite and divergences only appear at ${\cal
  O}(q^4)$. Here, we are only interested in these divergent pieces.
The complete expression including the finite pieces 
is given in ref.\cite{bkmplet}, where the
first corrections to the P--wave low--energy theorems~\cite{bkmpi0} 
in neutral pion photoproduction are worked out.
To be specific, consider the one--loop graph in Fig.\ref{fig2}a. Using
the Feynman rules given in~\cite{bkmrev}, its contribution to the
isoscalar magnetic form factor, which we denote by $I_M^s (\omega)$, is
\beq
I_M^s (\omega) = 3\, (1+ c_6 + 2c_7 ) \frac{g_A^2}{F_\pi^2} \frac{1}{2m} \, S_\mu \,
[S\cdot \epsilon , S\cdot k] \, S_\nu \, \frac{1}{i} \int \frac{d^d
  l}{(2\pi)^d} \, \frac{l_\mu l_\nu}{(v\cdot l - \omega) \,  v\cdot l
  \, (l^2 - M_\pi^2)}~,
\eeq
with $\omega = v \cdot k$ and $\epsilon$ the polarization vector of
the photon. The spin--matrices can be combined to give $[S\cdot
\epsilon ,  S\cdot k]/4$ and the d--dimensional integral for $\omega =
0$, i.e. for the magnetic moment, gives a contribution of the form
$(2\, M_\pi^2\, L + {\rm finite}) \, g_{\mu\nu}$. Putting pieces
together, we arrive at (we do not give the finite piece here)
\beq\label{isodiv}
I_M^{s, {\rm div}} (0) = \frac{3e}{4m} \, (1+ c_6 + 2c_7 ) \, 
\frac{g_A^2}{F_\pi^2} \, M_\pi^2 \, [S\cdot \epsilon , S\cdot k] \,
L~,
\eeq
with $L$ defined in Eq.(\ref{defL}).
We see that the divergent structure is proportional to $M_\pi^2$ and
the diagram is, of course, linear in the electromagnetic field
strength tensor. The pertinent fourth order counter term,
depicted in  Fig.\ref{fig2}b, is the one
numbered 55 in table~1 (or, since we are dealing with an example of
an eye graph, the operator L22 in table~\ref{eyetab}),
\beq
{\cal L}_{\pi N}^{(4)} = \bar{N} \bigl\{ \alpha \, v_\rho \,
\epsilon^{\rho\mu\nu\sigma} \, S_\sigma \, \langle F^+_{\mu\nu}\rangle
\, \langle \chi_+ \rangle \bigr\} \, N~,
\eeq
where the strength $\alpha$ will be specified shortly.  The pertinent
fourth order Feynman insertion from this part of the Lagrangian reads
\beq
\gamma NN-{\rm vertex:} \qquad i \, 16\, \alpha \, M_\pi^2 \, 
 [S\cdot \epsilon , S\cdot k] ~.
\eeq
Taking now $\alpha = \delta_{55}= (1+c_6+2c_7)(3g_A^2/64m)$ 
from table~1, the graph $I_M^{\rm  ct} (0)$ contributes as
\beqa \label{isoct}
I_M^{\rm ct} (0) &=& {\rm finite} - 16\,M_\pi^2 \cdot 3 \, (1+ c_6 + 2c_7 )
\, \frac{g_A^2}{F_\pi^2} \, \frac{e}{64 m}\,  [S\cdot \epsilon ,
S\cdot k] \, L \nonumber \\
&=&  {\rm finite}  -\frac{3e}{4m} \,  (1+ c_6 + 2c_7 ) \,
\frac{g_A^2}{F_\pi^2} \, M_\pi^2 \, [S\cdot \epsilon , S\cdot k] \,
L~,
\eeqa
so that by adding Eq.(\ref{isodiv}) and Eq.(\ref{isoct}), we see that
we are left with a finite piece. This is a particular simple example
to show how to use table~1 (or the tables in app.~\ref{app:tables})
and it checks exactly one operator.

\subsection{Example II: Scalar form factor of the nucleon}
\label{sec:CD}

The elastic pion--nucleon scattering amplitude has only been 
worked out to order $q^3$~\cite{moj,FMS} in HBCHPT. 
It already contains divergences at that
order. However, the so--called remainder at the Cheng--Dashen point,
which involves the scalar form factor of the nucleon, $\sigma_{\pi N}
(2M_\pi^2)$,  and the isospin--even $\pi N$ scattering amplitude
$\bar{D} (\nu , t)|_{\nu = 0 , t= 2M_\pi^2}$ (the `bar' means that the
nucleon pole graphs with pseudovector coupling are subtracted) have
been calculated at ${\cal O}(q^4)$~\cite{bkmcd}. Here, we consider
the scalar form factor of the nucleon for arbitrary values of the
squared momentum transfer $t$ at order $q^4$. For the particular 
kinematics at the Cheng--Dashen point, we recover of course the
results of ref.\cite{bkmcd}.

The various types of graphs are shown in fig.\ref{fig3}. These are most conveniently
evaluted in the Breit frame. However, one has to generalize the Breit frame 
kinematics appropriately to order $q^4$. Consequently, there are also 
contributions at this order from diagrams which formally start at order
$q^3$. To be precise, the product $v\cdot p$ picks up pieces that are
of order $q^2$ beyond the leading one $\sim q$
(which actually vanishes in the Breit frame). With this in mind,
it is straightforward to work out the contributions of graphs~\ref{fig3}b,c:
\beqa
I(3b) &=& 6 \, c_1\, \Delta_\pi \,\frac{M_\pi^2}{F_\pi^2} \,\, , \\
I(3c) &=& -6c_1 \frac{M_\pi^4}{F_\pi^2} I_0 (t) + 3\Big( c_2 - \frac{g_A^2}{8m}\Big)
I_2(t) + 3c_3\frac{M_\pi^2}{F_\pi^2} \Big( M_\pi^2 I_0(t) - \Delta_\pi + t I_1 (t)
\Big)~,
\eeqa
where the loop functions $\Delta_\pi, I_{0,1,2}(t)$ are given by
\beq
\frac{1}{i} \int \frac{d^d l}{(2\pi )^d} \frac{ \{1,l^2,l\cdot k,
  (v\cdot l)^2 \}}{[M_\pi^2 - l^2][M_\pi^2 - (l-k)^2]} =
\big\{ I_0 (t), -\Delta_\pi + M_\pi^2 I_0(t), t I_1(t) , I_2 (t)
\big\}~,
\eeq
with $t=k^2$. Putting the divergent pieces together, one gets
\beq\label{FFtad}
I^{(\rm div)}_{\rm tadpole} = 32\frac{M_\pi^4}{F_\pi^2} \Big( \frac{3}{4}\
c_1 - \frac{3}{32} \big( c_2 - \frac{g_A^2}{8m} \big) - \frac{3}{8}c_3 \Big)\, L
- 4\frac{M_\pi^2}{F_\pi^2}\, t \, \Big( -\frac{1}{8} \big( c_2 - \frac{g_A^2}{8m} \big)
- \frac{3}{4} c_3 \Big) \, L~.
\eeq
The pertinent tadpole counterterm structures are E1 and C3 (two derivatives) 
from table~\ref{tadtab}, which we list here together with the pertinent Feynman
insertions for a scalar--isoscalar source:
\beq
\langle \chi_+ \rangle \langle \chi_+ \rangle \, : \quad -32 i M_\pi^4~,\qquad
\langle [ \nabla_\mu, [\nabla^\mu  , \chi_+]] \rangle \, : \quad 4iM_\pi^2 t~.
\eeq
Injecting these in the genuine counterterm diagram Fig.\ref{fig3}a, one finds
\beq
I^{(\rm div)}_{\rm tadpole} = -I^{(\rm ct)}_{\rm tadpole}~,
\eeq
which is the desired result. Similarly, the divergent pieces of the
self--energy graphs (cf. fig~\ref{fig3}d,e) read
\beq
I^{(\rm div)}_{\rm self} = \Big(\frac{9}{4} \frac{g_A^2 M_\pi^4}{m F_\pi^2}
+\frac{1}{8} \frac{g_A^2 M_\pi^2}{m F_\pi^2} t \Big) \, L~.
\eeq  
Again, with the operators E1 and C3 (two derivatives) from table~\ref{selftab}
we find these divergences to be cancelled,
\beq
I^{(\rm div)}_{\rm self} = -I^{(\rm ct)}_{\rm self}~.
\eeq
Finally, consider the eye graphs (fig.~\ref{fig3}f,g,h,i). Their divergent parts read
\beqa\label{FFeye}
I(3f) &=& \Big[ -18 \frac{g_A^2 M_\pi^4}{F_\pi^2} c_1 - \frac{9}{16}
\frac{g_A^2 M_\pi^2}{m F_\pi^2} t \Big]\, L~, \nonumber \\
I(3g) &=& \Big[ -18 \frac{g_A^2 M_\pi^4}{F_\pi^2} c_1 \Big]\, L~, \\
I(3h+i) &=& \Big[ -\frac{45}{8} \frac{g_A^2 M_\pi^4}{F_\pi^2} + \frac{23}{16}
\frac{g_A^2 M_\pi^2}{m F_\pi^2} t \Big]\, L~, \nonumber
\eeqa
with the first contribution stems from the dimension three operator 
$O_{20}$ of Ecker's list~\cite{ecker}. At order $q^4$, it has the
Feynman insertion $-8M_\pi^2 \, b_{20}^{\rm div} \, v\cdot p$. The
operator $v\cdot p$ vanishes to leading order but has a finite piece
at second order so that this term can contribute here.\footnote{In principle,
the whole operator $O_{20}$ can be shifted to dimension four (and
higher) by using the nucleon equations of motion. We have checked that
adjusting the Z--factor and the other terms accordingly, one also
finds a cancellation of the divergences.} This remark also applies to
diagram~\ref{fig3}i, which already contributes at ${\cal O}(q^3)$ but
has an additional order $q^4$ piece due to the kinematics and thus is
of relevance here.  In addition to the already known operators
E1 and C3 (two derivatives) from table~\ref{eyetab}, we need 
the structure C5 from that table together with its pertinent Feynman insertion,
\beq
\stackrel{\leftarrow}{\nabla}_\mu \langle \chi_+ \rangle
\stackrel{\rightarrow}{\nabla}{}^\mu \, : \quad -4iM_\pi^2 p_1 \cdot p_2~,
\eeq
where $p_1 \, (p_2)$ is the incoming (outgoing) nucleon momentum and $p_1 \cdot
p_2 = -t/4$. Adding up these three terms amounts to
\beqa
I^{(\rm div)}_{\rm eye} &=& \Big[ (-i) \, (-32iM_\pi^4) \Big( -\frac{9}{8}
\frac{g_A^2}{F_\pi^2} c_1 - \frac{45}{256} \frac{g_A^2}{mF_\pi^2} \Big)\Big]
\, L    \nonumber \\
&+& 
 \Big[ (-i) \, (4iM_\pi^2 t) \Big( -\frac{5}{64} \frac{g_A^2}{m F_\pi^2} \Big)\Big]
\, L + \Big[ (-i) \, (iM_\pi^2 t) \Big( -\frac{9}{16} 
\frac{g_A^2}{m F_\pi^2} \Big)\Big]\, L
\nonumber \\
&=& \Big[ 36 \frac{g_A^2 M_\pi^4}{F_\pi^2} c_1 + \frac{45}{8}
\frac{g_A^2 M_\pi^4}{m F_\pi^2} - \frac{14}{16} \frac{g_A^2 M_\pi^2}{m F_\pi^2}t
\Big]\, L~,
\eeqa
which exactly cancels the terms in Eq.(\ref{FFeye}). Of course, one could have
also added all the divergences from the various types of graphs and then
use the operators 159, 161 and 181 
from the master table~1. This leads, of course, to the same
result. With that remark we conclude our discussion of the scalar form factor 
at order $q^4$. We have also checked the divergence structure of the
complete fourth order pion--nucleon scattering amplitude, which gives many additional
checks beyond the ones discussed here.\footnote{We are grateful to
  Nadia Fettes for providing us with the Feynman graph results.}

%

\section{Summary and conclusions}
\label{sec:summ}
\def\theequation{\arabic{section}.\arabic{equation}}
\setcounter{equation}{0}

In this paper, we have performed the chiral--invariant renormalization
of the effective two--flavor pion--nucleon field theory and
constructed the complete counterterm Lagrangian at next--to--leading one--loop 
order $q^4$. To incorporate the massive spin--1/2 degrees of freedom
(the nucleons), we have used
heavy baryon chiral perturbation theory in the path integral 
formulation \cite{bkkm}. This extends previous work by Ecker \cite{ecker},
who worked out the leading one--loop  divergences at order $q^3$.
The pertinent results of our study can be summarized as follows:

\begin{enumerate}

\item[(1)] At order $q^4$ in the chiral counting, one has to deal 
  with four irreducible diagrams involving  pion and
  nucleon propagators in the presence of external fields
  as shown in fig.~1. From these, the so--called
  eye graph involves a triple coincidence limit. 
  In section~\ref{sec:selfenergy2} we have developed a method to treat
  such multiple coincidence limites. All other contributions can be
  worked out using the standard double coincidence 
  limit~\cite{gl85,ecker,guido}. To deal with operators which are
  orthogonal to the direction defined by the nucleons' four--velocity
  (these only start to show up at order $q^4$), we have extended 
  the pertinent heat kernel methods as detailed in section~\ref{sec:hkex}.

\item[(2)] The method used destroys covariance in some 
  intermediate steps.  Of course, the final results are covariant. This is
  achieved by forming appropriate combinations of the operators.
  Furthermore, hermiticity is restored by again combining 
  appropriate terms. For completeness, some intermediate results for
  the eye--graph are given in app.~\ref{app:eye}. Combining all pieces
  results in the complete counterterm
  Lagrangian given in section~\ref{sec:LCT} in terms of the operators
  $\tilde{O}_i^{(4)}$, compare table~1. This table constitutes the
  central result of this work. To facilitate comparison and checks,
  we have also listed in app.~\ref{app:tables} the resulting operators and
  $\beta$--functions for the tadpole, self--energy and eye graphs, 
  respectively. The pertinent $\beta$--functions
  depend on the finite LECs $g_A$, $c_{1,\ldots,6}$ and the inverse 
  of the nucleon mass.

\item[(3)] We have performed a variety of checks on the rather
  involved manipulations leading to table~1. First, to third
  order we recover Ecker's result~\cite{ecker}. Second, since a variety of
  processes have been already calculated at order $q^4$ using
  Feynman diagram techniques including $\pi N \to \pi N$
  and $\gamma N \to N$, we have checked our results against
  these calculations as detailed in section~\ref{sec:sample}.
  In particular the complete fourth order amplitude for elastic 
  $\pi N$ scattering allows to check a large number of terms.

\end{enumerate}

\noindent To really address the question of isospin breaking 
alluded to in the
introduction, it is mandatory to construct also all
finite terms and to include virtual photons in the
pion--nucleon system  along similar lines.\footnote{For the case
of the purely pionic Lagrangian, see ref.\cite{mmspi}, and for the
pion--nucleon system to order $q^3$, see ref.\cite{ms}.} 
We hope to be able to report the results in the near future. 

\bigskip\bigskip


\section*{Acknowledgements}
We are grateful to Gerhard Ecker and Martin Moj\v zi\v s
for useful discussions and especially to Nadia Fettes for many independent
checks.  One of us (GM) thanks Prof. S. Weinberg and the Theory Group
at the University of Texas at Austin for discussions and hospitality during a 
stay when part of this work was done.

\bigskip\bigskip

\appendix
\def\theequation{\Alph{section}.\arabic{equation}}
\setcounter{equation}{0}
\section{Seeley--DeWitt coefficients}
\label{app:seeley}
In this appendix, we evaluate all possible Seeley--DeWitt coefficients up to
chiral dimension four (not all of them are used in the actual
calculation). The list presented below is the most
general which one gets for an elliptic differential operator of the form
\beq
- \, d^x_\mu \, d^x_\mu \,\, + \,\, a^x
\eeq
in Euclidean space, where we use the following definitions
 \beqa
d^x_\mu &=& \partial_\mu \,+ \,\gamma^x_\mu \,\,\,\, , \,\,\,\,
\stackrel{\leftarrow}{d}^y_\mu \,\,=\,\,
\stackrel{\leftarrow}{\partial}^y_\mu \,-\,
\stackrel{\leftarrow}{\gamma}^y_\mu \,\, , \\
\gamma_{\mu \nu} &=& \partial_\mu \, \gamma_\nu -\partial_\nu  \,\gamma_\mu + 
[ \, \gamma_\mu , \gamma_\nu \,] = [\, d_\mu \, , \,d_\nu\,] \,\,\, . 
\eeqa
The evaluation of the coefficients in the coincidence limit 
is described in \cite{guido} and is rather tedious. We introduce for a string
of derivatives acting on the  Seeley--DeWitt coefficients in the
coincident limit the notation
\beq
d_{\alpha}...d_\omega h_n| \,\, = \,\,
(\,d^x_{\alpha}...d^x_\omega h_n(x,y)\, ) \,\, |_{x\rightarrow y} 
\eeq
The result of such a calculation is always given by a string of commutators.
\beqa
\mbox{\underline{Seeley--DeWitt}} &-& \mbox{\underline{coefficients
    of dimension zero and one} }  
 \\
h_0| = 1\,\, && \,\, d_\mu h_0| = 0\,\, \,\,\,\,\,\,\,\,\,\,\,\, 
h_0 \stackrel{\leftarrow}{d}_\mu| =0  \\
\mbox{\underline{Seeley--DeWitt}} &-& \mbox{\underline{coefficients
    of dimension two} }  
 \\
d_\mu \, d_\nu \, h_0|  &=& d_\mu \, h_0 \,
\stackrel{\leftarrow}{d}_\nu| \, =\, h_0 \,
\stackrel{\leftarrow}{d}_\mu \, \stackrel{\leftarrow}{d}_\nu| \, = \, 
\frac{1}{2}\,\gamma_{\mu \nu} \nonumber \\
h_1| &=& - a \nonumber \\
\mbox{\underline{Seeley--DeWitt}} &-& \mbox{\underline{coefficients
    of dimension three} }  
 \\
d_\lambda \, d_\mu \, d_\nu \, h_0|  &=& d_\mu \, d_\lambda \, d_\nu
\, h_0| \,\, = \,\, \frac{1}{3}\, \biggl( [ \,
d_\lambda , \gamma_{\mu \nu}\,] + [\, d_\mu , \gamma_{\lambda \nu} \,]
\biggr) \,\, , \nonumber \\
d_\lambda \, d_\mu h_0 \, \stackrel{\leftarrow}{d}_\nu| &=&
d_\mu \, d_\lambda h_0 \, \stackrel{\leftarrow}{d}_\nu| 
 \,\,= \,\,\frac{1}{2} \,d_\lambda \, d_\mu \, d_\nu \, h_0| \,\,=
 \,\, \frac{1}{6}\, \biggl( [ \,
d_\lambda , \gamma_{\mu \nu}\,] + [\, d_\mu , \gamma_{\lambda \nu} \,]
\biggr)  \nonumber \\
d_\nu \, h_0 \, \stackrel{\leftarrow}{d}_\mu 
\, \stackrel{\leftarrow}{d}_\lambda| &=& d_\nu \, h_0 \,
\stackrel{\leftarrow}{d}_\lambda  \, \stackrel{\leftarrow}{d}_\mu|
\,\, = \,\, -\frac{1}{6} \, \biggl( \, [ 
d_\lambda , \gamma_{\mu \nu}] + [d_\mu\,, \gamma_{\lambda \nu}]
\biggr) \, \,\,  ,
\nonumber \\
h_0  \stackrel{\leftarrow}{d}_\nu  \stackrel{\leftarrow}{d}_\mu 
\, \stackrel{\leftarrow}{d}_\lambda| &=& 
h_0  \stackrel{\leftarrow}{d}_\nu  \stackrel{\leftarrow}{d}_\lambda 
\, \stackrel{\leftarrow}{d}_\mu| \,\,=\,\, 
- \frac{1}{3} \, \biggl( \, [ 
d_\lambda , \gamma_{\mu \nu}] + [d_\mu \,, \gamma_{\lambda \nu}] \,
\biggr) \nonumber  \\
d_\mu \, h_1| &=& -\frac{1}{2}\, [\, d_\mu ,a \,] + \frac{1}{6} \, [
\, d_\nu , \gamma_{\mu \nu}\,] \,\, , \nonumber \\ 
h_1 \, \stackrel{\leftarrow}{d}_\mu| &=& d_\mu \, h_1| - \frac{1}{3}
[d_\nu , \gamma_{\mu \nu}] = -\frac{1}{2}[d_\mu , a] - \frac{1}{6}
[d_\nu , \gamma_{\mu \nu}]  \nonumber \\ 
\mbox{\underline{Seeley--DeWitt}} &-& \mbox{\underline{coefficients
    of dimension four} }  
\\
d_\alpha \, d_\beta \, d_\gamma \, d_\delta \, h_0| &=& - \frac{1}{8}
 \biggl( \,
\{ \gamma_{\delta \alpha} \, , \, \gamma_{\beta \gamma} \} +
\{ \gamma_{\gamma \alpha} \, , \, \gamma_{\beta \delta} \} +
\{  \gamma_{\delta \gamma} \, , \,  \gamma_{\alpha \beta} \} \biggl)
 \nonumber \\
&& -  \frac{1}{4} \biggl( 
  [ d_\beta , [ d_\gamma , \gamma_{\delta \alpha} ]] 
 + [ d_\alpha , [ d_\beta , \gamma_{\delta \gamma} ]] 
 + [ d_\alpha , [ d_\gamma , \gamma_{\delta \beta} ]] \,  \biggl) \nonumber  \\
&=& d_\beta \, d_\alpha \, d_\gamma \, d_\delta \, h_0| \,\, = \,\,
 d_\alpha \, d_\gamma \, d_\beta \, d_\delta \, h_0| \,\, = \,\,
 d_\delta \, d_\gamma \, d_\beta \, d_\alpha \, h_0| \nonumber \\
d_\alpha \, d_\beta \, d_\gamma \, d_\gamma \, h_0| &=& - \frac{1}{4}
 \biggl( \,
\{ \gamma_{\gamma \alpha} \, , \, \gamma_{\beta \gamma} \} +
  [ d_\beta , [ d_\gamma , \gamma_{\gamma \alpha} ]] 
 + [ d_\alpha , [ d_\gamma , \gamma_{\gamma \beta} ]] \,
 \biggl) \nonumber \\
d_\alpha \, d_\alpha \, d_\gamma \, d_\gamma \, h_0| &=&  \frac{1}{2}
\gamma_{\alpha\gamma} \,  \, \gamma_{\alpha \gamma}  \nonumber \\
d_\alpha \, d_\beta \, d_\gamma h_0 \stackrel{\leftarrow}{d}_\delta|
&=&
\frac{1}{3} d_\alpha \, d_\beta \, d_\gamma \, d_\delta \, h_0| \,\,
 - \frac{1}{6} \biggl( \, 
 \gamma_{\gamma \beta} \,  \, \gamma_{\alpha \delta}  
+ \gamma_{\gamma \alpha} \,  \, \gamma_{\beta \delta}  
+ \gamma_{ \beta\alpha} \,  \, \gamma_{\gamma \delta}  
  \biggl) \nonumber  \\
d_\delta d_\gamma \, h_0 \stackrel{\leftarrow}{d}_\beta
  \stackrel{\leftarrow}{d}_\alpha|  
&=&
\frac{1}{2} \biggl( d_\delta \, h_0 \stackrel{\leftarrow}{d}_\gamma
  \stackrel{\leftarrow}{d}_\beta \stackrel{\leftarrow}{d}_\alpha|  
+ d_\gamma \, h_0 \stackrel{\leftarrow}{d}_\delta
  \stackrel{\leftarrow}{d}_\beta \stackrel{\leftarrow}{d}_\alpha| \,
  \biggl) \,\, - \,\,
\frac{1}{4} \gamma_{\delta\gamma}\, \gamma_{\beta\alpha} 
\nonumber \\
d_\delta \, h_0 \stackrel{\leftarrow}{d}_\gamma
  \stackrel{\leftarrow}{d}_\beta \stackrel{\leftarrow}{d}_\alpha|  
&=&
\frac{1}{3} h_0 \stackrel{\leftarrow}{d}_\delta
  \stackrel{\leftarrow}{d}_\gamma \stackrel{\leftarrow}{d}_\beta 
 \stackrel{\leftarrow}{d}_\alpha|  \,
 - \frac{1}{6} \biggl( \, 
 \gamma_{\delta \gamma} \,  \, \gamma_{\alpha \beta}  
+ \gamma_{\delta \beta} \,  \, \gamma_{\alpha \gamma}  
+ \gamma_{ \delta \alpha} \,  \, \gamma_{\beta\gamma}  
  \biggl) \nonumber  \\
h_0 \stackrel{\leftarrow}{d}_\delta  \stackrel{\leftarrow}{d}_\gamma
  \stackrel{\leftarrow}{d}_\beta 
 \stackrel{\leftarrow}{d}_\alpha| &=& - \frac{1}{8}
 \biggl( \,
\{ \gamma_{\delta \alpha} \, , \, \gamma_{\beta \gamma} \} +
\{ \gamma_{\gamma \alpha} \, , \, \gamma_{\beta \delta} \} +
\{ \gamma_{\alpha \beta} \, , \, \gamma_{\delta  \gamma} \} \biggr) \nonumber\\
&& +  \frac{1}{4} \biggl( 
  [ d_\beta , [ d_\gamma , \gamma_{\delta \alpha} ]] 
 + [ d_\alpha , [ d_\beta , \gamma_{\delta \gamma} ]] 
 + [ d_\alpha , [ d_\gamma , \gamma_{\delta \beta} ]] \,  \biggl) 
\nonumber  \\
h_0 \stackrel{\leftarrow}{d}_\gamma  \stackrel{\leftarrow}{d}_\gamma
  \stackrel{\leftarrow}{d}_\beta 
 \stackrel{\leftarrow}{d}_\alpha| &=& - \frac{1}{4} \biggl( \,
\{ \gamma_{\gamma \alpha} \, , \,  \gamma_{\beta \gamma} \}   
 - [ d_\beta , [ d_\gamma , \gamma_{\gamma \alpha} ]] 
 - [ d_\alpha , [ d_\gamma , \gamma_{\gamma \beta} ]] \,  \biggl) 
\nonumber  \\
h_0 \stackrel{\leftarrow}{d}_\gamma  \stackrel{\leftarrow}{d}_\gamma
  \stackrel{\leftarrow}{d}_\alpha 
 \stackrel{\leftarrow}{d}_\alpha| &=& \frac{1}{2} \,\, 
\gamma_{\gamma \alpha} \,   \gamma_{ \gamma \alpha}  
\nonumber  \\
d_\mu \, d_\nu \, h_1| &=& -\frac{1}{3}\, \biggl( \, [\, d_\mu ,[\,
d_\nu ,a \,]] + \, \gamma_{\mu\nu} \,  \, a \,  \,  + \,
\frac{1}{2} \, a \, \gamma_{\mu\nu} \, \biggl) \nonumber \\
&& 
+\frac{1}{12}\, \biggl( \, 
[\, d_\gamma ,[\, d_\mu , \gamma_{\nu\gamma} \,] + \, 
[\, d_\nu , \gamma_{\mu\gamma} \,]\,] \,
+\, \{ \, \gamma_{\mu\gamma}\, , \, \gamma_{\nu\gamma}  \, \} \, \biggl)
 \nonumber \\
d_\mu \, d_\mu \, h_1| &=& -\frac{1}{3}\, [\, d_\mu ,[\, d_\mu ,a \,]]
 + \frac{1}{6} \, \gamma_{\mu \alpha} \, \gamma_{\mu \alpha} \nonumber
 \\
d_\mu h_1  \stackrel{\leftarrow}{d}_\nu|  
&=& -\frac{1}{6}\, \biggl( \, [\, d_\mu ,[\,
d_\nu ,a \,]] + \, \gamma_{\mu\nu}  \, a  \,  + \,
 \, 2 \, a \, \gamma_{\mu\nu} \, \biggl) \nonumber \\
&& 
+\frac{1}{12}\, \biggl( \, 
[\, d_\gamma , [\, d_\gamma , \gamma_{\mu\nu} \,]\, ] \,
-\, \{ \, \gamma_{\mu\gamma}\, , \, \gamma_{\nu\gamma}  \, \} \, 
- 2 \, [ \, \gamma_{\mu\gamma}\, , \, \gamma_{\nu\gamma}  \, ] \, \biggl)
 \nonumber \\
d_\mu h_1  \stackrel{\leftarrow}{d}_\mu|  
&=& -\frac{1}{6}\,   [\, d_\mu ,[\,
d_\mu ,a \,]]   
-\frac{1}{6}\,  \gamma_{\mu\gamma}\,  \gamma_{\mu\gamma}  
\nonumber \\
h_1  \stackrel{\leftarrow}{d}_\nu  \stackrel{\leftarrow}{d}_\mu| 
&=& -\frac{1}{3}\, \biggl( \, [\, d_\mu ,[\,
d_\nu ,a \,]] + \frac{1}{2} \, \gamma_{\nu\mu}  \, a  \,  + \,
 \, a \, \gamma_{\nu\mu} \, \biggl) \nonumber \\
&& 
-\frac{1}{12}\, \biggl( \, 
[\, d_\gamma ,[\, d_\mu , \gamma_{\nu\gamma} \,] + \, 
[\, d_\nu , \gamma_{\mu\gamma} \,]\,] \,
-\, \{ \, \gamma_{\mu\gamma}\, , \, \gamma_{\nu\gamma}  \, \} \, \biggl)
 \nonumber \\
h_1  \stackrel{\leftarrow}{d}_\mu  \stackrel{\leftarrow}{d}_\mu 
&=& -\frac{1}{3}\,  \, [\, d_\mu ,[\,d_\mu ,a \,]]
+ \frac{1}{6} \, \gamma_{\mu\gamma}\, \, \gamma_{\mu\gamma}  
 \nonumber \\
h_2| &=& \frac{1}{2}\, a^2 - \frac{1}{6} \,[d_\mu,[d_\mu,a]]+
\frac{1}{12} \, (\gamma_{\mu \nu})^2 \,\, ,
\eeqa 

\def\theequation{\Alph{section}.\arabic{equation}}
\setcounter{equation}{0}
\section{Products of singular operators}
\label{app:singops}

In this appendix we consider the singularities arising from the
product of the baryon and meson propagators in the evaluation of the
self--energy diagram.
The corresponding singularities can best be extracted
in Euclidean $d$-dimensional Fourier space \cite{jacko}.
For some general remarks on the treatment of such singular products,
see appendix~B of \cite{guido} and references therein. Let $G_n (x,y)$
and $J_n (x,y)$ be the corresponding expansion coefficients for the 
meson and baryon propagators at proper time $t^n$, respectively. 
We thus have to deal with integrals of the type
\beq \label{intGJ}
\int d^dx \, G_n(x) \, J_m (x) \, {\rm e}^{ikx} \,\,\, .
\eeq
To evaluate it, we need a specific representation of the function
$g(x)$, which appears in the heat kernel of the one--dimensional
operator in the direction of the four--velocity $v$ \cite{ecker}
introduced to handle the self--energy graphs,
\beq
g(x) = \int \frac{d^d p}{(2\pi)^{d-1}} \, \delta(k \cdot v)  \, 
{\rm e}^{-ipx} \,\, \, .
\eeq
Straightforward algebra leads to the following list (which extends the
one given in ref.\cite{ecker})  that contains 
all singular products appearing in $\Sigma_1$ (with $\ve =
4-d$). Note that derivatives only act on the functions directly to
their right:
\beqa
G_0(x-y) J_0(x-y) &\sim& - \frac{4}{(4\pi)^2v^2} \frac{1}{\ve} \delta^d(x-y)\\
G_0(x-y)v \cdot \partial J_0(x-y) &\sim&  \frac{4}{(4\pi)^2v^2}
\frac{1}{\ve} v \cdot \partial \delta^d(x-y) \\
\partial_\mu G_0(x-y) J_0(x-y) &\sim& - \frac{8}{(4\pi)^2 }
\frac{1}{\ve} v_\mu v \cdot \partial \delta^d(x-y)\\
\partial_\mu G_0(x-y) v \cdot \partial J_0(x-y) &\sim& \frac{4}{(4\pi)^2 }
\frac{1}{\ve} v_\mu (v \cdot \partial)^2 \delta^d(x-y)\\
\partial_\mu G_0(x-y) v \cdot \partial J_1(x-y) &\sim& - \frac{4}{(4\pi)^2 }
\frac{1}{\ve} v_\mu   \delta^d(x-y)\\
\partial_\mu G_1(x-y) v \cdot \partial 
J_0(x-y) &\sim&  \frac{2}{(4\pi)^2 v^2}
\frac{1}{\ve} v_\mu   \delta^d(x-y)\\
\partial_\mu \partial_\nu G_0(x-y) J_0(x-y) &\sim&
 \frac{4}{(4\pi)^2}  \frac{1}{\ve}
 \delta_{\mu\nu} (v \cdot \partial)^2
\delta^d(x-y) \nonumber \\ 
&-& \frac{16}{(4\pi)^2v^2}  \frac{1}{\ve}
 v_\mu v_\nu  (v \cdot \partial)^2
\delta^d(x-y) \\
 \partial_\mu \partial_\nu G_0(x-y)v \cdot \partial J_0(x-y) &\sim&
- \frac{4}{3(4\pi)^2}  \frac{1}{\ve}
 \delta_{\mu\nu} (v \cdot \partial)^3
\delta^d(x-y) \nonumber \\
&+& \frac{16}{3(4\pi)^2v^2}  \frac{1}{\ve}
 v_\mu v_\nu  (v \cdot \partial)^3
\delta^d(x-y) \\
\partial_\mu \partial_\nu G_0(x-y) J_1(x-y) &\sim&
- \frac{4}{3(4\pi)^2} \frac{1}{\ve}
\delta_{\mu\nu} \delta^d(x-y) \nonumber \\
&+& \frac{16}{3(4\pi)^2v^2}  \frac{1}{\ve}
 v_\mu v_\nu \delta^d(x-y) \\
 \partial_\mu \partial_\nu G_0(x-y)v \cdot \partial  J_1(x-y) &\sim&
 \frac{4}{(4\pi)^2}  \frac{1}{\ve}
 \delta_{\mu\nu} v \cdot \partial
\delta^d(x-y) \nonumber \\
&-& \frac{16}{(4\pi)^2 v^2}  \frac{1}{\ve}
 v_\mu v_\nu  (v \cdot \partial)
\delta^d(x-y) \\
 \partial_\mu \partial_\nu G_1(x-y) J_0(x-y) &\sim&
 \frac{2}{(4\pi)^2v^2}  \frac{1}{\ve}
 \delta_{\mu\nu} \delta^d(x-y) \nonumber \\
&-& \frac{4}{(4\pi)^2 }  \frac{1}{\ve}
 v_\mu v_\nu  
\delta^d(x-y) \\
 \partial_\mu \partial_\nu G_1(x-y)v \cdot \partial J_0(x-y) &\sim&
- \frac{2}{(4\pi)^2v^2}  \frac{1}{\ve}
\delta_{\mu\nu} v \cdot \partial
\delta^d(x-y) \nonumber \\
&+& \frac{4}{(4\pi)^2}  \frac{1}{\ve}
 v_\mu v_\nu  (v \cdot \partial) 
\delta^d(x-y) \\
 \partial_\mu \partial_\nu \partial_\tau 
G_0(x-y) J_0(x-y) &\sim&
 \frac{16}{3(4\pi)^2v^2}  \frac{1}{\ve}
\delta_{\mu\nu} v_\tau (v \cdot \partial)^3
\delta^d(x-y) \nonumber \\
&-& \frac{32}{(4\pi)^2}  \frac{1}{\ve}
 v_\mu v_\nu v_\tau  (v \cdot \partial)^3 
\delta^d(x-y) \\
 \partial_\mu \partial_\nu \partial_\tau 
G_0(x-y) v \cdot \partial J_0(x-y) &\sim&
- \frac{4}{3(4\pi)^2v^2}  \frac{1}{\ve}
\delta_{\mu\nu} v_\tau (v \cdot \partial)^4
\delta^d(x-y) \nonumber \\
&+& \frac{8}{(4\pi)^2}  \frac{1}{\ve}
 v_\mu v_\nu v_\tau  (v \cdot \partial)^4 
\delta^d(x-y) \\
 \partial_\mu \partial_\nu \partial_\tau 
G_0(x-y)  J_1(x-y) &\sim&
- \frac{16}{3(4\pi)^2 v^2}  \frac{1}{\ve}
\delta_{\mu\nu} v_\tau (v \cdot \partial)
\delta^d(x-y) \nonumber \\
&+& \frac{32}{(4\pi)^2}  \frac{1}{\ve}
 v_\mu v_\nu v_\tau  (v \cdot \partial) 
\delta^d(x-y) \\
 \partial_\mu \partial_\nu \partial_\tau 
G_0(x-y) v \cdot \partial J_1(x-y) &\sim&
\frac{8}{(4\pi)^2v^2}  \frac{1}{\ve}
\delta_{\mu\nu} v_\tau (v \cdot \partial)^2
\delta^d(x-y) \nonumber \\
&-& \frac{48}{(4\pi)^2}  \frac{1}{\ve}
 v_\mu v_\nu v_\tau  (v \cdot \partial)^2 
\delta^d(x-y) \\
 \partial_\mu \partial_\nu \partial_\tau 
G_0(x-y) v \cdot \partial J_2(x-y) &\sim&
- \frac{8}{3(4\pi)^2v^2}  \frac{1}{\ve}
\delta_{\mu\nu} v_\tau 
\delta^d(x-y) \nonumber \\
&+& \frac{16}{(4\pi)^2}  \frac{1}{\ve}
 v_\mu v_\nu v_\tau  
\delta^d(x-y) \\
 \partial_\mu \partial_\nu \partial_\tau 
G_1(x-y) J_0(x-y) &\sim&
+ \frac{4}{(4\pi)^2}  \frac{1}{\ve}
\delta_{\mu\nu} v_\tau (v \cdot \partial)
\delta^d(x-y) \nonumber \\
&-& \frac{16}{(4\pi)^2}  \frac{1}{\ve}
 \frac{v_\mu v_\nu}{v^2} v_\tau  (v \cdot \partial) 
\delta^d(x-y) \\
 \partial_\mu \partial_\nu \partial_\tau 
G_1(x-y) v \cdot \partial J_0(x-y) &\sim&
- \frac{2}{(4\pi)^2}  \frac{1}{\ve}
\delta_{\mu\nu} v_\tau (v \cdot \partial)^2
\delta^d(x-y) \nonumber \\
&+& \frac{8}{(4\pi)^2 v^2}  \frac{1}{\ve}
 v_\mu v_\nu v_\tau  (v \cdot \partial)^2 
\delta^d(x-y) \\
 \partial_\mu \partial_\nu \partial_\tau 
G_1(x-y) v \cdot \partial J_1(x-y) &\sim&
 \frac{2}{(4\pi)^2}  \frac{1}{\ve}
\delta_{\mu\nu} v_\tau 
\delta^d(x-y) \nonumber \\
&-& \frac{8}{(4\pi)^2v^2}  \frac{1}{\ve}
 v_\mu v_\nu v_\tau   
\delta^d(x-y) \\
 \partial_\mu \partial_\nu \partial_\tau 
G_2(x-y) v \cdot \partial J_0(x-y) &\sim&
- \frac{1}{(4\pi)^2v^2}  \frac{1}{\ve}
\delta_{\mu\nu} v_\tau 
\delta^d(x-y) \nonumber \\
&+& \frac{2}{(4\pi)^2}  \frac{1}{\ve}
 v_\mu v_\nu v_\tau  
\delta^d(x-y) \,\,\, .          
\eeqa

\def\theequation{\Alph{section}.\arabic{equation}}
\setcounter{equation}{0}
\section{Triple products of singular operators}
\label{app:triple}

The integrals of the type Eq.(\ref{triplebasic}) can be worked out
in a variety of ways. One possibility is to reduce the triple
coincidence integral to a one--dimensional one,
\beqa \label{master}
I_3 &=& \int d^dx \, d^d y \,\, G_l(x+y) \, J_m (x) \, J_n (y) 
\, e^{i k \cdot x} \, e^{i q \cdot y}   \nonumber \\
&=& { \Gamma \bigl( \frac{3-d}{2} + l \bigr) \over
\sqrt{\pi} (4 \pi )^{d/2} \, (v^2)^{n+m+2} } \, \sum_{\nu = 0}^\infty 
\, \int_0^1 dt \, {d-3-2l \choose 2\nu } \, t^m \, [tk+(1-t)q]^{2\nu}
\, (1-t)^n \, \times \nonumber \\
&& \!\!\!\!\!\!\!\!\!\!\!\!\!\!\!\!\!\!\!
[t(1-t)(k-q)^2]^{-(3+m+n+l+\nu-d/2)} \,\, \Gamma \biggl( {d-2 \over
2} - l - \nu \biggr) \, \Gamma \biggl({6-d \over 2} +m+n +l + \nu \biggr)
\,\,\, .
\eeqa
It is then straightforward to extract the more complicated forms when
derivative operators are acting on one, two or three of the propagator
functions. In fact, only when a sufficient number of
derivatives acts on the various propagator functions, one encounters
singularities. For the eye graph under consideration, one has
up to six derivatives inside the integral. Still, all such terms can
be reduced to the form given above and it is thus appropriate to give
it here. 

Before listing the pertinent results, we give a short example to
illustrate the calculational procedure. Consider the product
\beqa
&& \int d^dy \, \int d^dx \, S \cdot \partial \, S \cdot \partial \,
G_l (x+y) \, J_m(x) \, J_n(y) \, {\rm e}^{ikx} \,{\rm e}^{iqy}
\nonumber \\
&=& -{1\over 2}\, S_\mu \, S_\nu \, \delta^{\mu \nu} \,  \int d^dy \, \int d^dx \,
G_{l-1} (x+y) \, J_m(x) \, J_n(y) \, {\rm e}^{ikx} \,{\rm e}^{iqy} \,\,\,
, \eeqa
where we have used that $S \cdot v =0$ to arrive at the second line.
This integral is only divergent for $l = m = n  =0$ so that use of 
Eq.(\ref{master}) leads to 
\beq
 \int d^dy \, \int d^dx \,S \cdot \partial \, S \cdot \partial \,
 G_0 (x+y) \, J_0(x) \, J_0(y) \, 
{\rm e}^{ikx} \,{\rm e}^{iqy} \sim {\Gamma(1-d/2) \over \sqrt{\pi}} \,
{1 \over (4\pi )^{d/2} \, v^4} \, \Gamma (d/2) \, \Gamma(\ve /2) \,\,\, .
\eeq
Using $v^4 =1$ and working out the product of the various
$\Gamma$--functions and reinstating all prefactors, 
one arrives at the result given in Eq.(\ref{rex}) below.

Straightforward algebra allows then to extract in a similar fashion
the pertinent triple products (in our case with one meson and 
two nucleon propagators):
\beq\label{rex}
S\cdot\partial\,S\cdot\partial\,G_0(x+y)\,J_0(x)\,J_0(y) \sim 
- \frac{4}{3} \frac{S_\mu S_\nu\,\delta_{\mu\nu}}{(4\pi)^{d/2}} 
\frac{1}{\epsilon} \delta^d(x)\,\delta^d(y)\eeq
\beq
G_0(x+y)\,v\cdot\partial\,J_0(x)\,v\cdot\partial\,J_0(y) \sim 
\frac{4}{v^2} \frac{1}{(4\pi)^{d/2}}
\frac{1}{\epsilon} \delta^d(x)\,\delta^d(y)\eeq
\beqa
S\cdot\partial\,S\cdot\partial\,G_0(x+y)\,v\cdot\partial\,J_0(x)\,J_0(y) \sim 
\frac{4}{3}\frac{S_\mu S_\nu\,\delta_{\mu\nu}}{(4\pi)^{d/2}} 
\frac{1}{\epsilon} \left[
v\cdot\partial\,\delta^d(x)\,\delta^d(y)\right. && \\
 \left.\nonumber
+2\,\delta^d(x)\,v\cdot\partial\,\delta^d(y)
\right]  && \eeqa
\beqa
S\cdot\partial\,S\cdot\partial\,G_0(x+y)\,v\cdot\partial\,J_0(x)\,v\cdot\partial\,J_0(y) \sim 
-\frac{4}{3}\frac{S_\mu S_\nu\,\delta_{\mu\nu}}{(4\pi)^{d/2}} 
\frac{1}{\epsilon} \left[
(v\cdot\partial)^2\,\delta^d(x)\,\delta^d(y) \right. &&\\
\left.\nonumber
+v\cdot\partial\,\delta^d(x)\,v\cdot\partial\,\delta^d(y)
+\delta^d(x)\,(v\cdot\partial)^2\,\delta^d(y)\right]&&\eeqa
\beq
S\cdot\partial\,S\cdot\partial\,G_1(x+y)\,v\cdot\partial\,J_0(x)\,v\cdot\partial\,J_0(y) \sim 
-\frac{2}{v^2}\frac{S_\mu S_\nu\,\delta_{\mu\nu}}{(4\pi)^{d/2}} 
\frac{1}{\epsilon} \delta^d(x)\,\delta^d(y) \eeq
\beq
S\cdot\partial\,S\cdot\partial\,G_0(x+y)\,v\cdot\partial\,J_1(x)\,v\cdot\partial\,J_0(y) \sim 
\frac{4}{3}\frac{S_\mu S_\nu\,\delta_{\mu\nu}}{(4\pi)^{d/2}} 
\frac{1}{\epsilon} \delta^d(x)\,\delta^d(y) \eeq
\beqa
\partial_\perp^2\,S\cdot\partial\,S\cdot\partial\,G_0(x+y)\,J_0(x)\,J_0(y)
\sim
\frac{4}{3\,v^2}\frac{S_\mu S_\nu\,\delta_{\mu\nu}}{(4\pi)^{d/2}}
\frac{1}{\epsilon} \left[
3(v\cdot\partial)^2\,\delta^d(x)\,\delta^d(y) \right. &&\\
\left.\nonumber
+4\,v\cdot\partial\,\delta^d(x)\,v\cdot\partial\,\delta^d(y)
+3\,\delta^d(x)\,(v\cdot\partial)^2\,\delta^d(y)\right]&&\eeqa
\beq
\partial_\perp^2\,S\cdot\partial\,S\cdot\partial\,G_1(x+y)\,J_0(x)\,J_0(y)
\sim
\frac{10}{3}\frac{S_\mu S_\nu\,\delta_{\mu\nu}}{(4\pi)^{d/2}}
\frac{1}{\epsilon} \delta^d(x)\,\delta^d(y) \eeq
\beq
\partial_\perp^2\,S\cdot\partial\,S\cdot\partial\,G_0(x+y)\,J_1(x)\,J_0(y)
\sim
-\frac{4}{3\,v^2}\frac{S_\mu S_\nu\,\delta_{\mu\nu}}{(4\pi)^{d/2}}
\frac{1}{\epsilon} \delta^d(x)\,\delta^d(y) \eeq
\beqa
\partial_\perp^2\,S\cdot\partial\,S\cdot\partial\,G_0(x+y)\,v\cdot\partial\,
J_0(x)\,J_0(y) \sim
-\frac{4}{3\,v^2}\frac{S_\mu S_\nu\,\delta_{\mu\nu}}{(4\pi)^{d/2}}
\frac{1}{\epsilon} \times && \\
\times \left[ \nonumber
(v\cdot\partial)^3\,\delta^d(x)\,\delta^d(y) 
+2\,(v\cdot\partial)^2\,\delta^d(x)\,v\cdot\partial\,\delta^d(y)
\right. && \\
 \left. \nonumber
+3\,v\cdot\partial\,\delta^d(x)\,(v\cdot\partial)^2\,\delta^d(y)
+4\,\delta^d(x)\,(v\cdot\partial)^3\,\delta^d(y)
\right] &&\\
\partial_\perp^2\,S\cdot\partial\,S\cdot\partial\,G_1(x+y)\,v\cdot\partial\,
J_0(x)\,J_0(y) \sim
-\frac{10}{3}\frac{S_\mu S_\nu\,\delta_{\mu\nu}}{(4\pi)^{d/2}}
\frac{1}{\epsilon} \left[ 
v\cdot\partial\,\delta^d(x)\,\delta^d(y) 
\right. && \\
 \left. \nonumber
+2\,\delta^d(x)\,v\cdot\partial\,\delta^d(y)
\right]&&\\
\partial_\perp^2\,S\cdot\partial\,S\cdot\partial\,G_0(x+y)\,v\cdot\partial\,
J_1(x)\,J_0(y) \sim
\frac{4}{3\,v^2}\frac{S_\mu S_\nu\,\delta_{\mu\nu}}{(4\pi)^{d/2}}
\frac{1}{\epsilon} \left[ 
3\,v\cdot\partial\,\delta^d(x)\,\delta^d(y) 
\right. && \\
 \left. \nonumber
+2\,\delta^d(x)\,v\cdot\partial\,\delta^d(y)
\right] &&\\
\partial_\perp^2\,S\cdot\partial\,S\cdot\partial\,G_0(x+y)\,v\cdot\partial\,
J_0(x)\,J_1(y) \sim
\frac{4}{3\,v^2}\frac{S_\mu S_\nu\,\delta_{\mu\nu}}{(4\pi)^{d/2}}
\frac{1}{\epsilon} \left[ 
v\cdot\partial\,\delta^d(x)\,\delta^d(y) 
\right. && \\
 \left. \nonumber
+4\,\delta^d(x)\,v\cdot\partial\,\delta^d(y)
\right] &&\\
\partial_\perp^2\,S\cdot\partial\,S\cdot\partial\,G_0(x+y)\,v\cdot\partial\,
J_0(x)\,v\cdot\partial\,J_0(y) \sim
\frac{4}{3\,v^2}\frac{S_\mu S_\nu\,\delta_{\mu\nu}}{(4\pi)^{d/2}}
\frac{1}{\epsilon} \left[ 
(v\cdot\partial)^4\,\delta^d(x)\,\delta^d(y) 
\right.  && \\
\left. \nonumber
+(v\cdot\partial)^3\,\delta^d(x)\,v\cdot\partial\,\delta^d(y)
\right. && \\
 \left. \nonumber
+(v\cdot\partial)^2\,\delta^d(x)\,(v\cdot\partial)^2\,\delta^d(y)
\right. && \\
 \left. \nonumber
+v\cdot\partial\,\delta^d(x)\,(v\cdot\partial)^3\,\delta^d(y)
+\delta^d(x)\,(v\cdot\partial)^4\,\delta^d(y)
\right] && \\
\partial_\perp^2\,S\cdot\partial\,S\cdot\partial\,G_1(x+y)\,v\cdot\partial\,
J_0(x)\,v\cdot\partial\,J_0(y) \sim
\frac{10}{3}\frac{S_\mu S_\nu\,\delta_{\mu\nu}}{(4\pi)^{d/2}}
\frac{1}{\epsilon} \left[ 
(v\cdot\partial)^2\,\delta^d(x)\,\delta^d(y) 
\right. && \\
 \left. \nonumber
+v\cdot\partial\,\delta^d(x)\,v\cdot\partial\,\delta^d(y)
+\delta^d(x)\,(v\cdot\partial)^2\,\delta^d(y)
\right] &&\\
\partial_\perp^2\,S\cdot\partial\,S\cdot\partial\,G_0(x+y)\,v\cdot\partial\,
J_1(x)\,v\cdot\partial\,J_0(y) \sim
-\frac{4}{3\,v^2}\frac{S_\mu S_\nu\,\delta_{\mu\nu}}{(4\pi)^{d/2}}
\frac{1}{\epsilon} \left[ 
6\,(v\cdot\partial)^2\,\delta^d(x)\,\delta^d(y) 
\right. && \\
 \left. \nonumber
+3\,v\cdot\partial\,\delta^d(x)\,v\cdot\partial\,\delta^d(y)
+\delta^d(x)\,(v\cdot\partial)^2\,\delta^d(y)
\right] &&\\
\partial_\perp^2\,S\cdot\partial\,S\cdot\partial\,G_2(x+y)\,v\cdot\partial\,
J_0(x)\,v\cdot\partial\,J_0(y) \sim
\frac{5}{v^2}\frac{S_\mu S_\nu\,\delta_{\mu\nu}}{(4\pi)^{d/2}}
\frac{1}{\epsilon}\delta^d(x)\,\delta^d(y)
&& \\
\partial_\perp^2\,S\cdot\partial\,S\cdot\partial\,G_1(x+y)\,v\cdot\partial\,
J_1(x)\,v\cdot\partial\,J_0(y) \sim
-\frac{10}{3}\frac{S_\mu S_\nu\,\delta_{\mu\nu}}{(4\pi)^{d/2}}
\frac{1}{\epsilon}\delta^d(x)\,\delta^d(y)
&& \\
\partial_\perp^2\,S\cdot\partial\,S\cdot\partial\,G_0(x+y)\,v\cdot\partial\,
J_2(x)\,v\cdot\partial\,J_0(y) \sim
\frac{8}{3\,v^2}\frac{S_\mu S_\nu\,\delta_{\mu\nu}}{(4\pi)^{d/2}}
\frac{1}{\epsilon}\delta^d(x)\,\delta^d(y)
&& \\
\partial_\perp^2\,S\cdot\partial\,S\cdot\partial\,G_0(x+y)\,v\cdot\partial\,
J_1(x)\,v\cdot\partial\,J_1(y) \sim
\frac{4}{3\,v^2}\frac{S_\mu S_\nu\,\delta_{\mu\nu}}{(4\pi)^{d/2}}
\frac{1}{\epsilon}\delta^d(x)\,\delta^d(y) &&
\eeqa
where the perpendicular derivative  is defined as follows:
\beqa
\partial_\perp^2 &:=& \left(\delta_{\mu\nu} - \frac{v_\mu\,v_\nu}{v^2}\right)
\partial_\mu\,\partial_\nu
\eeqa

\def\theequation{\Alph{section}.\arabic{equation}}
\setcounter{equation}{0}
\section{Eye graph contributions}
\label{app:eye}

In this appendix,
 we list the resulting contribution of the eye graphs to the divergent part
of the generating functional. First,
we consider the part of the second order insertion  $ T_{(2)}
 $ which  does not contain any derivative acting on the nucleon
 propagator. These are the monomials  proportional to $ c_1, ... ,
 c_7$. Furthermore we introduce the notation $T_{(2)} \equiv T_{(2)} S $
which shows explicitly the spin--dependence, with 
$S \in \{ 1, S^\mu ,[S^\mu, S^\nu] \} $. Thus,
the so redefined $T_{(2)} $ does not contain any spin--matrices. 
We find after some algebra the final result:
\beqa
\hat{\Sigma}^{(2)}_{1,1} &=& - v \cdot u \, \langle v \cdot u  \, 
T_{(2)}\rangle   + {1 \over 2} \langle ( v \cdot u)^2 \rangle \, 
\langle T_{(2)} \rangle  \,\,  
\nonumber \\ &&
+ g_A^2 \Biggl[ 
-4 \, S^\mu  \Biggl( \, \langle
\Gamma_{\mu\nu} \,  T_{(2)} \rangle \, \,  -
\Gamma_{\mu\nu}  \, \langle T_{(2)} \rangle \, \Biggl)
S^\nu  \,\,   +2 \,  S^\nu \, \eta( T_{(2)} ) \, S_\nu  
\nonumber \\ &&
 + {4 \over 3}   S^\nu \, \Biggl( 
- 6 v \cdot \stackrel{\leftarrow}{\nabla}
 \langle  T_{(2)} \rangle \, v\cdot \nabla 
+3  v \cdot \stackrel{\leftarrow}{\nabla} T_{(2)}  \, v\cdot \nabla 
\nonumber \\ &&
+2 \langle  [v\cdot \nabla , [ v \cdot \nabla , T_{(2)} ]] \rangle \, 
-  [v\cdot \nabla , [ v\cdot \nabla , T_{(2)} ] ]
\Biggl) \,S_\nu  \Biggl]
 \nonumber \\  &&
 \frac{-2}{3} i g_A^3 \Biggl[ S^\kappa \{S,S^\rho \} S_\kappa \Biggl( 
3 \langle \{ T_{(2)} , u_\rho \} \rangle \, v \cdot \nabla + \mbox{
  h.c. }
\nonumber \\ &&
 - \frac{3}{2} \{ T_{(2)} , u_\rho \} \, v \cdot \nabla + \mbox{ h.c. } 
- \frac{1}{2} [u_\rho , [ v  \cdot \nabla , T_{(2)} ]]
+ \frac{1}{2} [ [v \cdot \nabla , u_\rho ] , T_{(2)} ]  \Biggl)
\nonumber \\ && + 
S^\kappa [S,S^\rho ] S_\kappa  \Biggl( - \frac{3}{2} [  T_{(2)} ,
u_\rho ] \, v \cdot \nabla + \mbox{ h.c. } 
- \langle \{ u_\rho, [ v\cdot \nabla , T_{(2)} ] \}  \rangle \nonumber \\ &&
+ \frac{1}{2} \{  u_\rho, [ v\cdot \nabla , T_{(2)} ] \}  
+ \langle \{  T_{(2)} , [ v \cdot \nabla , u_\rho ] \} \rangle 
- \frac{1}{2}  \{  T_{(2)} , [ v \cdot \nabla , u_\rho ] \} \Biggl) \Biggl] 
\nonumber \\ &&
+g_A^4 \Biggl[   {4\over  3}   S^\nu \,  \Biggl( -2
\langle  T_{(2)} \, (S \cdot u )^2 + 
( S \cdot u)^2  \, T_{(2)} \rangle +
 T_{(2)} \, (S \cdot u )^2 + 
( S \cdot u)^2  \, T_{(2)}   \Biggl) \, S_\nu
\nonumber \\ &&
- {4\over  3}  \, S^\nu \, \Biggl( \
2 \langle S \cdot u \, T_{(2)} S \cdot u \rangle 
- S \cdot u \, T_{(2)} S \cdot u \Biggl) \, S_\nu~.
\eeqa

\noindent
We are considering now the part which contains exactly one
derivative. The operator 
$$
\frac{-ig_A}{2m} \,\,  \{ S\cdot \nabla , v \cdot u \}
$$ 
leads to the following contributions:
\beqa
\hat{\Sigma}^{(2)}_{1,2} &=& \frac{i g_A}{2m} \Biggl[ \langle (v \cdot
u)^2 \rangle v \cdot u \, S \cdot \nabla + \mbox{ h.c. } \Biggl]
\nonumber \\
 && + \frac{g_A^2}{2m} \Biggl[ 
{3 \over 4} \langle (v \cdot u)^2 \rangle \,\, \eta(1) 
- {3 \over 4} \langle v \cdot u \eta(v \cdot u) \rangle 
-6 v \cdot \stackrel{\leftarrow}{\nabla} \langle (v \cdot u)^2 \rangle
v \cdot \nabla  \nonumber \\
&& 
+5 \langle v \cdot u \, [ v \cdot \nabla , [ v \cdot \nabla, v \cdot
u]] \rangle 
+3 \langle [ v\cdot \nabla, v \cdot u ]\, [ v \cdot \nabla , v \cdot
u] \rangle \Biggl] \nonumber \\
&& + \frac{g_A^2}{2m} [S^\mu,S^\nu] \Biggl[ 
-2 \langle (v \cdot u)^2 \rangle  \,\, \Gamma_{\mu\nu} 
-2 \langle \Gamma_{\mu\nu} v\cdot u \rangle \,\, v \cdot u \Biggl] 
\nonumber \\
&& + \frac{ig_A^3}{2m} \Biggl[ 
2 \langle v \cdot u \, S \cdot u \rangle \,\, v \cdot u \, v\cdot
\nabla + \mbox{ h.c. } 
+ \frac{1}{3} \, [ v \cdot \nabla \, , [  v \cdot \nabla  \, ,  v \cdot u
]] \, S\cdot \nabla + \mbox{ h.c. } \nonumber  \\
&&
- v \cdot \stackrel{\leftarrow}{\nabla} \, \{ S\cdot\nabla , v \cdot u
\} \, v \cdot \nabla  
- \frac{2}{3} [v \cdot \nabla , v \cdot u ] \, \langle S^\mu
\Gamma_{\mu\nu} v^\nu \rangle  \nonumber \\
&&  
- \frac{1}{3} v \cdot u \, \langle [v\cdot \nabla , S^\mu
\Gamma_{\mu\nu} v^\nu ] \rangle  
- 3  \langle v\cdot u \, [ v\cdot \nabla , S^\mu
\Gamma_{\mu\nu} v^\nu ] \rangle  \nonumber \\
&&
- \frac{1}{2} \eta(v \cdot u) S \cdot \nabla + \mbox{ h.c. } 
+ \frac{2}{3} \langle [ \nabla^\mu , \, \Gamma_{\mu\nu} S^\nu]  \, v
 \cdot u \rangle \Biggl] \nonumber \\
&&
+ \frac{g_A^3}{2m} 
\Biggl[- \frac{1}{2} \, v_\rho \epsilon^{\rho\mu\nu\sigma} \, 
\langle \Gamma_{\mu\nu} \, v \cdot u \rangle \, 
\nabla_\sigma  + \mbox{h.c.}  \Biggl] \nonumber \\
&&
+ \frac{g_A^4}{2m} \Biggl[  
- \frac{1}{6} \langle v \cdot u \, u_\mu \rangle ^2  
-\frac{5}{12} \langle (v \cdot u)^2 \rangle  
+ \frac{7}{12} \langle (v \cdot u)^2 \rangle \, \langle u \cdot u
\rangle  \nonumber \\
&&
+ \frac{ \{S^\mu,S^\nu \} }{4} 
\Biggl[ -18  v \cdot \stackrel{\leftarrow}{\nabla} \langle v \cdot u
\, u_\nu \rangle \, \nabla_\mu 
-18  \stackrel{\leftarrow}{\nabla}_\mu \langle v \cdot u
\, u_\nu \rangle \, v \cdot \nabla  \nonumber \\
&&
+ [ [ v\cdot \nabla , v\cdot u] , u_\nu ] \, \nabla_\mu + \mbox{
  h.c. } 
- [ v\cdot u, [v \cdot \nabla , u_\nu]]\, \nabla_\mu + \mbox{h.c.}
\nonumber \\ 
&& 
-3 [ v \cdot u , [ \nabla_\mu ,u_\nu]]  v \cdot \nabla +  \mbox{h.c.} 
- [v \cdot u, u_\nu] \, \langle \Gamma_{\mu\tau} v^\tau \rangle 
-7 \langle [v \cdot u, u_\nu] \,\Gamma_{\mu\tau} v^\tau \rangle 
\nonumber \\
&&
+6 \langle [v\cdot \nabla , v \cdot u ]\, [\nabla_\mu,u_\nu] \rangle 
+9 \langle [\nabla_\mu , v \cdot u ]\, [v \cdot \nabla , u_\nu] \rangle 
+12 \langle v\cdot u [v\cdot\nabla, [\nabla_\mu, u_\nu]] \rangle 
\nonumber \\
&&
+9 \langle [v \cdot \nabla, [\nabla_\mu, v \cdot u ]]\, u_\nu \rangle 
\Biggl]  \Biggl] \nonumber \\
&&
+ \frac{g_A^4}{2m} \, [S^\mu, S^\nu] \Biggl[ 
\frac{1}{2} \langle [u_\mu, u_\nu] \, v \cdot u \rangle \, v \cdot u 
+\frac{1}{2} \Gamma_{\mu\tau}v^\tau \, \langle u_\nu \, v \cdot u
\rangle 
-\frac{1}{4} \langle  \Gamma_{\mu\tau}v^\tau \rangle 
\, \langle u_\nu \, v \cdot u \rangle
\nonumber \\
&&
- \frac{1}{2}   v \cdot \stackrel{\leftarrow}{\nabla} \, [v \cdot u
\, ,  u_\nu ]  \, \nabla_\mu 
-\frac{1}{2} \stackrel{\leftarrow}{\nabla}_\mu  \, [ v \cdot u
\, , u_\nu ] \, v \cdot \nabla  \nonumber \\
&&
+\frac{1}{4}  \langle  [ v\cdot \nabla , v\cdot u] \, u_\nu  \rangle
\, \nabla_\mu + \mbox{ h.c. } 
- \frac{1}{4}  \langle  v\cdot u \,  [v \cdot \nabla , u_\nu] \rangle \,
 \nabla_\mu + \mbox{h.c.}
\nonumber \\ 
&& 
-\frac{3}{4} \langle  v \cdot u \, [ \nabla_\mu ,u_\nu] \rangle 
  v \cdot \nabla +  \mbox{h.c.} 
+ \frac{1}{6} [ [v\cdot \nabla  , v \cdot u] , [\nabla_\mu,u_\nu]]
+ \frac{1}{4} [ [\nabla_\mu  , v \cdot u] , [v \cdot \nabla , u_\nu]]
\nonumber \\
&& 
+ \frac{1}{4} [[v\cdot \nabla, [\nabla_\mu, v\cdot u ]], u_\nu] 
- \frac{1}{6} [[v\cdot \nabla, [\nabla_\mu, u_\nu ]], v \cdot u ] 
- \frac{1}{6} [[\nabla_\mu , [v\cdot \nabla, u_\nu ]], v \cdot u ] 
\Biggl] 
\nonumber  \\
&&
+ \frac{i g_A^5}{2m} \Biggl[
-\frac{1}{24} \langle (v\cdot u )^2 \rangle \, v\cdot u \, S\cdot
\nabla + \mbox{ h.c. }  
+ \frac{1}{8} \langle u \cdot u \rangle \, v\cdot u \, S\cdot
\nabla + \mbox{ h.c. }  
\nonumber \\
&&
\frac{1}{4} \langle v\cdot u \, u_\mu  \rangle \, S\cdot u \, 
\nabla^\mu + \mbox{ h.c. }  
- \frac{1}{4} \langle (v \cdot u)^2 \rangle \, S\cdot u \, v\cdot
\nabla + \mbox{ h.c. }  
\nonumber \\
&& 
-\frac{1}{12} \langle v\cdot u \, S \cdot  u  \rangle \,  u_\mu  \, 
\nabla^\mu + \mbox{ h.c. }  
+ \frac{1}{6} \langle v \cdot u S \cdot u  \rangle \, v\cdot u \, v\cdot
\nabla + \mbox{ h.c. }  
\nonumber \\
&&
-\frac{1}{12} \langle  u_\mu \, S \cdot  u  \rangle \, v \cdot  u \, 
\nabla^\mu + \mbox{ h.c. }  
- \frac{1}{12} \langle v \cdot u  u_\mu  \rangle \,  u^\mu \, S\cdot
\nabla + \mbox{ h.c. }  
\nonumber \\
&&
-\frac{1}{8} \langle [v\cdot u ,u_\mu ] \, [\nabla^\mu , S \cdot u ] \rangle 
+\frac{3}{8} \langle [v\cdot u ,S \cdot u ] \, [\nabla^\mu , u_\mu ] \rangle 
\nonumber \\
&&
-\frac{3}{8} \langle [v\cdot u ,S \cdot u ] \, [v \cdot \nabla ,
v\cdot u  ] \rangle 
+\frac{1}{8} \langle [ u_\mu ,v \cdot u ] \, [S \cdot \nabla ,
 u^\mu  ] \rangle  \Biggl]
\nonumber \\
&&
+ \frac{g_A^5}{2m} v_\rho \, \epsilon^{\rho\mu\nu\sigma} \Biggl[
-\frac{9}{32} 
\langle [u_\mu, u_\nu] \, v\cdot u \rangle \, \nabla_\sigma + \mbox{
  h.c. }  
- \frac{1}{8} [\nabla_\mu, u_\nu ] \, \langle u_\sigma \, v \cdot u
\rangle  
\nonumber \\
&&
+ \frac{3}{8} \langle   [\nabla_\mu, u_\nu ]  \, v\cdot u \rangle 
u_\sigma 
- \frac{1}{8} \langle   [\nabla_\mu, u_\nu ]  \,  u_\sigma \rangle 
 v \cdot u   \Biggl]~.
\eeqa

\noindent 
Consider next the part which contains exactly two
derivatives. The operator 
$$
\frac{1}{2m} \,\,  [ (v\cdot \nabla)^2 -  \nabla^2 ]
$$ 
leads to the following contributions:

\beqa
\hat{\Sigma}^{(2)}_{1,3} &=& \frac{1}{2m} \Biggl[ 
\stackrel{\leftarrow}{\nabla}_\mu \, \langle (v \cdot u)^2 \rangle
\nabla^\mu  
-4 v \cdot\stackrel{\leftarrow}{\nabla} \langle (v \cdot u)^2 \rangle
v \cdot \nabla  
\nonumber \\
&&
- \frac{1}{2} [ v \cdot u , [ \nabla_\mu , v \cdot u ]] \, \nabla^\mu
+ \mbox{ h.c. } 
+2  [ v \cdot u , [ v \cdot \nabla , v \cdot u ]] \, v \cdot  \nabla
+ \mbox{ h.c. } 
\nonumber  \\
&&
- \langle v \cdot u [\nabla^\mu, [ \nabla_\mu , v \cdot u ]] \rangle 
+ 4 \langle v \cdot u \, [ v \cdot \nabla , [ v \cdot \nabla , v \cdot
u ]] \rangle  
\nonumber \\
&&
+ \frac{3}{8} \langle ( v\cdot u)^2 \rangle \rangle \, \eta(1) 
- \frac{3}{8} \langle v \cdot u \, \eta(v \cdot u) \rangle 
\Biggl] 
\nonumber \\
&&
+ \frac{i g_A}{2m} \Biggl[
3 \langle v\cdot u \, S \cdot u \rangle \, v \cdot u \, v \cdot \nabla
+ \mbox{ h.c. } 
- \frac{3}{2} \langle [ v \cdot u , S \cdot u ] \, [ v \cdot \nabla ,
v \cdot u ] \rangle 
+ \eta(v \cdot u) \, S \cdot \nabla + \mbox{ h.c. } 
\nonumber \\
&& 
- v \cdot u \, \eta(1) \, S \cdot \nabla + \mbox{ h.c. } 
-8 [ v \cdot u , S^\mu \Gamma_{\mu\nu} v^\nu ]  \, v \cdot \nabla +
\mbox{ h.c. } 
+2 [ v \cdot u , S^\mu \Gamma_{\mu\nu}  ]  \,  \nabla^\nu +
\mbox{ h.c. }  
\nonumber \\
&&
-\frac{14}{3} \langle v \cdot u \, [\nabla^\mu , \Gamma_{\mu\nu} S^\nu
] \rangle  
- \frac{64}{3} \langle v \cdot u \, [v \cdot \nabla , S^\mu
\Gamma_{\mu\nu} v^\nu ] \rangle 
+20 \langle S^\mu \Gamma_{\mu\nu}v^\nu \, [v\cdot \nabla , v \cdot u ]
\rangle  
\nonumber \\
&&
-4 \langle S^\mu \Gamma_{\mu\nu} \, [\nabla^\nu , v \cdot u ] \rangle 
+4 [ v \cdot \nabla , v \cdot u ] \, \langle S^\mu \Gamma_{\mu\nu}
v^\nu \rangle 
+8 v \cdot u \, \langle [v\cdot \nabla, S^\mu \Gamma_{\mu\nu} v^\nu ]
\rangle 
\nonumber \\
&&
+8 v \cdot\stackrel{\leftarrow}{\nabla} \, \{S \cdot \nabla , v \cdot
u \} \, v \cdot \nabla 
-4 [S \cdot \nabla, [ v \cdot \nabla, v \cdot u ]]\, v \cdot \nabla 
+ \mbox{ h.c. } 
\nonumber \\
&& 
-4 [v \cdot \nabla , [ v \cdot \nabla , v \cdot u ]]\, S \cdot \nabla
+   \mbox{ h.c. } 
\Biggl]
\nonumber \\
&&
+ \frac{g_A^2}{2m} \Biggl[
-24   (v \cdot\stackrel{\leftarrow}{\nabla})^2 \, (v \cdot \nabla)^2 
+\frac{9}{2}v \cdot\stackrel{\leftarrow}{\nabla}
 \stackrel{\leftarrow}{\nabla}_\mu \, \nabla^\mu v \cdot \nabla 
+\frac{9}{2} \stackrel{\leftarrow}{\nabla}_\mu 
 v \cdot \stackrel{\leftarrow}{\nabla} \, v \cdot \nabla \nabla^\mu 
\nonumber \\
&&
+ \frac{9}{8} \langle [\nabla^\mu, \Gamma_{\mu\nu} v^\nu ]\rangle \, 
v \cdot \nabla + \mbox{ h.c. }  
- \frac{3}{8} \langle [ v \cdot \nabla , \Gamma_{\mu\nu} v^\nu]
\rangle \, \nabla^\mu +  \mbox{ h.c. }  
\nonumber \\
&&
 -\frac{7}{6} \langle [v \cdot u, u^\mu ] v^\nu \Gamma_{\nu\mu}
\rangle 
- \frac{3}{8} \langle (v \cdot u)^2 \rangle^2 
+ \frac{3}{8} \langle (v \cdot u)^2 \rangle \, 
\langle u \cdot u \rangle  
+ \frac{9}{2} \langle \Gamma_{\mu\nu} v^\nu \rangle^2
\nonumber \\
&&
+5 \Gamma_{\mu\nu} v^\nu \, \langle \Gamma_{\mu\tau} v^\tau \rangle 
- \frac{15}{4} \Biggl(
 \frac{3}{32} \langle \chi_+ \rangle^2 
+ \frac{1}{8} \langle \chi_+ \rangle \langle u \cdot u \rangle 
+ \frac{1}{8} \langle u \cdot u \rangle^2 
+ \frac{1}{8} \langle u_\mu u_\nu \rangle^2 
\Biggl)
\nonumber \\
&&
- \frac{5}{24} \eta(1)_{\mu\mu} 
- \frac{35}{12} v^\mu \eta(1)_{\mu\nu} v^\nu 
+ \frac{49}{12} \langle \Gamma_{\mu\nu} \, \Gamma^{\mu\nu} \rangle 
- \frac{46}{3} \langle v^\mu \Gamma_{\mu\kappa} \, v_\nu
\Gamma^{\nu\kappa} \rangle  
\nonumber \\
&&
- \frac{31}{4} [\nabla^\mu, \Gamma_{\mu\nu} v^\nu ] \, v \cdot \nabla + 
 \mbox{ h.c. }  
+ \frac{7}{3} [ \nabla^\mu , \Gamma_{\mu\nu} ] \, \nabla^\nu +  \mbox{ h.c. }
+ \frac{5}{4} [ v \cdot \nabla , \Gamma_{\mu\nu} v^\nu ] \, \nabla^\mu
+  \mbox{ h.c. } 
\nonumber \\
&&
-4  v \cdot\stackrel{\leftarrow}{\nabla} \, \langle (v \cdot u)^2
\rangle \, v \cdot \nabla 
-2 v \cdot\stackrel{\leftarrow}{\nabla} \, \langle v \cdot u \, u_\mu
\rangle \, \nabla^\mu
-2 \stackrel{\leftarrow}{\nabla}_\mu \, \langle u^\mu \, v \cdot u
\rangle \, v \cdot \nabla 
\nonumber \\
&&
+ \frac{10}{3} \langle v \cdot u \, [ v \cdot \nabla , [ v \cdot
\nabla , v \cdot u ]] \rangle 
+2 \langle [v \cdot \nabla , v \cdot u ] \, [v \cdot \nabla , v \cdot
u] \rangle 
- \frac{4}{3} \langle v \cdot u \, [ v \cdot \nabla , [ \nabla^\mu ,
u_\mu]] \rangle 
\nonumber \\ 
&&
- 2  \langle u^\mu  \, [ v \cdot \nabla , [ \nabla_\mu , v \cdot u ]] \rangle 
- \langle [v \cdot \nabla , v \cdot u ] \, [\nabla^\mu , u_\mu ] \rangle 
- \langle [ \nabla^\mu , v \cdot u ] \, [ v \cdot \nabla , u_\mu ] \rangle 
\nonumber \\
&& 
+ 9  v \cdot\stackrel{\leftarrow}{\nabla} \, \eta(1)  v \cdot \nabla 
- \frac{3}{2} \stackrel{\leftarrow}{\nabla}_\mu  \, \eta(1) 
 \nabla^\mu 
\Biggl]
\nonumber \\ 
&&
+ \frac{g_A^2}{2m} \, [ S^\mu, S^\nu] \Biggl[
-32  v \cdot\stackrel{\leftarrow}{\nabla} \, \Gamma_{\mu\nu} \, 
 v \cdot \nabla 
+ 2 \stackrel{\leftarrow}{\nabla}_\kappa \, \Gamma_{\mu\nu} \, 
 \nabla^\kappa 
+ 2 \langle [\nabla^\mu , v \cdot u ] \, u_\nu \rangle \, v \cdot \nabla
+  \mbox{ h.c. } 
\nonumber \\
&& 
-16 v \cdot \stackrel{\leftarrow}{\nabla} \, \Gamma_{\mu\kappa}
v^\kappa \, \nabla_\nu 
-16  \stackrel{\leftarrow}{\nabla}_\nu \, \Gamma_{\mu\kappa} v^\kappa
v \cdot \nabla 
+ 2 \langle [v \cdot \nabla , v \cdot u ] \, u_\nu \rangle \,
\nabla_\mu + \mbox{ h.c. }
\nonumber \\
&& 
+ 6 v \cdot u \, \langle u_\mu \, \Gamma_{\nu\kappa} v^\kappa \rangle 
+ 6 \Gamma_{\nu\kappa} v^\kappa \, \langle u_\mu v \cdot u \rangle 
+ \frac{3}{4} v \cdot u \, \langle v \cdot u \, [ u_\mu , u_\nu ]
\rangle 
\nonumber \\
&&
+ \frac{7}{2} \eta(1) \, \Gamma_{\mu\nu}
- \frac{7}{2} \eta(\Gamma_{\mu\nu})
+ \eta(1)_\mu \, \nabla_\nu +  \mbox{ h.c. } 
- \frac{14}{3} [ \Gamma_{\mu\kappa} , \, \Gamma^{\nu\kappa} ]
\nonumber \\
&&
+ 0 [ v^\kappa \Gamma_{\kappa\mu} , \, v^\tau
\Gamma_{\tau\nu} ] 
+ \frac{1}{3} [\nabla^\kappa ,[ \nabla_\kappa , \Gamma_{\mu\nu} ]]
+ 16 [ v\cdot \nabla , [ v \cdot \nabla , \Gamma_{\mu\nu} ]]
\Biggl] 
\nonumber \\
&&
+ \frac{ig_A^3}{2m} \Biggl[
\frac{1}{6} \langle [u_\mu, S \cdot u ]\,[\nabla^\mu , v\cdot u]
\rangle
- \frac{1}{6} \langle [v \cdot u, S \cdot u ]\,[v\cdot \nabla , v\cdot u]
\rangle
+ \frac{1}{3} \langle [v \cdot u, u_\mu ]\,[S\cdot \nabla , u^\mu ]
\rangle
\nonumber \\
&&
+ \langle ( v\cdot u)^2 \rangle \, v \cdot u \, S \cdot \nabla 
+   \mbox{ h.c. }  
- \langle  u\cdot u  \rangle \, v \cdot u \, S \cdot \nabla 
+   \mbox{ h.c. }  
\nonumber \\
&&
- v \cdot\stackrel{\leftarrow}{\nabla} \, \{v \cdot \nabla , S \cdot
u \} \, v \cdot \nabla 
+ \stackrel{\leftarrow}{\nabla}_\mu \, \{v \cdot \nabla , S \cdot
u \} \, \nabla^\mu 
\nonumber \\
&&
-\frac{1}{6} [S\cdot u , \Gamma_{\mu\nu} v^\nu ] \nabla^\mu 
+ \mbox{ h.c. } 
+ \frac{2}{3} [v \cdot \nabla , [ v \cdot \nabla , S \cdot u ]] \, 
v \cdot \nabla +  \mbox{ h.c. } 
\nonumber \\
&&
- \frac{1}{6} [v\cdot \nabla , [\nabla_\mu , S \cdot u ]] \,
\nabla^\mu +  \mbox{ h.c. } 
-\frac{1}{2} [\nabla^\mu , [ \nabla_\mu , S \cdot u ]] \, v \cdot \nabla
+  \mbox{ h.c. } 
\nonumber \\
&& 
- \frac{1}{2} S \cdot u \, \langle [\nabla^\mu, \Gamma_{\mu\nu} v^\nu]
\rangle  
- \frac{2}{3} [\nabla^\mu , S \cdot u ] \, \langle
\Gamma_{\mu\nu}v^\nu \rangle 
+ \frac{1}{6} \langle S \cdot u \, [\nabla^\mu, \Gamma_{\mu\nu} v^\nu]
\rangle 
\nonumber \\
&&
- \frac{4}{3} \langle u^\mu \, [S \cdot \nabla , \Gamma_{\mu\nu} v^\nu
] \rangle 
+\frac{7}{3} \langle S \cdot u \, [ \nabla^\mu , \Gamma_{\mu\nu}
v^\nu]  \rangle 
 \nonumber \\
&&
- \frac{5}{2} \eta(S \cdot u) \, v \cdot \nabla +  \mbox{ h.c. } 
+ \frac{10}{3}  S \cdot u \, (v \cdot \nabla)^3 + \mbox{ h.c. } 
-\frac{5}{3} [v \cdot \nabla , [v \cdot \nabla , S \cdot u ]]\, 
v \cdot \nabla + \mbox{ h.c. }
\Biggl]
\nonumber \\
&&
+ \frac{g_A^3}{2m} v_\rho \epsilon^{\rho\mu\nu\sigma} 
\Biggl[ 
- \frac{1}{4} \langle [u_\mu, u_\nu] \, v \cdot u \rangle \,
\nabla_\sigma +  \mbox{ h.c. } 
- \frac{1}{3}  v \cdot u \, \langle u_\mu \, [\nabla_\nu, u_\sigma]
\rangle 
\nonumber \\
&&
+ \frac{7}{2} \langle u_\mu \Gamma_{\nu\sigma} \rangle \, v \cdot
\nabla + \mbox{ h.c. } 
-2 \langle u_\mu \Gamma_{\nu\tau} v^\tau  \rangle \, 
\nabla_\sigma + \mbox{ h.c. } 
\Biggl]
\nonumber \\
&&
+ \frac{g_A^4}{2m}
\Biggl[ 
\frac{5}{16} \langle (v \cdot u)^2 \rangle \, \eta(1)
- \frac{5}{16} \langle u \cdot u  \rangle \, \eta(1)
+ \frac{7}{4} \langle \Gamma_{\mu\nu} \, [u^\mu, u^\nu] \rangle 
- \frac{7}{2} \langle \Gamma_{\mu\nu} v^\nu  \, [u^\mu, v \cdot u] \rangle 
\nonumber \\
&&
+ \{S^\mu , S^\nu \} 
\Biggl[ 
- \frac{99}{4}  v \cdot\stackrel{\leftarrow}{\nabla} \langle u_\mu
u_\nu \rangle   v \cdot \nabla 
+ \frac{9}{4}  \stackrel{\leftarrow}{\nabla}^\kappa \langle u_\mu
u_\nu \rangle    \nabla_\kappa 
\nonumber \\
&& 
+ \frac{11}{4} [[ v\cdot \nabla , u_\mu], u_\nu ] \, v \cdot \nabla 
  + \mbox{ h.c. } 
-\frac{1}{4} [ [\nabla_\kappa, u_\mu] , u_\nu ] \, \nabla^\kappa +   
\mbox{ h.c. }  
\nonumber \\
&& 
+\frac{51}{4} \langle [v\cdot \nabla , u_\mu ] \, [ v \cdot \nabla ,
u_\nu ] \rangle  
+\frac{69}{4} \langle u_\mu \, [v \cdot \nabla , [ v \cdot \nabla ,
u_\nu]] \rangle 
\nonumber \\
&&
- \frac{3}{2} \langle [\nabla_\kappa, u_\mu]\, [\nabla^\kappa , u_\nu]
\rangle 
-\frac{9}{4} \langle u_\mu [\nabla_\kappa, [ \nabla^\kappa , u_\nu]]
\rangle  
\Biggl] 
\nonumber \\
&&
+ \frac{g_A^4}{2m} [S^\mu,S^\nu]  \Biggl[
-\frac{5}{24} \eta([u_\mu,u_\nu]) 
-\frac{7}{2} \Gamma_{\mu\nu} \langle (v \cdot u)^2 \rangle 
+\frac{7}{2} \Gamma_{\mu\nu} \langle u \cdot u \rangle 
\nonumber \\
&&
-\frac{33}{12}  v \cdot\stackrel{\leftarrow}{\nabla} \, [u_\mu ,u_\nu]
 \, v \cdot \nabla  
+\frac{1}{4}  \stackrel{\leftarrow}{\nabla}_\kappa \, [u_\mu ,u_\nu]
 \,  \nabla^\kappa   
+ \frac{33}{12} \langle [v \cdot \nabla ,u_\mu ] \, u_\nu \rangle \, v
\cdot \nabla + \mbox{ h.c. } 
\nonumber \\
&& 
- \frac{1}{4} \langle [\nabla_\kappa, u_\mu] u_\nu \rangle \,
\nabla^\kappa +  \mbox{ h.c. } 
+\frac{17}{12} [[v \cdot \nabla , u_\mu ] , [ v \cdot \nabla , u_\nu
]] 
\nonumber \\
&&
+ \frac{23}{12} [u_\mu, [ v \cdot \nabla , [ v \cdot \nabla , u_\nu
]]]
- \frac{1}{6} [[\nabla_\kappa, u_\mu], [ \nabla^\kappa , u_\nu ]] 
- \frac{1}{4} [ u_\mu , [ \nabla^\kappa, [\nabla_\kappa, u_\nu ]]] 
\Biggl] 
\nonumber \\
&& 
+\frac{i g_A^5}{2m} \Biggl[
\frac{40}{3} S^\mu \Biggl( 2 \langle (S\cdot u)^3 \rangle - (S\cdot
u)^3 \Biggl)  S_\mu \, v \cdot \nabla +   \mbox{ h.c. } 
\nonumber \\
&&
+\frac{20}{3}   S^\mu \Biggl( 2 \langle (S\cdot u)^2 [v \cdot \nabla ,
S \cdot u ] \rangle - (S\cdot
u)^2 \,  [v \cdot \nabla , S \cdot u]  \Biggl)  S_\mu 
\nonumber \\
&&
-\frac{20}{3}   S^\mu \Biggl( 2 \langle [ v \cdot \nabla , S \cdot u ]
\, (S\cdot u)^2  \rangle - [ v \cdot \nabla , S \cdot u ] \,  (S\cdot
u)^2 \,   \Biggl)  S_\mu 
\Biggl]
\nonumber \\
&&
+ \frac{g_A^6}{2m} \Biggl[ 
\frac{20}{3}  \, S^\mu \Biggl( 2 \langle (S\cdot u)^4 \rangle -
 (S\cdot u)^4 \Biggl)  S_\mu \Biggl]~.
\eeqa

\def\theequation{\Alph{section}.\arabic{equation}}
\setcounter{equation}{0}
\section{Check on some eye graph contributions}
\label{app:eyecheck}

\noindent In this appendix, we discuss one possible check on the calculation
of certain contributions from the eye graph. To be precise,
we note that  part of the eye graph can be
worked out along the methods used for the tadpole and vertex--corrected
selfenergy graphs, and only some remaining
pieces  then involve a triple coincidence limit. 
To be precise, let us take a closer look at the singular products of
the meson and nucleon propagators. This reveals that terms of the type
$v \cdot \partial J_m$ never appear, only terms $\sim J_m$ (compare
appendix~\ref{app:singops}). Terms of the former type can only appear
if one has just one nucleon line, where as the latter allows for two 
propagators, as it should be due to the insertion on the intermediate
nucleon line. However, we  now proof that Ecker's method can
also be used here for some parts of the diagram. 
For that, define\footnote{Again, the subscripts
$(i)$ ($i=1,2$) denote the chiral dimension.}
\beq
D_{(2)} \equiv i A_{(2)} \,\,\, ,
\eeq 
in analogy to Eq.(\ref{D1}), thus $D = i(A_{(1)} +  A_{(2)})$. We seek the
inverse of the operator $[A_{(1)} +  A_{(2)}]$. One can write
\beq
[A_{(1)} +  A_{(2)}]^{-1} = i D^\dagger [ D D^\dagger]^{-1}
\equiv i D^\dagger \, \Delta^{-1} \,\,\, ,
\eeq
with $\Delta^{-1}$ being a hermitian and positive definite operator.
The chiral expansion of this operator takes the form
\beqa \label{delexp}
\Delta^{-1} &=& \Delta^{-1}_{(-2)}  -  \Delta^{-1}_{(-2)} 
[ D_{(1)} D_{(2)}^\dagger +  D_{(2)} D_{(1)}^\dagger ] \Delta^{-1}_{(-2)}
 -  \Delta^{-1}_{(-2)} [ D_{(2)} D_{(2)}^\dagger ] \Delta^{-1}_{(-2)}
 \,\, , \nonumber \\
&\equiv& \Delta^{-1}_{(-2)} + \Delta^{-1}_{(-1)} + {\cal O}(1) \,\,\, .
\eeqa 
The first term on the right hand side of Eq.({\ref{delexp}) is nothing
but the inverse of the elliptic operator considered in the previous
paragraph. The two operators in the square brackets defining
$\Delta^{-1}_{(-1)}$  are positive
definite and hermitean. These have to be handled by the
multi--coincidence method described below. Consequently, the
inverse of $[A_{(1)} +  A_{(2)}]$ can be written as
\beq
[A_{(1)} +  A_{(2)}]^{-1} = i A_{(1)}^\dagger \, \Delta^{-1}_{(-2)} 
+  i A_{(2)}^\dagger \, \Delta^{-1}_{(-2)} 
+  i A_{(1)}^\dagger \, \Delta^{-1}_{(-1)} 
+ {\cal O}(1) \,\,\, ,
\eeq
the first term giving the already calculated counterterms at order
$q^3$. The terms of order $q^4$ are thus generated by the terms 
$ i A_{(2)}^\dagger \, \Delta^{-1}_{(-2)} +  i A_{(1)}^\dagger \, 
\Delta^{-1}_{(-1)} $. Consider only the first term of this sum.
It can be interpreted as follows: By constructing the inverse
of $[A_{(1)} +  A_{(2)}]$, we have shifted the dependence on the
intermediate point $z$, i.e. on the coordinates of the dimension two 
insertion, into one of the two vertices which connects the meson loop
to the nucleon line. Stated differently,  part of the eye graph has been
transformed into a vertex--corrected type of self--energy diagram and
thus can be treated along the lines outlined before.
As a check, one can verify that it reproduces the correct
result for the free operator $A_{(2)}$, i.e. it leads to the corrected
nucleon propagator \cite{bkmrev}
\beq
S_N^{(2)} = {i \over 2m} \biggl[ 1 - \frac{\ell^2}{(v\cdot \ell + i
  \eta )^2} \biggr] \,\,\, ,
\eeq
with $\ell$ the small nucleon off--shell momentum and $\eta \to 0^+$.

\vfill\eject

\def\theequation{\Alph{section}.\arabic{equation}}
\setcounter{equation}{0}
\section{Tables for the tadpole, self--energy and eye graphs}
\label{app:tables}

In this appendix, we list separately the fourth order monomials and their
$\beta$--functions arising from the tadpole, the self--energy and the eye
graphs, in order. To make the comparison easier, we give the operators in
a certain order, like the Ai (i = 1, 2, $\ldots$) only contain 
combinations of $v_\mu$'s and $u_\mu$'s,
all operators with exactly one $\chi_+$ are grouped together (the Ci) and so
on. Furthermore, all operators are also ordered according to the number
of nucleon covariant derivatives $\nabla_\mu$, which is zero, one, two $\ldots\,$. 
In the tables, we use the definition $\tilde A = A - \langle A \rangle
/2$ for $A = \chi_+, F_{\mu\nu}^+$.



\begin{table}[hb]
\caption{{\bf Tadpole counterterms}}
$$
\begin{tabular}{|r|c|c|} \hline  \label{tadtab}
i  & $\tilde{O}_i^{(4)}$ & $\delta_i$ \\ \hline
A1  & $ \langle u_\mu \, u_\nu \rangle \, \langle u^\mu \, u^\nu \rangle
$ & $ - 1/16m +  2c_3 /3  -  (c_2-g_A^2/8m)/6 - 2c_3/3 $ \\
A2  & $ \langle u^2 \rangle \, \langle u^2 \rangle
$ & $ -1/16m +   c_3 /3  - (c_2-g_A^2/8m)/12 - c_3/3 $ \\
A3  & $ \langle (v\cdot u)^2 \rangle \langle u \cdot u \rangle $  &
$ 1/16m + (c_2- g_A^2/8m)/3  $ \\ 
A4  & $ \langle  u^\mu v\cdot u \rangle \langle  u_\mu v\cdot u \rangle $  &
$  1 /16m + 2(c_2- g_A^2/8m)/3  $ \\ 
 \hline
B1 & $ i\,\langle [u^\mu , u^\nu] \, \tilde{F}_{\mu\nu}^+ \, \rangle 
$ & $ - c_3 /3  + (c_2-g_A^2/8m)/24 + c_3/6 $ \\ 
B3 & $ i\,\langle [u^\mu , v \cdot u] \, \tilde{F}_{\mu\nu}^+ \, v^\nu\rangle 
$ & $ -(c_2 - g_A^2/8m) /3 $ \\
\hline
C1 & $ \langle u \cdot u  \rangle   \, \langle  \chi_+ \rangle
$ & $  c_1 -1/16m + c_3/2  - (c_2-g_A^2/8m)/4 - c_3 $\\
\hline
\end{tabular}
$$
\end{table}

$\,$

\addtocounter{table}{-1}

$\,$
\begin{table}
\caption{continued, one derivative}
$$
\begin{tabular}{|r|c|c|} \hline
i  & $\tilde{O}_i^{(4)}$ & $\delta_i$ \\ \hline
C2 & $ \langle [u_\mu , S \cdot u  ] [ i \nabla^\mu , v \cdot u   ]
 \rangle  $& $  g_A / 24m $ \\
C4 & $ \langle [v \cdot u , u_\mu ] [ i \nabla^\mu , S \cdot u   ]
 \rangle  $& $ g_A / 24m $ \\
C7 &  $ \langle v\cdot u \, u_\mu  \rangle \, u^\mu  i S \cdot \nabla
$ + h.c. &  
$ - g_A/4m $ \\
C8 &  $ \langle u \cdot u  \rangle \, v\cdot u  \, i S \cdot \nabla  $
+ h.c. &   $ - g_A/4m $ \\ 
\hline
D12 & $ \langle  [ \nabla^\mu , \tilde{F}^+_{\mu\nu}  S^\nu  ]\, v \cdot u
 \rangle  $& $ g_A / 12m $ \\
D13 & $ \langle  [ \nabla^\mu , \tilde{F}^+_{\mu\nu}  v^\nu  ]\, S \cdot u
 \rangle  $& $ -g_A / 12m $ \\
\hline
E4 &  $ \langle \chi_+ \rangle v\cdot u  i S \cdot \nabla  $ + h.c. & 
$ - g_A/4m $ \\
\hline
\end{tabular}
$$
\end{table}


\renewcommand{\arraystretch}{1.1} 

\begin{table}
\caption{{\bf Self--energy counterterms}}
$$
\begin{tabular}{|r|c|c|} \hline \label{selftab} 
i  & $\tilde{O}_i^{(4)}$ & $\delta_i$ \\ \hline
A1  & $ \langle u_\mu \, u_\nu \rangle \, \langle u^\mu \, u^\nu \rangle
$ & $ g_A^2 \, c_3 / 2 - g_A^2 (c_4 + 1/4m) /4  + c_3 g_A^2/2 + c_3
g_A^4/6 $ \\
 & &  $ + (c_4 + 1/4m) g_A^2/2 -(c_4 + 1/4m) g_A^4/3 $ \\  
 &&  $ + (g_A^4/12)/2m +  (g_A^2/6 - g_A^4/4 -3g_A^6/16)/2m $  \\ 
A2  & $ \langle u^2 \rangle \, \langle u^2 \rangle
$ & $ -g_A^2 \, c_3 /2 + g_A^2 (c_4 + 1/4m) /4 - c_3 g_A^2/2 + c_3
g_A^4/4 $ \\ 
 & &  $  + (c_4 + 1/4m) g_A^2/2 + (c_4 + 1/4m) g_A^4/2 $ \\  
 && $ +  (-3g_A^2/8 -g_A^4/8)/2m + (5g_A^2/24 +5g_A^4/8 + 9g_A^6/32)/2m  $ \\
A3  & $ \langle (v\cdot u)^2 \rangle \langle u \cdot u \rangle $  &
$  g_A^2(c_3-c_2+ g_A^2/8m)/2 - g_A^2 (c_4 + 1/4m) /2 $ \\
\hline
\end{tabular}
$$
\end{table}

\addtocounter{table}{-1}

\begin{table}
\caption{continued}
$$
\begin{tabular}{|r|c|c|} \hline 
i  & $\tilde{O}_i^{(4)}$ & $\delta_i$ \\ \hline
F1 &  $ i v_\rho \, \epsilon^{\rho\mu\nu\sigma} \, [u_\mu, u_\nu] \,  
 \langle u_\sigma \, S \cdot u  \,  \rangle $ & 
$  -g_A^2 (c_4 + 1/4m) / 2 $ \\
F2 &  $ i v_\rho \, \epsilon^{\rho\mu\nu\sigma} \,  
 \langle [u_\nu , u_\sigma] \,  S \cdot u  \, \rangle  \, u_\mu $ & 
$ - g_A^2 (c_4 + 1/4m) + (g_A^6/24)/2m   $ \\
F3 &  $ i v_\rho \, \epsilon^{\rho\mu\nu\sigma} \,  
 \langle \, [u_\mu , u_\nu] u_\sigma \rangle  \, S \cdot u $ & 
$ c_3 g_A^4 /6  + (-g_A^4/12)/2m  + (-g_A^6/24)/2m  $ \\
 && $ -  (c_4 + 1/4m) g_A^4/3 $ \\  
\hline
G1 &  $  v_\rho \, \epsilon^{\rho\mu\nu\sigma} \,
 \tilde{F}_{\mu\nu}^+ \langle u_\sigma  \, S \cdot u \rangle $ & 
$  -  g_A^2 (c_4+1/4m)   $ \\
G3 &  $  v_\rho \, \epsilon^{\rho\mu\nu\sigma} \,
 \langle \tilde{F}_{\mu\nu}^+  \,  S \cdot u \rangle  \,  u_\sigma  $ & 
$   -  g_A^2 (c_4+1/4m)   $ \\
G5 & $  v_\rho \, \epsilon^{\rho\mu\nu\sigma} \,
 \langle S^\tau \tilde{F}_{\tau\sigma}^+ \, u_\mu \rangle \,    u_\nu  $ & 
$  -  g_A^2 (c_4+1/4m)    $ \\
\hline
\end{tabular}
$$
\end{table}

\addtocounter{table}{-1}

\begin{table}
\caption{continued, one derivative }
$$
\begin{tabular}{|r|c|c|} \hline
M1 & $ v_\rho \, \epsilon^{\rho\mu\nu\sigma} \langle  \tilde{F}^-_{\mu
  \nu}   \, v \cdot u \rangle \,  u_\sigma   $& $ -g_A (1+c_6) / 16m $ \\
M2 & $ v_\rho \, \epsilon^{\rho\mu\nu\sigma} \langle u_\sigma \tilde{F}^-_{\mu
   \nu} \rangle  \,  \, v \cdot u   $& $ -{g_A}( 1+c_6) / 8m
 + g_A( 1+c_6) / 16m  $ \\
M3 & $ v_\rho \, \epsilon^{\rho\mu\nu\sigma} \langle u_\sigma v\cdot u
 \rangle \,  \tilde{F}^-_{\mu \nu} \,   $& $ - {g_A} (1+c_6) / 8m$ \\
M4 & $ i v_\rho \, \epsilon^{\rho\mu\nu\sigma} [ \tilde{F}^-_{\mu \nu} \, , \, 
 \tilde{F}^+_{\sigma\tau} v^\tau ]   $& $ -{g_A (1+ c_6)} / {16m} $ \\
M5 & $ i \langle  S^\mu F^-_{\mu \tau } \rangle   \, [
 u^\tau \, , \, v \cdot u ]   $& $ {g_A (1+c_6)} / {4m} $ \\
M6 & $ i \langle  S^\mu \tilde{F}^-_{\mu \tau }   \, [
 u^\tau \, , \, v \cdot u ] \rangle  $& $ -{g_A (1+c_6)}/ {2m} $ \\ 
M7 & $   \langle S^\mu  F^-_{\mu \tau} \rangle  \, 
 F^+_{\tau\rho} v^\rho   $& $ {g_A (1+c_6)} / {2m} $ \\
M8 & $   \langle S^\mu  \tilde{F}^-_{\mu \tau}   \, 
 \tilde{F}^+_{\tau\rho} v^\rho \rangle  $& $ -{g_A (1+c_6)} / {2m} $  \\
\hline
\end{tabular}
$$
\end{table}

\addtocounter{table}{-1}

\begin{table}
\caption{continued, one derivative }
$$
\begin{tabular}{|r|c|c|} \hline
C9  &  $ \langle S \cdot u u_\mu  \rangle  \, u^\mu  \, i v \cdot
  \nabla $ + h.c.  &  
$  -4c_3 g_A -  2 c_3 g_A^3 + (c_4 +1/4m) g_A $ \\
 && $ - 2 (c_4 + 1/4m) g_A   $ \\
&& $  +   (g_A^5/3)/2m  $ \\
C10  &  $ \langle v \cdot u \, u_\mu  \rangle  \, S \cdot u  \, i
\nabla^\mu  $ + h.c.  &  
$ + ( -g_A  ) /2m $ \\  
C11  &  $ \langle v \cdot u \, S \cdot u  \rangle  \, u_\mu  \, i
\nabla^\mu  $ + h.c.  &  
$ (-g_A) / 2m $ \\
C12  &  $ \langle  u_\mu  \, S \cdot u   \rangle  \, v \cdot u  \, i
\nabla^\mu  $ + h.c.  &  
$ 0 $ \\
C13  &  $ \langle (v \cdot u)^2 \rangle  \, S \cdot u  \, i v \cdot
  \nabla $ + h.c.  &  
  $  + (c_4+1/4m) g_A  + 2 g_A (c_2-g_A^2/8m) -2 g_A^3 c_3  $ \\
 && $ + 2 (c_4 + 1/4m) g_A^3 - 4 g_A(c_2- g_A^2/8m +c_3)  + (2 g_A)/2m   $  \\
\hline
\end{tabular}
$$
\end{table}

\addtocounter{table}{-1}

\begin{table}
\caption{continued, two derivatives}
$$
\begin{tabular}{|r|c|c|} \hline
F1 & $ \langle v \cdot u \,  [ v \cdot \nabla , [ \nabla^\mu , 
 u_\mu ]] \rangle  $ & 
$  (-1)/2m  +  (g_A^2)/2m    $ \\
F2 & $ \langle [ v \cdot \nabla , v \cdot u ] \, [ \nabla^\mu ,  u_\mu ]
\rangle $ & 
$  (-1)/2m +   (2g_A^2) / 2m  $ \\
F3 & $  [ [\nabla^\mu ,u_\mu ] , v \cdot u ]   \, v \cdot \nabla $ + h.c. &
  $ -2 c_1- c_5  + (-g_A^2/3)/2m + (5g_A^2/12)/2m     $ \\  
F4  & $  [ [ \nabla^\mu  ,  u_\mu ] , \,   u_\nu  ]  \, \nabla^\nu
$  + h.c.
  & $ (-g_A^2/6 ) /2m   $\\   
\hline 
G1 & $ v_\rho \epsilon^{\rho\mu\nu\sigma} S_\sigma \langle v \cdot u
  \tilde{F}^-_{\mu\nu} \rangle v \cdot \nabla  $ + h.c. &
  $ (1+c_6)/2m $ \\
G2 & $ v_\rho \epsilon^{\rho\mu\nu\sigma} S_\sigma  \langle
   F^-_{\mu\nu} \rangle v \cdot u \,  v \cdot \nabla  $ + h.c. &
  $ -(1+c_6)/2m $ \\
G3 & $ v_\rho \epsilon^{\rho\mu\nu\sigma}  S_\sigma 
 [  \tilde{F}^-_{\mu\nu} , [ v \cdot \nabla , v \cdot u ]]  $  &
  $ (1+c_6)/4m $ \\
G4 & $ v_\rho \epsilon^{\rho\mu\nu\sigma}   S_\sigma
 [ v \cdot u , [ v \cdot \nabla , \tilde{F}^-_{\mu\nu} ]]  $ &
  $ (1+c_6)/4m $ \\
\hline
\end{tabular}
$$
\end{table}

\addtocounter{table}{-1}

$ \, $
\begin{table}
\caption{continued, two derivatives }
$$
\begin{tabular}{|r|c|c|} \hline
i  & $\tilde{O}_i^{(4)}$ & $\delta_i$ \\ \hline
A19 &  $ v \cdot \stackrel{\leftarrow}{\nabla} \langle (v \cdot u)^2
 \rangle  \, v \cdot \nabla  $  &
 $ -2 g_A^2 c_3 +   4  (c_4 + 1/4m) g_A^2 + $ \\
&& $ (-4 -g_A^2)/2m + (4g_A^2)/2m + (9g_A^4/2)/2m $  \\
A21 & $ v \cdot \stackrel{\leftarrow}{\nabla} 
\, \langle v\cdot u \, u_\mu \rangle  \, \nabla^\mu  $ + h.c. & $ 
 (2)/2m + (-2g_A^2)/2m $ \\ 
A22 &  $ v \cdot \stackrel{\leftarrow}{\nabla} \langle u \cdot u
 \rangle \, v \cdot \nabla   $   &
 $ 2 g_A^2 c_3  -4 (c_4 + 1/4m) g_A^2 $ \\
&& $  +(g_A^2)/2m + (-3g_A^2 - 9g_A^4/2)/2m    $  \\
\hline
B3 & $  [ v \cdot  \nabla , \,  \tilde{F}^+_{\mu\nu} v^\nu ] \, i
\nabla^\mu    $ + h.c.  & 
 $  (g_A^2) / 2m $ \\ 
B4 & $  [ \nabla^\mu , \,  \tilde{F}^+_{\mu\nu} v^\nu ] \, i
v \cdot \nabla    $ + h.c.  & 
 $  (2g_A^2/3) / 2m  +  (-5g_A^2/6) /2m $ \\
\hline
\end{tabular}
$$
\end{table}

\addtocounter{table}{-1}

$ \, $
\begin{table}
\caption{continued, three and more derivatives }
$$
\begin{tabular}{|r|c|c|} \hline
i  & $\tilde{O}_i^{(4)}$ & $\delta_i$ \\ \hline
E1 & $  v_\rho \epsilon^{\rho\mu\nu\sigma} S_\sigma \, [ v \cdot
\nabla , [ v \cdot \nabla , \tilde{F}^+_{\mu\nu}]]   $   & 
$  (-g_A^2) /2m $  \\
E2 & $  v_\rho \epsilon^{\rho\mu\nu\sigma} S_\sigma \, [
\nabla^\kappa , \, [ \nabla_\kappa  , \tilde{F}^+_{\mu\nu} ] ]  $  & 
$ ( -g_A^2/3 ) / 2m $  \\
E4 & $  v_\rho \epsilon^{\rho\mu\nu\sigma} S_\sigma \, v \cdot
\stackrel{\leftarrow}{\nabla} \tilde{F}^+_{\mu\nu} v \cdot \nabla  $ & 
$  (4 g_A^2) / 2m $ \\
E6 & $  v_\rho \epsilon^{\rho\mu\nu\sigma} S_\sigma \, 
v \cdot \stackrel{\leftarrow}{\nabla}\,  \tilde{F}^+_{\mu\kappa}
v^\kappa \,
\nabla_\nu  $  + h.c.  & 
$  (4 g_A^2) / 2m $ \\
E7 & $  v_\rho \epsilon^{\rho\mu\nu\sigma} S_\sigma \, [ 
\nabla_\mu , [ v \cdot \nabla , \tilde{F}^+_{\mu\kappa} v^\kappa ]]    $   & 
$  (4g_A^2) /2m $  \\
E8 & $  v_\rho \epsilon^{\rho\mu\nu\sigma} S_\sigma \,  [ 
\nabla_\mu , [ \nabla^\kappa , \tilde{F}^+_{\mu\kappa} ]]    $   & 
$  (-2g_A^2/3) /2m $  \\
\hline
F1 & $  v_\rho \epsilon^{\rho\mu\nu\sigma} S_\sigma \, 
 \langle  [ \nabla_\mu , \chi_+ ]  \rangle \, \nabla_\nu   $ + h.c.  & 
$ (-3 g_A^2/4 ) / 2m $  \\
\hline
\end{tabular}
$$
\end{table}




\begin{table}
\caption{{\bf Eye graph counterterms }}
$$
\begin{tabular}{|r|c|c|} \hline  \label{eyetab}
i  & $\tilde{O}_i^{(4)}$ & $\delta_i$ \\ \hline
A1  & $ \langle u_\mu \, u_\nu \rangle \, \langle u^\mu \, u^\nu \rangle
$ & $ -(c_4 + 1/4m) g_A^2/4  -11 (c_4+1/4m) g_A^4/16   $  \\
&& $ +(g_A^2/24 + 7g_A^4/8 + 15g_A^6/32)/2m $ \\
A2  & $ \langle u^2 \rangle \, \langle u^2 \rangle
$ & $ +(c_4+1/4m) g_A^2/4 - 3 c_3 g_A^4 /16 - 3c_3g_A^2/4 $ \\
 && $  - 3c_3 g_A^4/8 +11(c_4+1/4m)g_A^4/16 $ \\
 && $ + (-47g_A^2/48 - 29g_A^4/16  -45g_A^6/64) / 2m $ \\
A3  & $ \langle (v\cdot u)^2 \rangle \langle u \cdot u \rangle $  &
$ +  c_3 / 2 - (c_4 +1/4m) g_A^2/2  - 11(c_4+1/4m) g_A^4/8 $ \\ 
   & & $ -3 g_A^4 ( c_2 - g_A^2/8m)/8  $ \\
\hline
\end{tabular}
$$
\end{table}

$\,$

\addtocounter{table}{-1}

$\,$
\begin{table}
\caption{continued}
$$
\begin{tabular}{|r|c|c|} \hline
i  & $\tilde{O}_i^{(4)}$ & $\delta_i$ \\ \hline
D1 & $ \, \langle \, \tilde{F}_{\mu\kappa}^+  \,   \tilde{F}^+_{\mu\kappa}
\, \rangle 
$ & $ +(1+c_6) g_A^2/8m - (49 g_A^2 / 48) / 2m  $ \\
D2 & $  \, \tilde{F}_{\mu\kappa}^+  \, \langle  F^+_{\mu\kappa}
\, \rangle 
$ & $ -(1+c_6+2c_7) g_A^2/8m  $ \\
D3 & $ \, \langle v^\mu \, \tilde{F}_{\mu\kappa}^+  \, v^\nu
\tilde{F}^+_{\nu\kappa} 
\, \rangle 
$ & $ -(1+c_6) g_A^2/4m  + (23 g_A^2 /6 ) /2m   $ \\
D4 & $ \,  v^\mu \, \tilde{F}_{\mu\kappa}^+  \,  \langle v^\nu  F^+_{\nu\kappa}
\, \rangle 
$ & $ +(1+c_6+2c_7) g_A^2/4m  - (5g_A^2 / 4 ) / 2m  $ \\
D5 & $ \,  \langle v^\mu \, {F}_{\mu\kappa}^+ \rangle 
 \,  \langle v^\nu  F^+_{\nu\kappa}
\, \rangle 
$ & $ -  (9g_A^2 / 8 ) / 2m  $ \\
\hline
E1 & $ \langle \chi_+ \rangle  \langle \chi_+ \rangle $ & 
$ -9 c_1 g_A^2/8  - (45 g_A^2 / 128 ) / 2m   $ \\
E2 & $ \tilde\chi_+ \langle \chi_+ \rangle  $ & 
$ +3 c_5 g_A^2 /8 $ \\
\hline
\end{tabular}
$$
\end{table}

\addtocounter{table}{-1}

$\,$
\begin{table}
\caption{continued}
$$
\begin{tabular}{|r|c|c|} \hline
i  & $\tilde{O}_i^{(4)}$ & $\delta_i$ \\ \hline
I1 &  $  v_\rho \, \epsilon^{\rho\mu\nu\sigma} \, S_\sigma \,
 \, \tilde{F}_{\mu\nu}^+ \, \langle u \cdot u \, \rangle $ & 
$   + c_3 g_A^2  -(1+c_6)g_A^4/32m $ \\
& & $   + (7 g_A^2/4 + 7 g_A^4/4) / 2m  $ \\
I2 &  $  v_\rho \, \epsilon^{\rho\mu\nu\sigma} \, S_\sigma \,
 \, \langle F_{\mu\nu}^+ \rangle  \, \langle u \cdot u \, \rangle $ & 
$    (1+c_6+2c_7) g_A^4 /64m + (1+c_6+2c_7)g_A^2/16m $ \\
I3 &  $  v_\rho \, \epsilon^{\rho\mu\nu\sigma} \, S_\sigma \,
u^\kappa \langle   \,\tilde{F}_{\mu\nu}^+ \, u_\kappa  \, \rangle $ & 
$ -(1+c_6)g_A^2/8m  +(1+c_6)g_A^4/48m  +(-7g_A^2/4) / 2m   $ \\
I4 &  $  v_\rho \, \epsilon^{\rho\mu\nu\sigma} \, S_\sigma \,
u_\mu \langle   \,\tilde{F}_{\nu\kappa}^+ \, u^\kappa  \, \rangle $ & 
$   - (7g_A^2/3) / 2m   $ \\
I5 &  $  v_\rho \, \epsilon^{\rho\mu\nu\sigma} \, S_\sigma \,
u^\kappa \langle   \,\tilde{F}_{\nu\kappa}^+  \,  u_\mu  \,
\rangle $ &  
$  (7g_A^2/3) /2m $ \\
I7 &  $  v_\rho \, \epsilon^{\rho\mu\nu\sigma} \, S_\sigma \,
 \, \tilde{F}_{\mu\nu}^+ \, \langle (v \cdot u)^2 \, \rangle $ & 
$  +(c_2-g_A^2/8m) g_A^2  + (1+c_6)g_A^4/32m $  \\
\hline
\end{tabular}
$$
\end{table}

\addtocounter{table}{-1}

$ \, $
\begin{table}
\caption{continued, one derivative}
$$
\begin{tabular}{|r|c|c|} \hline
i  & $\tilde{O}_i^{(4)}$ & $\delta_i$ \\ \hline
A1 & $ v_\rho \epsilon^{\rho\mu\nu\sigma}  \langle [u_\mu , u_\nu ] \,
v \cdot u \rangle  \,   \nabla_\sigma  $ + h.c. &
$ (-7g_A^3/8 - 9 g_A^5/32) / 2m $ \\
A2 & $ v_\rho \epsilon^{\rho\mu\nu\sigma}  \langle [u_\mu , u_\nu ] \,
 u_\sigma  \rangle  \,   v \cdot \nabla   $ + h.c. &
$ + 9 (c_4+1/4m) g_A^3 /16 + (7g_A^3/8 - 15g_A^5/16) / 2m $  \\
A3 & $ v_\rho \epsilon^{\rho\mu\nu\sigma}  [\nabla_\mu , u_\nu ] \,
 \langle  u_\sigma \,   v \cdot u \rangle    $  &
$ (-g_A^5/8) / 2m $ \\
A4 & $ v_\rho \epsilon^{\rho\mu\nu\sigma}  \langle [\nabla_\mu , u_\nu ] \,
v \cdot  u \,   \rangle  u_\sigma    $  &
$ (3g_A^5/8) / 2m $ \\
A5 & $ v_\rho \epsilon^{\rho\mu\nu\sigma}  \langle [\nabla_\mu , u_\nu ] \,
 u_\sigma  \,   \rangle  v \cdot u    $  &
$ (-g_A^3/3 - g_A^5/8) / 2m $ \\
A6 & $ v_\rho \epsilon^{\rho\mu\nu\sigma}  \langle u_\mu \,  [v \cdot
\nabla , \,  u_\nu ] \,   \rangle  \,  u_\sigma    $  &
$ (5g_A^5/4) / 2m -  (c_4+1/4m)g_A^3/2  $ \\
\hline
B1 & $ v_\rho \epsilon^{\rho\mu\nu\sigma}  [ \tilde{F}_{\mu\nu}\, , [
i  v \cdot  \nabla ,  u_\sigma ] \, ]  $ &
$  (1+c_6)g_A^3/32m  $ \\
\hline
\end{tabular}
$$
\end{table}

\addtocounter{table}{-1}

$ \, $
\begin{table}
\caption{continued, one derivative }
$$
\begin{tabular}{|r|c|c|} \hline
i  & $\tilde{O}_i^{(4)}$ & $\delta_i$ \\ \hline
D1 & $  \langle [v \cdot \nabla, F^+_{\mu\nu} S^\nu ] \,
\rangle \, u^\mu    $ &
$ -(1+c_6+2c_7) g_A^3/24m $ \\
D2 & $  \langle [v \cdot \nabla,  v^\mu  F^+_{\mu\nu} S^\nu ] \,
\rangle \, v \cdot u    $ &
$ (1+c_6+2c_7) g_A^3/24m  + (-4g_A +g_A^3/6) / 2m   $ \\
D3 & $  \langle [\nabla^\mu ,  F^+_{\mu\nu} v^\nu ] \,
\rangle \, S \cdot u    $ &
$  -(g_A^3/4) / 2m   $ \\
D4 & $  \langle   F^+_{\mu\nu} v^\nu \,
\rangle \, [ \nabla^\mu , \,  S \cdot u ]     $ &
$  -(g_A^3/3) / 2m   $ \\
D5 & $  \langle F^+_{\mu\nu} S^\nu   \,
\rangle \, [ v \cdot \nabla , u^\mu ]   $ &
$ (1+c_6+2c_7) g_A^3/24m $ \\
D6 & $  \langle  v^\mu  F^+_{\mu\nu} S^\nu  \,
\rangle \, [ v \cdot \nabla ,  v \cdot u ]   $ &
$ -(1+c_6+2c_7) g_A^3/24m + (-2g_A + g_A^3/3) / 2m  $ \\
D7 & $ \langle  [ v \cdot \nabla , v^\mu \tilde{F}^+_{\mu\nu} S^\nu ]
\, v \cdot u \rangle $ & 
  $ -(1+c_6) g_A^3/8m + (32g_A/3 + 3g_A^3/2) / 2m   $ \\
\hline
\end{tabular}
$$
\end{table}

\addtocounter{table}{-1}

$ \, $
\begin{table}
\caption{continued, two derivatives }
$$
\begin{tabular}{|r|c|c|} \hline
i  & $\tilde{O}_i^{(4)}$ & $\delta_i$ \\ \hline
A1 & $ \langle [ v \cdot \nabla , v \cdot u ] \, [ v \cdot \nabla , v
\cdot u ] \rangle  $ & 
$ ( 5g_A^2 + 33g_A^4/4) / 2m  - 3g_A^2 (c_2-g_A^2/8m)  $ \\
A2 & $ \langle v \cdot u \,  [ v \cdot \nabla , [ v \cdot \nabla , v
\cdot u ]] \rangle  $ & 
$ ( 4 + 25g_A^2/3  + 45g_A^4/4) / 2m    - 3g_A^2 (c_2-g_A^2/8m)  $ \\
A3 & $ \langle u^\mu  \,  [ v \cdot \nabla , [ \nabla_\mu , 
 v \cdot u  ]] \rangle  $ & 
$ ( - 2g_A^2  -9 g_A^4/8) / 2m   $ \\
A4 & $ \langle v \cdot u   \,  [ \nabla^\mu , [ \nabla_\mu , 
 v \cdot u  ]] \rangle  $ & 
$ ( - 1  -9 g_A^4/8) / 2m   $ \\
A5 & $ \langle [ v \cdot \nabla , u_\mu ] \, [ \nabla^\mu , v \cdot u ]
\rangle $ & 
$ (-g_A^2 - 9 g_A^4/8) / 2m  $ \\
A6 & $ \langle [ v \cdot \nabla ,  u_\mu ] \, [ v \cdot \nabla ,  u^\mu ]
\rangle $ & 
$ (-35g_A^2/6 - 51 g_A^4/8 ) / 2m - 3 c_3 g_A^2 $ \\
A7 & $ \langle u_\mu \,  [ v \cdot \nabla , [ v \cdot \nabla ,  u^\mu
]]  \rangle $ & 
$ (-35g_A^2/6 - 69 g_A^4/8 ) / 2m - 3 c_3 g_A^2  $ \\
\hline
\end{tabular}
$$
\end{table}

\addtocounter{table}{-1}

$ \, $
\begin{table}
\caption{continued, two derivatives }
$$
\begin{tabular}{|r|c|c|} \hline
i  & $\tilde{O}_i^{(4)}$ & $\delta_i$ \\ \hline
C1 & $  [ v \cdot \nabla , [ v \cdot \nabla , \tilde{\chi}_+]]  $  &
 $  c_5 g_A^2 $  \\
C2 & $ \langle [ v \cdot \nabla , \, [ v \cdot \nabla , \, \chi_+ ]]
 \rangle  $ & $-(35 g_A^2/16) / 2m - 3 c_1 g_A^2 $ \\
C3 & $ \langle [ \nabla^\mu , \, [ \nabla_\mu , \, \chi_+ ]]
 \rangle  $ & $ -(5 g_A^2/32) / 2m $ \\
C4 & $ v \cdot  \stackrel{\leftarrow}{\nabla} 
\, \langle \chi_+ \rangle  v \cdot \nabla $ & $ 9
 c_1 g_A^2  + (27 g_A^2 / 4 ) / 2m  $ \\
C5 & $   \stackrel{\leftarrow}{\nabla}_\mu 
\, \langle \chi_+ \rangle \, \nabla^\mu $ & $ - (9 g_A^2 / 8 ) / 2m  $ \\
C6 & $ v \cdot  \stackrel{\leftarrow}{\nabla} 
\, \tilde\chi_+ \,   v \cdot \nabla $ & $ -3
 c_5 g_A^2  $ \\
\hline
D1 & $  i v_\rho \epsilon^{\rho\mu\nu\sigma}  S_\sigma 
    [ [ v \cdot \nabla , v\cdot u ] \, , [ \nabla_\mu  , u_\nu ] ] $ &
 $ (g_A^4/6) /2m  $ \\
\hline
\end{tabular}
$$
\end{table}

\addtocounter{table}{-1}

$ \, $
\begin{table}
\caption{continued, two derivatives }
$$
\begin{tabular}{|r|c|c|} \hline
i  & $\tilde{O}_i^{(4)}$ & $\delta_i$ \\ \hline
E1 & $  v_\rho \epsilon^{\rho\mu\nu\sigma} S_\sigma \, [ v \cdot
\nabla , \, [ v \cdot \nabla , \tilde{F}^+_{\mu\nu} ] ]  $  & 
$ -(1+c_6)g_A^2/12m  + (8 g_A^2 ) / 2m $  \\
E2 & $  v_\rho \epsilon^{\rho\mu\nu\sigma} S_\sigma \, [
\nabla^\kappa , \, [ \nabla_\kappa  , \tilde{F}^+_{\mu\nu} ] ]  $  & 
$ ( g_A^2/6 ) / 2m $  \\
E3 & $  v_\rho \epsilon^{\rho\mu\nu\sigma} S_\sigma \, \langle [ v \cdot
\nabla , [ v \cdot \nabla , {F}^+_{\mu\nu}]] \rangle   $   & 
$ (1+c_6+ 2c_7) g_A^2 /8m $  \\
E4 & $  v_\rho \epsilon^{\rho\mu\nu\sigma} S_\sigma \, v \cdot
\stackrel{\leftarrow}{\nabla} \tilde{F}^+_{\mu\nu} v \cdot \nabla  $ & 
$ (1+c_6) g_A^2/4m  - (16 g_A^2) / 2m $ \\
E5 & $  v_\rho \epsilon^{\rho\mu\nu\sigma} S_\sigma \, 
\stackrel{\leftarrow}{\nabla}_\kappa \tilde{F}^+_{\mu\nu} \,
\nabla^\kappa  $ & 
$  (2 g_A^2) / 2m $ \\
E6 & $  v_\rho \epsilon^{\rho\mu\nu\sigma} S_\sigma \, 
v \cdot \stackrel{\leftarrow}{\nabla}\,  \tilde{F}^+_{\mu\kappa}
v^\kappa \,
\nabla_\nu  $  + h.c.  & 
$  (-8 g_A^2) / 2m $ \\
E7 & $  v_\rho \epsilon^{\rho\mu\nu\sigma} S_\sigma \, v \cdot
\stackrel{\leftarrow}{\nabla} \langle {F}^+_{\mu\nu}  \rangle \, 
v \cdot \nabla  $ & 
$ -3(1+c_6+2c_7) g_A^2/8m   $ \\
\hline
F1 & $ i  v_\rho \epsilon^{\rho\mu\nu\sigma} S_\sigma \, 
 \langle  [ \nabla_\mu , \chi_+ ]  \rangle \, \nabla_\nu   $ + h.c.  & 
$ (3 g_A^2/4 ) / 2m $  \\
\hline
\end{tabular}
$$
\end{table}

\addtocounter{table}{-1}

$ \, $
\begin{table}
\caption{continued, three and more derivatives }
$$
\begin{tabular}{|r|c|c|} \hline
i  & $\tilde{O}_i^{(4)}$ & $\delta_i$ \\ \hline
A1 & $  [ v \cdot \nabla , \, [ v \cdot \nabla , v \cdot u ]] \,  
i S \cdot \nabla  $ + h.c.   &    $ (-4 g_A + g_A^3/3 ) / 2m $  \\
A2 & $  [ S \cdot \nabla , \, [ v \cdot \nabla , v \cdot u ]] \,  
i v \cdot \nabla  $ + h.c.   &    $ -4 g_A / 2m $  \\
A3 & $  [ v \cdot \nabla , \, [ v \cdot \nabla , S \cdot u ]] \,  
i v \cdot \nabla  $ + h.c.   &    $ - g_A^3 / 2m $  \\
A4 & $  [ v \cdot \nabla , \, [ \nabla_\mu ,  S \cdot u ]] \,  
i  \nabla^\mu  $ + h.c.   &    $ - (g_A^3/6)  / 2m $  \\
A5 & $  [ \nabla^\mu , \, [ \nabla_\mu ,  S \cdot u ]] \,  
i v \cdot  \nabla  $ + h.c.   &    $ - (g_A^3/2)  / 2m $  \\
A6 & $   S \cdot u  \, \,  i (v \cdot  \nabla)^3  $ + h.c.   &    $
  (10g_A^3/3)  / 2m $   \\
A7 & $  v \cdot  \stackrel{\leftarrow}{\nabla} \, \{ i S \cdot \nabla ,
v \cdot u \} \,  v \cdot \nabla  $  &  $ (8g_A - g_A^3) / 2m $  \\
A8 & $    \stackrel{\leftarrow}{\nabla}_\mu \, \{ i v \cdot \nabla ,
S \cdot u \} \,  \nabla^\mu  $  &  $  g_A^3 / 2m $  \\
A9 & $  v \cdot  \stackrel{\leftarrow}{\nabla} \, \{ i v \cdot \nabla ,
S \cdot u \} \,  v \cdot \nabla  $  &  $ ( - g_A^3) / 2m $  \\
\hline
B1 & $  (v \cdot  \stackrel{\leftarrow}{\nabla})^2 \, ( v \cdot
\nabla)^2 $  &  $ (-24g_A^2) / 2m $  \\
B2 & $   \stackrel{\leftarrow}{\nabla}_\mu  \, v \cdot
\stackrel{\leftarrow}{\nabla}  \,  v \cdot
\nabla \, \nabla^\mu  $  &  $ (9g_A^2/2) / 2m $  \\  
B3 & $   v \cdot \stackrel{\leftarrow}{\nabla} \, 
\stackrel{\leftarrow}{\nabla}_\mu  \,  
\nabla^\mu  \, v \cdot \nabla  $  &  $ (9g_A^2/2) / 2m $  \\  
\hline
\end{tabular}
$$
\end{table}

\renewcommand{\arraystretch}{1.0}

\vfill\eject


\newpage

\section*{Figures}

\vspace{3cm}

\begin{figure}[hbt]
   \vspace{0.5cm}
   \epsfysize=10cm
   \centerline{\epsffile{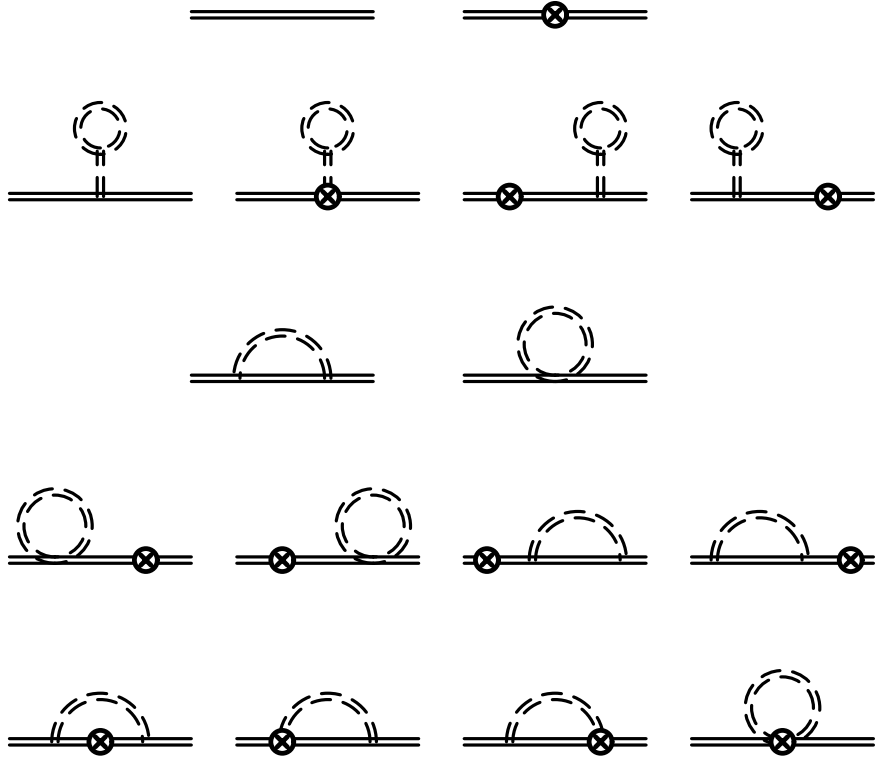}}
   \vspace{1.0cm}
   \centerline{\parbox{15cm}{\caption{\label{fig1}
  Contributions to the one--loop generating functional to order $q^4$.
  The solid (dashed) double lines denote the baryon
  (meson) propagator in the presence of external fields,
  respectively. The circle--cross  denotes an insertion from ${\cal
   L}_{\pi N}^{(2)}$. First line: Terms of order $\xi^0$. Second line:
  Terms of order $\xi$.
  Third line:  Irreducible graphs of order $\xi^2$ and $q^3$ in the chiral
   expansion. Fourth line:  Reducible graphs of order $\xi^2$ and
   $q^4$ in the chiral expansion. Fifth line: Irreducible graphs of
   order $\xi^2$ and $q^4$ in the chiral expansion.
  }}}
\end{figure}

\begin{figure}[htb]
   \vspace{0.5cm}
   \epsfysize=5cm
   \centerline{\epsffile{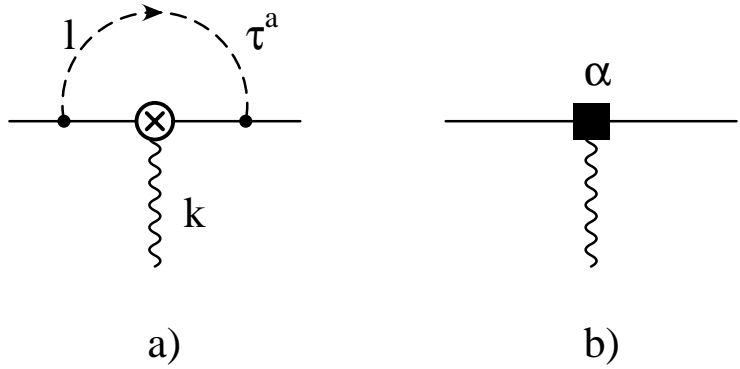}}
   \vspace{1.0cm}
   \centerline{\parbox{15cm}{\caption{\label{fig2}
   (a) One--loop graph contributing to the isoscalar magnetic moment.
   The dot and the circle--cross denote an insertion from the
   dimension one and two Lagrangian, in order. Nucleons, pions and 
   photons are depicted by solid, dashed and wiggly lines,
   respectively. (b) Fourth order counterterm with strength $\alpha$
   that renormalizes the divergence of the loop graph (a).
  }}}
\end{figure}

\begin{figure}[htb]
   \vspace{0.5cm}
   \epsfysize=12cm
   \centerline{\epsffile{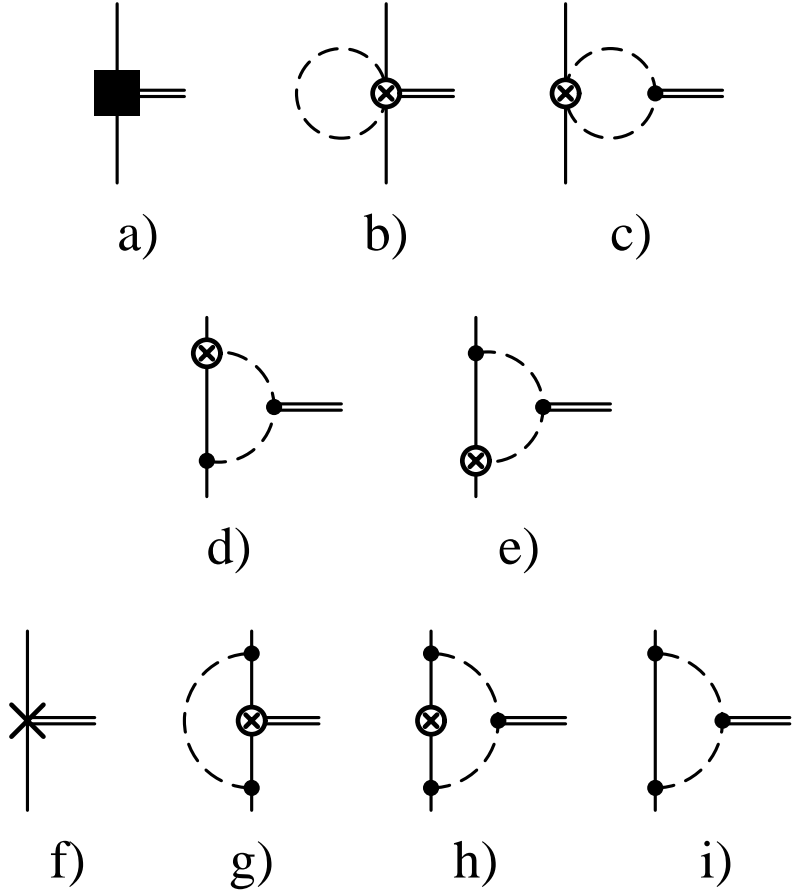}}
   \vspace{1.0cm}
   \centerline{\parbox{15cm}{\caption{\label{fig3}
   Divergent graphs contributing to the scalar form factor of the nucleon. (a)
   is a genuine counterterm diagram at order $q^4$. (b) and (c) are
   the tadpole, (d) and (e) the self--energy and (g)--(i) the eye
   graphs. (f) is a counterterm from the dimension three Lagrangian,
   as indicated by the cross, which also contributes to this order as
   explained in the text. Note also that graph (i) starts at order
   $q^3$ but picks up a fourth order piece due to the kinematics.
   The dot and the circle--cross denote an insertion from the
   dimension one and two Lagrangian, in order. Nucleons and pions  
   are depicted by solid and dashed  lines, respectively. The
   double--line denotes the scalar--isoscalar source.
  }}}
\end{figure}

\end{document}